\documentclass[prd,aps,twocolumn,superscriptaddress,tightenlines,nofootinbib,floatfix,preprintnumbers,10pt,longbibliography]{revtex4-1}
\usepackage{amsmath,amssymb}
\usepackage{bm}
\usepackage{comment}
\usepackage{graphicx}
\usepackage{color}
\usepackage{cancel}
\usepackage{tikz}
\usetikzlibrary{shapes.misc}
\usepackage{CJKutf8}

\definecolor{enrico}{rgb}{0.65, 0, 0.58}

\usepackage{hyperref}
\hypersetup{
    colorlinks=true,       
    linkcolor=blue,          
    citecolor=blue,        
    filecolor=blue,      
    urlcolor=blue           
}
\allowdisplaybreaks

\def\a{{\alpha}}

\def\d{{\delta}}
\def\D{{\Delta}}
\def\e{{\epsilon}}
\def\g{{\gamma}}

\def\L{{\Lambda}}

\def\s{{\sigma}}
\def\t{{\tau}}

\def\mc#1{{\mathcal #1}}

\def\eqref#1{{(\ref{#1})}}

\newcommand{\glasgow}{
	School of Physics and Astronomy, 
    University of Glasgow, 
    Glasgow G12 8QQ, UK
	}
\newcommand{\jlab}{
	Theory Center, 
	Thomas Jefferson National Accelerator Facility, 
	Newport News, VA 23606, USA
	}
\newcommand{\jlabcomp}{
	Scientific Computing Group, 
	Thomas Jefferson National Accelerator Facility, 
	Newport News, VA 23606, USA
	}
\newcommand{\julich}{
	Institut f\"{u}r Kernphysik and Institute for Advanced Simulation,
	Forschungszentrum J\"{u}lich, 54245 J\"{u}lich Germany
	}
\newcommand{\lblnsd}{
	Nuclear Science Division,
    Lawrence Berkeley National Laboratory,
	Berkeley, CA 94720, USA
	}
\newcommand{\lblnersc}{
	NERSC,
    Lawrence Berkeley National Laboratory,
	Berkeley, CA 94720, USA
	}
\newcommand{\llnl}{
	Nuclear and Chemical Sciences Division,
	Lawrence Livermore National Laboratory,
	Livermore, CA 94550, USA
	}
\newcommand{\nvidia}{
    NVIDIA Corporation,
    2701 San Tomas Expressway, Santa Clara, CA 95050, USA
    }
\newcommand{\riken}{
	RIKEN-BNL Research Center, 
	Brookhaven National Laboratory, 
	Upton, NY 11973, USA
	}
\newcommand{\rutgers}{
	New High Energy Theory Center and Department of Physics and Astronomy,
	Rutgers,
	The State University of New Jersey, Piscataway, NJ 08854, USA
	}
\newcommand{\ucb}{
	Department of Physics,
	University of California,
	Berkeley, CA 94720, USA
	}
\newcommand{\wm}{
	Department of Physics,
	The College of William \& Mary,
	Williamsburg, VA 23187, USA
	}

\newcount\hour \newcount\hourminute \newcount\minute
\hour=\time \divide \hour by 60
\hourminute=\hour \multiply \hourminute by 60
\minute=\time \advance \minute by -\hourminute

\begin{document}
\preprint{LLNL-JRNL-719521, RBRC-1227}

\title{M\"{o}bius domain-wall fermions on gradient-flowed dynamical HISQ ensembles}

\author{Evan~Berkowitz}
\affiliation{\julich}
\affiliation{\llnl}

\author{Chris~Bouchard}
\affiliation{\glasgow}
\affiliation{\wm}

\author{Chia~Cheng~Chang \begin{CJK*}{UTF8}{bsmi}(張家丞)\end{CJK*}}
\affiliation{\lblnsd}

\author{M.~A.~Clark}
\affiliation{\nvidia}

\author{B\'{a}lint~Jo\'{o}}
\affiliation{\jlabcomp}

\author{Thorsten~Kurth}
\affiliation{\lblnersc}

\author{Christopher~Monahan}
\affiliation{\rutgers}

\author{Amy~Nicholson}
\affiliation{\ucb}
\affiliation{\lblnsd}

\author{Kostas~Orginos}
\affiliation{\wm}
\affiliation{\jlab}

\author{Enrico~Rinaldi}
\affiliation{\riken}
\affiliation{\llnl}

\author{Pavlos~Vranas}
\affiliation{\llnl}
\affiliation{\lblnsd}

\author{Andr\'{e}~Walker-Loud}
\affiliation{\lblnsd}
\affiliation{\llnl}


\begin{abstract}
We report on salient features of a mixed lattice QCD action using valence M\"{o}bius domain-wall fermions solved on the dynamical $N_f=2+1+1$ HISQ ensembles generated by the MILC Collaboration.
The approximate chiral symmetry properties of the valence fermions are shown to be significantly improved by utilizing the gradient-flow scheme to first smear the HISQ configurations.
The greater numerical cost of the M\"{o}bius domain-wall inversions is mitigated by the highly efficient QUDA library optimized for NVIDIA GPU accelerated compute nodes.
We have created an interface to this optimized QUDA solver in Chroma.
We provide tuned parameters of the action and performance of QUDA using ensembles with the lattice spacings $a \simeq \{0.15, 0.12, 0.09\}$~fm and pion masses $m_\pi \simeq \{310, 220,130\}$~MeV.
We have additionally generated two new ensembles with $a\sim0.12$~fm and $m_\pi\sim\{400, 350\}$~MeV.
With a fixed flow-time of $t_{gf}=1$ in lattice units, the residual chiral symmetry breaking of the valence fermions is kept below 10\% of the light quark mass on all ensembles, $m_{res} \lesssim 0.1\times m_l$, with moderate values of the fifth dimension $L_5$ and a domain-wall height $M_5 \leq 1.3$.
As a benchmark calculation, we perform a continuum, infinite volume, physical pion and kaon mass extrapolation of $F_{K^\pm}/F_{\pi^\pm}$ and demonstrate our results are independent of flow-time, and consistent with the FLAG determination of this quantity at the level of less than one standard deviation.
\end{abstract}
\maketitle



%
\section{Introduction\label{sec:intro}}
QCD (Quantum Chromodynamics)~\cite{Fritzsch:1972jv,Fritzsch:1973pi} is the fundamental theory of the strong interaction, and one of the three gauge theories of the SM (Standard Model) of particle physics.
QCD encodes the interactions between quarks and gluons, the constituents of strongly interacting matter, which both carry \textit{color charges} of QCD.
At short distances, the quarks and gluons perturbatively  interact with a coupling strength that runs to zero in the UV (ultraviolet) limit~\cite{Gross:1973id,Politzer:1973fx}.
Conversely, at long distance/low energy, the IR (infrared) regime, the coupling becomes $\mathcal{O}(1)$, and QCD becomes a strongly coupled theory.
Consequently, the quarks and gluons are confined into the \textit{colorless} hadrons we observe in nature, such as the proton, neutron, pions, etc.
In order to compute properties of nucleons, nuclei, and other strongly interacting matter directly from QCD, we must therefore use a nonperturbative regularization scheme.

Asymptotic freedom, the property in which the gauge coupling becomes perturbative in the UV, makes the theory perfectly amenable to a numerical approach.
QCD can be constructed on a discrete, Euclidean spacetime lattice, with a technique known as LQCD (lattice QCD).
As the discretization scale is made sufficiently fine and the coupling becomes perturbative, the lattice action can be matched onto the continuum action to a desired order in perturbation theory.
To aid the matching, EFT (Effective Field Theory)~\cite{Weinberg:1978kz} can be used to perform an expansion of the lattice action in powers of the discretization scale, typically denoted $a$, which is referred to as the Symanzik expansion~\cite{Symanzik:1983dc,Symanzik:1983gh}.
There are many different choices for constructing the discretized action, each of which corresponds to a different lattice action.
As the continuum limit is taken, the difference between these lattice actions vanishes as the only dimension-4 operators allowed by the symmetries are those of QCD: the discretization effects, which include Lorentz violating interactions, are all described by \textit{irrelevant} operators in the Symanzik expansion.
An important test of this universality is to perform calculations of various physical quantities, with different lattice actions, and show consistency between them in the continuum limit.
This is now routinely done for mesonic quantities and reviewed every two to three years by the FLAG Working Group, with the latest review in Ref.~\cite{Aoki:2016frl}.

Lattice gauge theory began with the formulation of gauge fields on a spacetime lattice as originally proposed by Wilson~\cite{Wilson:1974sk}.
The inclusion of fermions presents further challenges.
The \textit{naive} discretization of the fermion action leads to the fermion doubling problem, in which there are $2^D$ fermions in $D$ dimensions for each fermion field implemented.
These doublers arise from the periodicity of the lattice action in momentum space and the single derivative in the Dirac equation.
Wilson proposed the original method, now known as the Wilson fermion action, to remove these doublers by adding an irrelevant operator to the action which provides an additive mass to the doublers which scales as $1/a$.
This irrelevant operator breaks chiral symmetry and requires fine-tuning the bare fermion mass to simulate a theory with light fermions, such as QCD with light $up$ and $down$ quarks.
Despite (or because of) its simplicity, the Wilson fermion action is still one of the most popular in use.
These days, the leading $\mathcal{O}(a)$ discretization corrections are removed perturbatively or nonperturbatively through an additional dimension-5 operator, the clover operator $c_{SW}\, a \bar{q} \s_{\mu\nu} G_{\mu\nu} q$, in what is known as the Wilson-Clover or \textit{Clover} fermion action.
The parameter $c_{SW}$ is the Sheikholeslami-Wohlert coefficient~\cite{Sheikholeslami:1985ij} which can be tuned to remove the $\mathcal{O}(a)$ discretization effects from correlation functions.
The idea has also been extended to twisted mass Wilson fermions~\cite{Frezzotti:2000nk}, in which a complex quark-mass term is used, allowing for automatic $\mc{O}(a)$ improvement of physical observables provided the theory is computed at \textit{maximal twist}~\cite{Frezzotti:2003ni}.

Another common lattice action is known as the Kogut-Susskind or \textit{staggered} fermion action~\cite{Kogut:1974ag,Susskind:1976jm}.
This action reduces the number of fermion doublers by exploiting a symmetry of the \textit{naive} fermion action.
A suitable spacetime-dependent phase rotation of the fermion fields allows for the Dirac equation to be diagonalized, thereby reducing the number of doublers from 16 to 4, in four spacetime dimensions.
To perform numerical simulations with just one or two light fermion flavors, a fourth or square root of the fermion determinant is used~\cite{Marinari:1981qf}.
This rooting leads to nonlocal interactions at finite lattice spacing~\cite{Bernard:2006vv,Bernard:2006ee,Creutz:2007yg}; however, perturbation theory~\cite{Bernard:2004ab,Bernard:2006zw}, the renormalization group~\cite{Shamir:2004zc,Shamir:2006nj,Bernard:2007ma}, and numerical simulations~\cite{Durr:2004ta,Durr:2006ze,Hasenfratz:2006nw}, have been used to argue that these nonlocal effects vanish in the continuum limit.
While this has not been proven nonperturbatively, some of the potential sicknesses of the theory can be shown to be the same as those of \textit{partially quenched} lattice QCD~\cite{Bernard:2007eh}, which we will discuss briefly in short order.
While not universally accepted, all numerical evidence suggests that rooted-staggered LQCD is in the same universality class as QCD as the continuum limit is taken~\cite{Sharpe:2006re,Kronfeld:2007ek,Bazavov:2009bb,Aoki:2016frl}.

Determining a nonperturbative regulator that both preserves chiral symmetry and has the correct number of light degrees of freedom is challenging.  It has been shown that in four spacetime dimensions, one cannot simultaneously have all four of the conditions: chiral symmetry, ultralocal action, undoubled fermions, and the correct continuum limit.  This is known as the Nielsen-Ninomiya no-go theorem~\cite{Nielsen:1980rz,Nielsen:1981xu,Nielsen:1981hk}.
However, one can extend the definition of chiral symmetry at finite lattice spacing: if the lattice Dirac operator, $D$, satisfies the Ginsparg-Wilson relation~\cite{Ginsparg:1981bj}
\begin{equation}\label{eq:gw}
\left\{ \g_5, D \right\} = a D \g_5 D\, ,
\end{equation}
it will respect chiral symmetry even at finite lattice spacing~\cite{Luscher:1998pqa}.
One consequence is the theory will be automatically $\mathcal{O}(a)$ improved as the only nontrivial dimension-5 operator that cannot be removed through field redefinitions and equations of motion is the clover operator, which explicitly breaks chiral symmetry and is thus not allowed.
There are two lattice actions which satisfy the Ginsparg-Wilson relation: the DW (domain-wall) fermion action~\cite{Kaplan:1992bt,Shamir:1993zy,Furman:1994ky} and the overlap fermion action~\cite{Narayanan:1993sk,Narayanan:1993ss,Narayanan:1994gw}.  The DW fermion action is formulated with a finite fifth dimension of extent $L_5$, where the left and right chiral modes are bound to opposite ends of the fifth dimension.
The gluon action is a trivial copy of the 4D action on each fifth-dimensional slice with unit link variable between the slices, and so the fermions have only a simple kinetic action in the fifth dimension.
At finite $L_5$, the left and right modes have a nonvanishing overlap due to fermion modes which propagate into the fifth dimension.
The massive modes decay exponentially in the fifth dimension, while the fermion zero modes have only a power-law falloff.
This small overlap leads to a small, residual breaking of chiral symmetry at finite $L_5$, characterized by a quantity known as $m_{res}$.
The overlap fermion action can be shown to be equivalent to the domain-wall action as $L_5 \rightarrow \infty$~\cite{Borici:1999zw,Borici:1999da} and respects chiral symmetry to a desired numerical precision.

The numerical cost of generating lattice ensembles with domain-wall and overlap actions is 1 or more orders of magnitude greater than the cost of generating ensembles with Wilson-type or staggered fermion actions~\cite{Kennedy:2004ae}.
This has led to interest in, and the development of, mixed lattice actions or MA (mixed-actions)~\cite{Bar:2002nr}, in which the valence and sea-quark lattice actions are not the same at finite lattice spacing.
In the most common MALQCD calculations, the dynamical sea-quark action is generated with a numerically less expensive discretization scheme, such as staggered- or Wilson-type fermions, while the valence-quark action, which is used to construct correlation functions, is implemented with domain-wall or overlap fermions, thus retaining the full chiral symmetry in the valence sector.
The first implementation of a MALQCD calculation was performed by the LHP Collaboration~\cite{Renner:2004ck} utilizing DW fermions on the publicly available asqtad ($a^2$ tadpole improved)~\cite{Orginos:1998ue,Orginos:1999cr} rooted staggered ensembles generated by the MILC Collaboration~\cite{Bernard:2001av,Bazavov:2009bb}.
A number of important  results  were obtained with this particular MALQCD setup, including
the first dynamical calculation of the nucleon axial charge with light pion masses~\cite{Edwards:2005ym} and more general nucleon structure~\cite{Hagler:2007xi,Bratt:2010jn},
the first dynamical calculation of two-nucleon elastic scattering~\cite{Beane:2006mx},
a precise calculation of the $I=2\ \pi\pi$ scattering length~\cite{Beane:2007xs},
a detailed study of the quark-mass dependence of the light baryon spectrum~\cite{WalkerLoud:2008bp},
a calculation of the kaon bag parameter with fully controlled uncertainties~\cite{Aubin:2009jh},
and many more.

The predominant reason for the success of these MALQCD calculations is the good chiral symmetry properties of the DW action, which significantly suppresses chiral symmetry breaking from the staggered sea fermions and discretization effects.
EFT can be used to understand the salient features of such MALQCD calculations. $\chi$PT (Chiral Perturbation Theory)~\cite{Langacker:1973hh,Gasser:1983yg,Leutwyler:1993iq} can be extended to incorporate discretization effects into the analytic formulae describing the quark-mass dependence of various hadronic quantities.
The procedure is to first construct the local Symanzik action for a given lattice action and then to use spurion analysis to construct all operators in the low-energy EFT describing such a lattice action, including contributions from higher-dimension operators~\cite{Sharpe:1998xm}.
The MAEFT~\cite{Bar:2003mh} for DW valence fermions on dynamical rooted staggered fermions is well developed~\cite{Bar:2005tu,Tiburzi:2005is,Chen:2005ab,Chen:2006wf,Orginos:2007tw,Jiang:2007sn,Chen:2007ug,Chen:2009su}.
The use of valence fermions which respect chiral symmetry leads to a universal form of the MAEFT extrapolation formulae at NLO (next-to-leading order) in the dual quark-mass and lattice spacing expansions~\cite{Chen:2006wf,Chen:2007ug}.
This universal behavior follows from the suppression of chiral symmetry breaking discretization effects from the sea sector when constructing correlation functions from valence fermions.
Further, quantities which are protected by chiral symmetry are free of new LECs (low-energy constants) at NLO provided on-shell renormalized quantities are used in the extrapolation formulae~\cite{Chen:2005ab,Chen:2006wf}.
This universality allows for the derivation of NLO MAEFT formula directly from their PQ$\chi$PT (partially quenched $\chi$PT)~\cite{Bernard:1993sv,Sharpe:2000bc,Sharpe:2001fh,Chen:2001yi,Sharpe:2003vy,Arndt:2003ww,WalkerLoud:2004hf,Bernard:2010qc,Bernard:2013kwa} counterparts, provided they are known~\cite{Beane:2002vq,Beane:2002np,Arndt:2003we,Arndt:2003vd,Tiburzi:2004rh,Tiburzi:2004kd,Tiburzi:2005na,OConnell:2005mfp}.
MALQCD calculations with DW valence quarks on the asqtad rooted staggered ensembles have been stress tested through a comparison of quantities which are directly sensitive to the unitarity violations present in MALQCD calculations, in particular the $a_0$ meson correlation function~\cite{Prelovsek:2005rf,Aubin:2008wk}.
There are a few other MA constructions that have been tested, but only a few others that are actively used.
The HPQCD Collaboration utilizes HISQ (highly improved staggered quark) valence fermions on the asqtad ensembles; for example, see Refs.~\cite{Na:2015kha,Donald:2013sra}.
The $\chi$QCD Collaboration utilizes overlap valence fermions on the dynamical $N_f=2+1$ domain-wall ensembles~\cite{Li:2010pw,Lujan:2012wg,Gong:2013vja} generated by the RBC/UKQCD Collaboration~\cite{Allton:2007hx,Aoki:2010dy}.
The work in Refs.~\cite{Basak:2012py,Basak:2013oya,Basak:2014kma,Mathur:2016hsm} uses valence overlap fermions on the $N_f=2+1+1$ HISQ ensembles~\cite{Bhattacharya:2013ehc}.
The PNDME Collaboration has utilized clover improved valence fermions on the $N_f=2+1+1$ HISQ ensembles~\cite{Bhattacharya:2013ehc,Bhattacharya:2015wna}.
While this MA choice is economical, it does not benefit from the suppression of chiral symmetry breaking discretization effects as with the DW on asqtad or overlap on DW MALQCD calculations.

Given the successes described above, MALQCD provides an economical means of performing LQCD calculations in which chiral symmetry breaking effects are highly suppressed by utilizing a valence fermion action that respects chiral symmetry in combination with a set of LQCD ensembles that do not, but are less numerically expensive to generate.
In this article, we motivate a new MALQCD action and present numerical evidence for salient features of the action.

%
\section{M\"{o}bius Domain-Wall fermions on gradient-flowed HISQ ensembles\label{sec:mdwf_hisq}}

\begin{table*}
\caption{\label{tab:hisq} The HISQ ensembles used in this work and planned for future MALQCD calculations.  In addition to the pion mass and lattice spacing, we list the number of configurations used in the present work, $N_{cfg}$ as well as the Monte Carlo time, $\D\t_{MC}$, by which the configurations were separated in this work.  The short name, introduced in Ref.~\cite{Bhattacharya:2015wna}, is for brevity. The last two HISQ ensembles were generated at LLNL targeting heavier pion masses to test the radius of convergence of the chiral extrapolation in future MALQCD calculations.}
\begin{ruledtabular}
\begin{tabular}{llcccccccc}
Short& Ensemble& $am_\pi^{HISQ-5}$& $am_{ss}^{HISQ-5}$& Volume& $\sim a$ & $\sim m_\pi$& $m_\pi L$& $N_{cfg}$& $\D\t_{MC}$ \\
name&&&&&                  [fm]& [MeV]\\
\hline
a15m310& l1648f211b580m013m065m838a& 0.23646(17)& 0.51858(17)&
    $16^3\times48$& 0.15& 310& 3.78& 196& 50 \\
a12m310& l2464f211b600m0102m0509m635a& 0.18931(10)& 0.41818(10)&
    $24^3\times64$& 0.12& 310& 4.54& 199& 25 \\
a09m310& l3296f211b630m0074m037m440e& 0.14066(13)& 0.31133(12)&
    $32^3\times96$& 0.09& 310& 4.50& 196& 24 \\
a15m220& l2448f211b580m0064m0640m828a& 0.16612(08)& 0.51237(10)&
    $24^3\times48$& 0.15& 220& 3.99& 199& 25 \\
a12m220& l3264f211b600m00507m0507m628a& 0.13407(06)& 0.41559(07)&
    $32^3\times64$& 0.12& 220& 4.29& 199& 25 \\
a09m220& l4896f211b630m00363m0363m430a& 0.09849(07)& 0.30667(07)&
    $48^3\times96$& 0.09& 220& 4.73& --& -- \\
a15m130& l3248f211b580m00235m0647m831a& 0.10161(06)& 0.51427(05)&
    $32^3\times48$& 0.15& 130& 3.25& --& -- \\
a12m130& l4864f211b600m00184m0507m628a& 0.08153(04)& 0.41475(05)&
    $48^3\times64$& 0.12& 130& 3.91& --& -- \\
\hline
a12m400& l2464f211b600m0170m0509m635a& 0.24398(12) & 0.41970(12) & $24^3\times64$& 0.12& 400& 5.86& --& -- \\
a12m350& l2464f211b600m0130m0509m635a& 0.21376(13) & 0.41923(13) & $24^3\times64$& 0.12& 350& 5.13& --& -- \\
\end{tabular}
\end{ruledtabular}
\end{table*}

Present-day LQCD calculations for mesonic quantities are performed with multiple lattice spacings, multiple volumes and physical pion masses, allowing for complete control over all LQCD systematics, see Ref.~\cite{Aoki:2016frl} for many examples.
The simplest single baryon properties are also computed with multiple lattice spacings/volumes and near-physical and sometimes physical pion masses~\cite{Durr:2011mp,Durr:2015dna,Yang:2015uis,Sufian:2016pex}, including the first calculation of the nucleon axial charge with both physical pion masses and a continuum limit~\cite{Bhattacharya:2016zcn} and isospin violating corrections~\cite{Borsanyi:2013lga,Borsanyi:2014jba,Bhattacharya:2016zcn,Brantley:2016our}.
If one is interested in a set of ensembles allowing for this much control over LQCD systematics, there are only two such sets publicly available, both of which are generated and provided by the MILC Collaboration:
the $N_f=2+1$ asqtad ensembles~\cite{Bazavov:2009bb}
and the $N_f=2+1+1$ HISQ~\cite{Follana:2006rc} ensembles generated more recently~\cite{Bazavov:2010ru,Bazavov:2012xda}.
The HISQ ensembles have taste splittings in the pseudoscalar sector that are one generation finer in discretization~\cite{Bazavov:2012xda}, such that the $a\sim0.15$~fm HISQ ensemble taste violations are similar in size to the $a\sim0.12$~fm asqtad ensembles.
There is a vast set of HISQ ensembles with $130 \lesssim m_\pi \lesssim 310$~MeV, strange and charm quark masses tuned near their physical values and lattice spacings of $a\sim\{0.15,0.12,0.09,0.06,0.042,0.03\}$~fm, including multiple spatial volumes and lighter than physical strange quark masses.
In addition to the publicly available HISQ ensembles, we have generated two additional sets at $a\sim0.12$~fm and $m_{\pi}\approx 350,\ 400$~MeV with fixed volume in lattice units such that $m_{\pi}L\geq5.1$.
In Table~\ref{tab:hisq}, we list the HISQ ensembles utilized in the present work as well as ensembles for which we have tuned the MDWF parameters for future work.

Given the great success of the MA DW fermion on asqtad LQCD calculations~\cite{Edwards:2005ym,Hagler:2007xi,Bratt:2010jn,Beane:2006mx,Beane:2007xs,WalkerLoud:2008bp,Aubin:2009jh}, we have chosen to use DW fermions for the present MALQCD calculations as well.
In the present work, we have chosen to use the MDWF (M\"{o}bius DW fermion) action~\cite{Brower:2004xi,Brower:2005qw,Brower:2012vk} which offers reduced residual chiral symmetry breaking at fixed fifth-dimensional extent, $L_5$.
With the introduction of two new parameters, $b_5$ and $c_5$, the M\"{o}bius kernel can be smoothly interpolated between the Shamir~\cite{Shamir:1993zy} and the Neuberger/Bori\c{c}i~\cite{Neuberger:1997bg,Neuberger:1997fp,Borici:1999zw,Borici:1999da} kernels.
Following Ref.~\cite{Brower:2012vk}, the M\"{o}bius kernel can be expressed as
\begin{equation}
D^\textrm{M\"{o}bius}(M_5) =
    \frac{(b_5 + c_5)D^\textrm{Wilson}(M_5)}
    {2 +  (b_5 - c_5) D^\textrm{Wilson}(M_5)}\, .
\end{equation}
Alternatives include a polar decomposition to the sign function~\cite{Vranas:1997da,Kikukawa:1999sy,Edwards:2000qv} or other methods of approximating the sign function~\cite{Kennedy:2006ax}.
In this work, we have always chosen values of $b_5$ and $c_5$ with the constraint $b_5-c_5 = 1$, such that the M\"{o}bius kernel is a rescaled version of the Shamir kernel
\begin{equation}
D^\textrm{M\"{o}bius}(M_5) =
    \frac{\alpha D^\textrm{Wilson}(M_5)}
    {2 + D^\textrm{Wilson}(M_5)}
\equiv \alpha D^\textrm{Shamir}(M_5) .
\end{equation}
It was demonstrated in Ref.~\cite{Brower:2012vk} that this rescaling factor, $\alpha$, exponentially enhances the suppression of residual chiral symmetry breaking as
\begin{equation}
m_{res} \sim e^{-\alpha L_5}\, ,
\end{equation}
provided the action is in a regime where these exponentially damped terms are the dominant contribution to $m_{res}$ and
$\alpha$ is not too large, but of the order $\alpha \sim 2-4$.
With the constraint $b_5-c_5=1$, the rescaling factor is given by $\alpha=b_5+c_5$.

%
\section{Gradient-flow smearing \label{sec:gflow}}
From the DW on asqtad action~\cite{Edwards:2004sx}, it is known that the asqtad gauge fields required additional levels of smearing to reduce the residual chiral symmetry breaking.
For that action, HYP smearing~\cite{Hasenfratz:2001hp,DeGrand:2002vu,DeGrand:2003in,Durr:2004as} was utilized for this purpose.
In this work, we choose to investigate the use of the gradient flow~\cite{Narayanan:2006rf,Luscher:2011bx,Luscher:2013cpa} as a smearing method.
The gradient flow is a nonperturbative, classical evolution of the original fields in a new parameter, the \textit{flow-time}, that drives those fields toward a classical minimum. In real space, this corresponds to smearing out the degrees of freedom through an infinitesimal \textit{stout-smearing} procedure \cite{Morningstar:2003gk}.

Gradient flow smearing introduces a new scale, of the order $l_{gf} \sim \sqrt{8 t_{gf}}\, a$, where $t_{gf}$ is the (dimensionless) flow-time.
Correlation functions depend upon this new scale, which can serve as a nonperturbative, rotationally invariant UV regulator that provides the possibility for improved renormalization procedures for various LQCD matrix elements~\cite{DelDebbio:2013zaa,Suzuki:2013gza,Monahan:2013lwa,Luscher:2014kea,Endo:2015iea,Monahan:2015lha,Monahan:2016bvm}.
Here, however, we are interested in the gradient flow as a smearing algorithm \cite{Luscher:2010iy,Lohmayer:2011si}.

To ensure that the continuum limit of LQCD matrix elements is free of any flow-time dependence, one must use a fixed flow-time in lattice units such that all flow-time dependence extrapolates to zero as the continuum limit is taken.

In this work, we have found that moderate values of the flow-time allow for a reduction of the residual chiral symmetry breaking such that $m_{res} < 0.1 \times m_l^{dwf}$ for moderate values of $L_5$.
The resulting flow-time dependence of $m_{res}$ at fixed pion mass demonstrates that the gradient-flow highly suppresses the zero-mode contributions to $m_{res}$, such that an exponential dependence of $m_{res}$ on $L_5$ is recovered.
Further, we have observed that gradient flow smearing has allowed us to use small values of the DW height, with $M_5 \leq 1.3$ on all ensembles used in this work.
This is important because with the larger values of $M_5$ used in the DW on asqtad calculations, there was strong contamination of the UV modes with an oscillatory time behavior, modes which are known to decouple as $M_5\rightarrow1$~\cite{Syritsyn:2007mp}.
With the values of $M_5$ used in this work, there is no discernible contamination from these modes at larger flow-times.

We finally settled on a gradient flow-time of $t_{gf}=1.0$, which provided significant suppression of residual chiral symmetry breaking without introducing a large flow-time length scale.
In the next section, we present detailed calculations showing the flow-time dependence of various quantities.
This action has been used to compute the $\pi^-\rightarrow\pi^+$ matrix element relevant for neutrinoless double beta decay~\cite{Nicholson:2016byl} and also to perform an exploratory calculation of an improved method of computing hadronic matrix elements~\cite{Bouchard:2016heu} and an application to $g_A$~\cite{Berkowitz:2017gql}.

%
\subsection{Tuning the action \label{sec:action_params}}

Before showing results, we describe how to match the valence MDWF action and the HISQ action. With a given flow-time, our general algorithm for choosing values of the MDWF action parameters is:
\begin{enumerate}
\item For a fixed value of $L_5$, optimize $M_5$ to minimize the resulting value of $m_{res}$.

\item Vary the values of $L_5$, $b_5$ and $c_5$ under the constraints $b_5-c_5=1$ and $m_{res} \leq 0.1 m_l^{dwf}$ while minimizing $L_5$.

\item Tune $m_l^{dwf}$ and $m_s^{dwf}$ such that $m_\pi^{dwf} \simeq m_\pi^{HISQ-5}$ and $m_{ss}^{dwf} \simeq m_{ss}^{HISQ-5}$ within $\mc{O}(2\%)$ or less where $HISQ-5$ denotes the taste-5 pseudoscalar mass of the dynamical HISQ action and $m_{ss}$ is the mass of the connected $\bar{s}\g_5 s$ pseudoscalar meson.

\end{enumerate}
This procedure required just a few iterations to converge to the desired results.
For this work, we have used the definition of $m_{res}$ from the Shamir kernel as the residual chiral symmetry breaking between Shamir and M\"{o}bius becomes the same in the continuum limit~\cite{Brower:2012vk},
\begin{equation}
m_{res}(t) = \frac{\sum_{\mathbf{x}} \langle \bar{Q}(t,\mathbf{x}) \g_5 Q(t,\mathbf{x})\, \bar{q}(0,\mathbf{0})\g_5 q(0,\mathbf{0}))\rangle}
    {\sum_{\mathbf{x}} \langle \bar{q}(t,\mathbf{x}) \g_5 q(t,\mathbf{x})\, \bar{q}(0,\mathbf{0})\g_5 q(0,\mathbf{0}))\rangle}\, ,
\label{eq:mres}
\end{equation}
where $Q$ is a quark field in the midpoint of the fifth dimension and $q$ is a quark field bound to the domain wall.

In Table~\ref{tab:tunedMDWF}, we list the resulting MDWF parameters at the chosen gradient flow-time of $t_{gf}=1$.
These parameters were used in Refs.~\cite{Nicholson:2016byl,Berkowitz:2017gql}.
\begin{table}
\caption{\label{tab:tunedMDWF} Tuned MDWF parameters for our MALQCD calculations. Some of the ensembles are used for example in Refs.~\cite{Nicholson:2016byl,Berkowitz:2017gql}.
}
\begin{ruledtabular}
\begin{tabular}{cccccccc}
Dnsemble& $M_5$& $L_5$& $b_5$& $c_5$& $t_{gf}$&
    $am_l^{mdwf}$& $am_s^{mdwf}$\\
\hline
a12m400& 1.2& \phantom{1}8& 1.25& 0.25& 1.0& 0.02190&  0.0693 \\
a12m350& 1.2& \phantom{1}8& 1.25& 0.25& 1.0& 0.01660&  0.0693 \\
a15m310& 1.3& 12& 1.50& 0.50& 1.0& 0.01580&  0.0902 \\
a12m310& 1.2& \phantom{1}8&  1.25& 0.25& 1.0& 0.01260&  0.0693 \\
a09m310& 1.1& \phantom{1}6&  1.25& 0.25& 1.0& 0.00951& 0.0491 \\
a15m220& 1.3& 16& 1.75& 0.75& 1.0& 0.00712& 0.0902 \\
a12m220& 1.2& 12& 1.50& 0.50& 1.0& 0.00600& 0.0693 \\
a09m220& 1.1& \phantom{1}8&  1.25& 0.25& 1.0& 0.00449& 0.0491 \\
a15m130& 1.3& 24& 2.25& 1.25& 1.0& 0.00216& 0.0902 \\
a12m130& 1.2& 20& 2.00& 1.00& 1.0& 0.00195& 0.0693 \\
\end{tabular}
\end{ruledtabular}
\end{table}

%
\section{Flow-time dependence of various quantities\label{sec:flowtime}}

\begin{figure}
\includegraphics[width=\columnwidth]{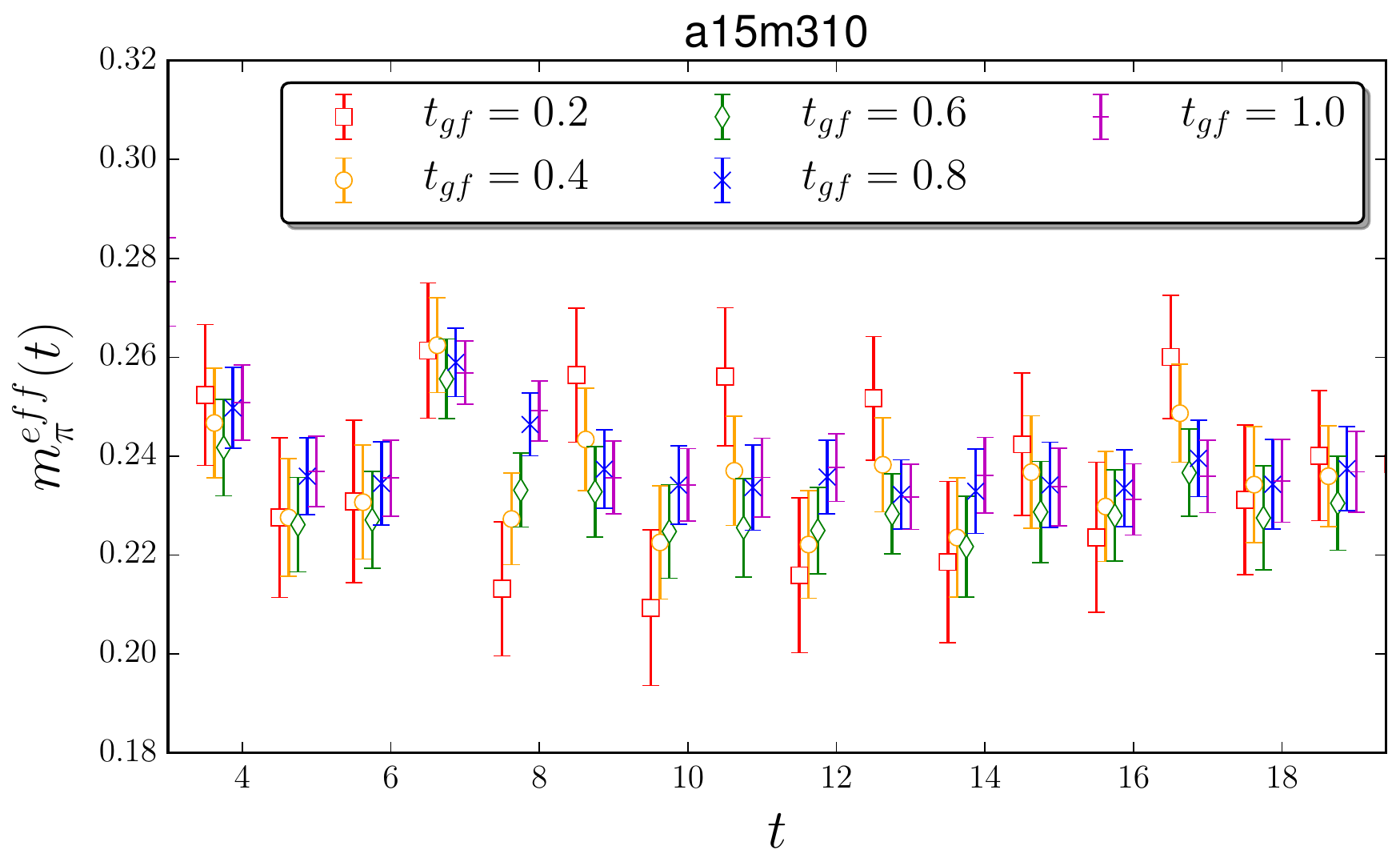}\vspace{5pt}
\includegraphics[width=\columnwidth]{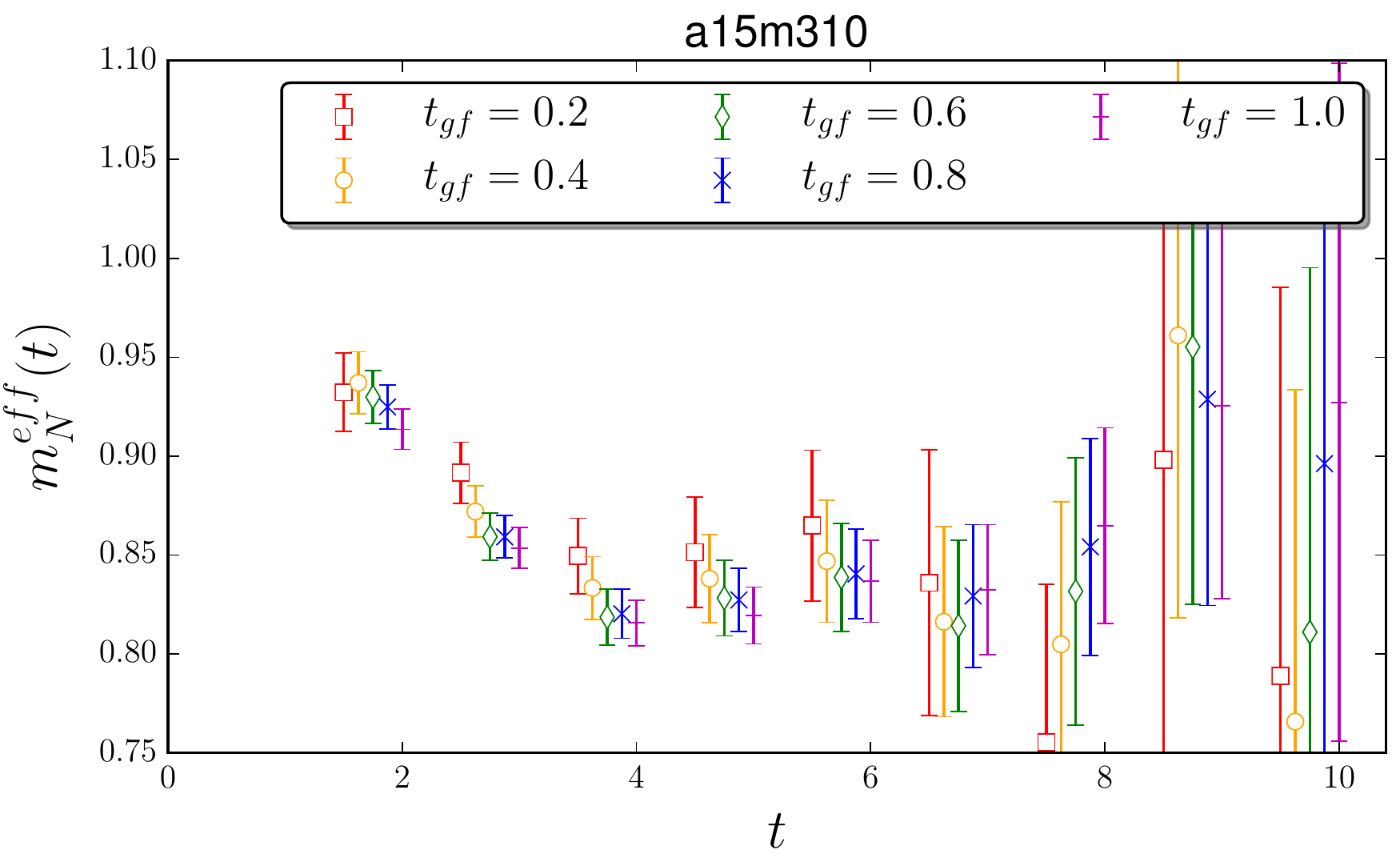}
\caption{\label{fig:mpi_tgf} Effective mass of the pion (top) and proton (bottom) as a function of the Euclidean time $t$, at different flow-times on the a15m310 ensemble.
The different flow-time values are slightly shifted horizontally for visual clarity.}
\end{figure}

To study the efficacy of this action, we compute the flow-time dependence of various quantities. In the next section we will show that the continuum limits of various ratios of physical quantities are flow-time independent.
In order to test the flow-time dependence, we tune the input quark masses to hold the pion mass and the connected $ss$ pseudoscalar meson masses fixed within $\mc{O}(2\%)$.
In the Appendix (Table~\ref{tab:flowtune}), we list the tuned values of the input quark masses for various flow-times on the ensembles used in this work.
We also list the resulting values of the plaquette, $m_{res}$, and the values of $Z_A$ determined as described below.
In Fig.~\ref{fig:mpi_tgf}, we show the effective masses of the pion and nucleon, respectively, on the a15m310 ensemble for all flow-times. We observe that the contamination from oscillatory modes is suppressed at larger flow-times.

From the input quark masses used at fixed pseudoscalar masses, and the average values of the plaquettes, one observes a substantial flow-time dependence of UV quantities.
This is expected as the gradient flow smearing filters out the UV modes of the gauge fields.
It is important to check the flow-time dependence of hadronic quantities
and verify the continuum limit is flow-time independent.
This can easily be checked with ratios of hadronic quantities.
In Table~\ref{tab:flow_hadron}, we list values of the meson masses, $m_\pi$, $m_K$ and $m_{ss}$ as well as the decay constants $F_\pi$ and $F_K$ and the nucleon mass $m_N$.
We also provide the ratios of $F_K / F_\pi$ and $m_N / F_\pi$.

\subsection{Fit functions\label{sec:fitfunc}}
To determine the value of $m_{res}$, we fit the correlation function described by Eq.~(\ref{eq:mres}) to a constant.

The meson correlation functions were folded in time to double the statistics while the nucleon correlation functions were averaged between the forward propagating positive parity interpolating operator and the backward propagating negative parity interpolating operator, constructed as in Refs.~\cite{Basak:2005aq,Basak:2005ir}. The fit Ansatz describing a $\bar{q_1}q_2$-meson correlation functions is given by
\begin{align}
C_{\textrm{2pt}}^{{q_1q_2}}(t) = & \sum_n z^{q_1q_2}_n z^{q_1q_2\dagger}_n \left(e^{-E^{q_1q_2}_n t} + e^{-E^{q_1q_2}_n (T-t)}\right)\nonumber\\
& + (-1)^t z^{\textrm{osc.}}z^{\textrm{osc.}\dagger}\left(e^{-E^{\textrm{osc.}} t} + e^{-E^{\textrm{osc.}}(T-t)}\right),
\label{eq:c2pt}
\end{align}
where we define $z_n$ as the overlap factor of the $n$th state with energy $E_n$ and the superscript osc. denotes the overlap and energy of the oscillating mode.

In order to determine the pseudoscalar decay constants, we utilize the 5D Ward Identity relating the renormalized decay constants to various correlation functions including those used to determine the values of $m_{res}$~\cite{Blum:2000kn,Aoki:2002vt},
\begin{equation}
 F^{q_1 q_2} = z_p^{q_1q_2}
    \frac{m^{q_1} + m^{q_1}_{res} + m^{q_2} + m^{q_2}_{res}}{\sqrt[3]{E^{q_1q_2}_0}}\, ,
\end{equation}
where $z_p$ denotes the point-sink overlap factor.
This normalization is such that the physical pion decay constant is $F_\pi=92.2$~MeV.

In order to determine the axial renormalization constants, we can also compute the bare values of $F^{q_1 q_2}$ using the 4D axial-vector current,
\begin{align}
C_{\textrm{axial}}^{q_1q_2}=& \partial_4 \langle 0| A_4(t) P_S(0) |0\rangle\nonumber\\
    =& -\sum_n f^{q_1q_2}_n z_{s,n}^{q_1q_2} \left(e^{-E_n^{q_1q_2} t}
    + e^{-E_n^{q_1q_2} (T-t)}\right)
\nonumber\\
&-(-1)^t f^{\textrm{osc.}} z^{\textrm{osc.}}_s \left(e^{-E^{\textrm{osc.}} t}
    + e^{-E^{\textrm{osc.}} (T-t)}\right)
\end{align}
where $f_0^{q_1q_2}=\sqrt{E_0^{q_1q_2}}F^{q_1q_2}/Z_A$ with renormalization coefficient $Z_A$ and $z_s$ is the same ground-state overlap factor determined in the two-point function.

For the nucleon two-point correlation function, we use the fit Ansatz analogous to Eq.~(\ref{eq:c2pt}) without the oscillating state and wraparound terms.

\subsection{Analysis strategy\label{sec:analysis}}
The correlator analysis is performed using the Python package \texttt{lsqfit}~\cite{lsqfit}. We perform a chained fit~\cite{Bouchard:2014ypa} to the light and strange $m_{res}$ correlator; the pion, kaon, and $\bar{s}\g_5 s$-meson two-point and axial correlators; and the nucleon two-point correlator. In particular, as part of the chained fit, we perform a simultaneous fit to the pseudoscalar two-point (point- and smeared-sink) and axial correlators and to the nucleon point- and smeared-sink correlators. The chained fit implementation in \texttt{lsqfit} preserves all correlations by numerically implementing the propagation of error under the assumption that all parameters are Gaussian distributed. We use the resulting correlated posterior distributions to propagate all subsequent uncertainties (\textit{e.g.} ratios) without performing any bootstrap resampling.

For the pseudoscalar correlators, we truncate the fit Ansatz at 2+1 states, where the +1 denotes the oscillating state. For the nucleon correlator, we perform a two-state fit.
For the pseudoscalar correlators, in an independent analysis, using similar fit regions, we observe using three states without oscillating modes results in a consistent determination of the ground-state masses and overlap factors.  Further, using an unconstrained, single-state fit in the late time region also results in consistent ground state parameters.

We choose unconstraining ground-state priors such that the prior widths are at least an order of magnitude wider than the width of the posterior distribution. The oscillating-state energy splitting is chosen to be at the lattice cutoff scale. The first excited-state energy splitting is chosen to be at the two-pion threshold. Details on our prior choices are given in Table~\ref{tab:priors}.

The fit region is chosen such that $t_{\textrm{min}}\sim 1$~fm and $t_{\textrm{max}}\sim 2.3$~fm for all pseudoscalar correlators.  For the nucleon correlator analysis, $t_{\textrm{min}} \sim 0.6$~fm and $t_{\textrm{max}} \sim 1.4$~fm are chosen for all ensembles. It is necessary to fit the nucleon correlator closer to the origin due to the poorer signal-to-noise ratio when compared to the pseudoscalar observables. Explicit fit regions in lattice units are given in Table~\ref{tab:fit_region}. We observe that all final correlator fits are in the region of stability for varying $t_{\textrm{min}}$ and $t_{\textrm{max}}$, including the more aggressive nucleon analysis, indicating that the results are free of excited-state contamination.

\subsection{Observations about flow-time dependence}
From our calculations, there are a few substantial benefits one observes from the use of the gradient-flow smearing.
Before discussing these, we first comment on the strong oscillations observed at small flow-time in the pseudoscalar correlators.
In Fig.~\ref{fig:mpi_tgf}, we observe a strong signal for an oscillating excited state with $(-1)^t$ behavior (where $t$ is the Euclidean time) at small flow-times, most notably for $t_{gf}=0.2$.
These oscillating modes become completely damped out for $t_{gf} \geq 0.6$, with the statistics used in this work.

\begin{figure}
\includegraphics[width=\columnwidth]{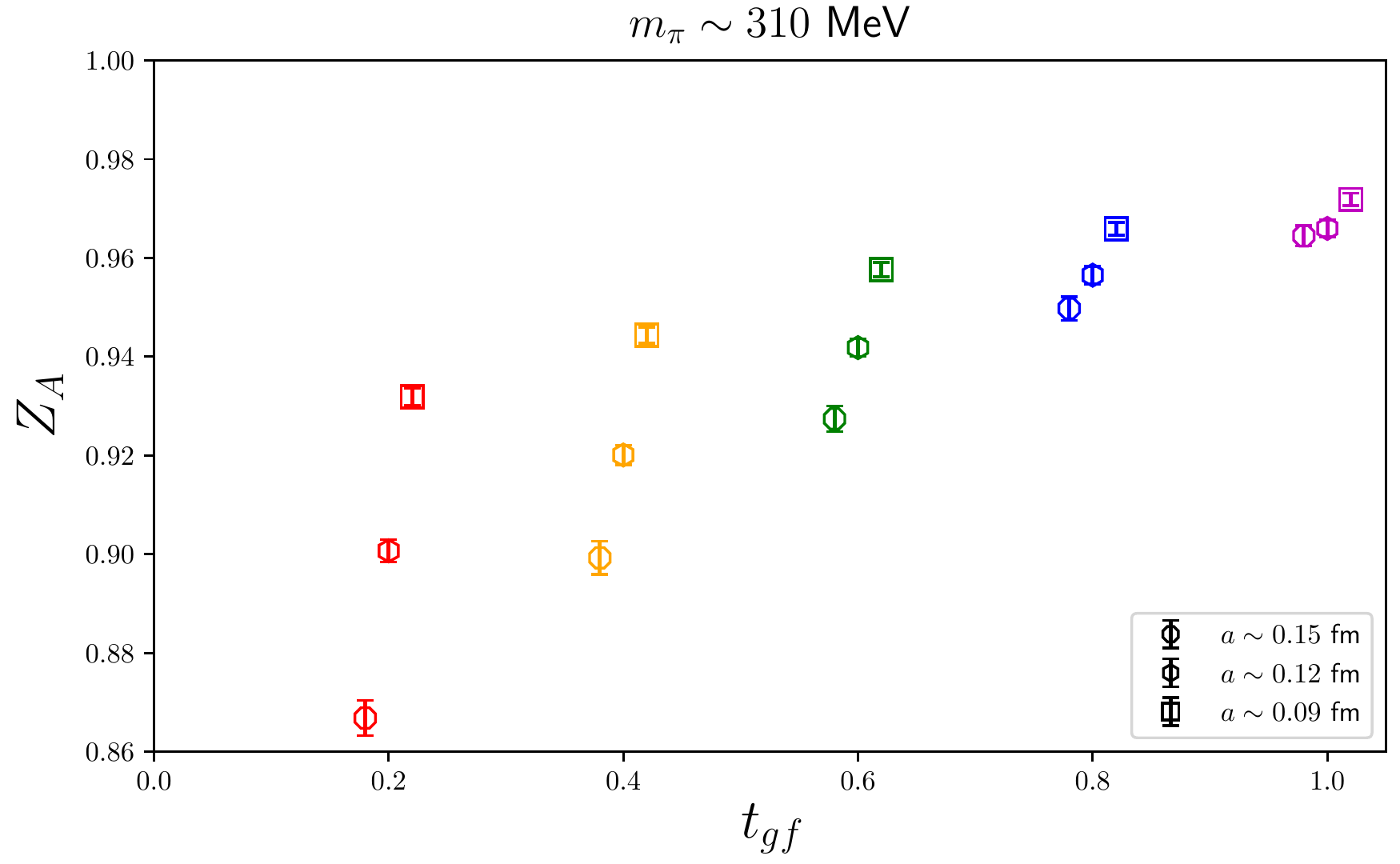}\vspace{5pt}
\includegraphics[width=\columnwidth]{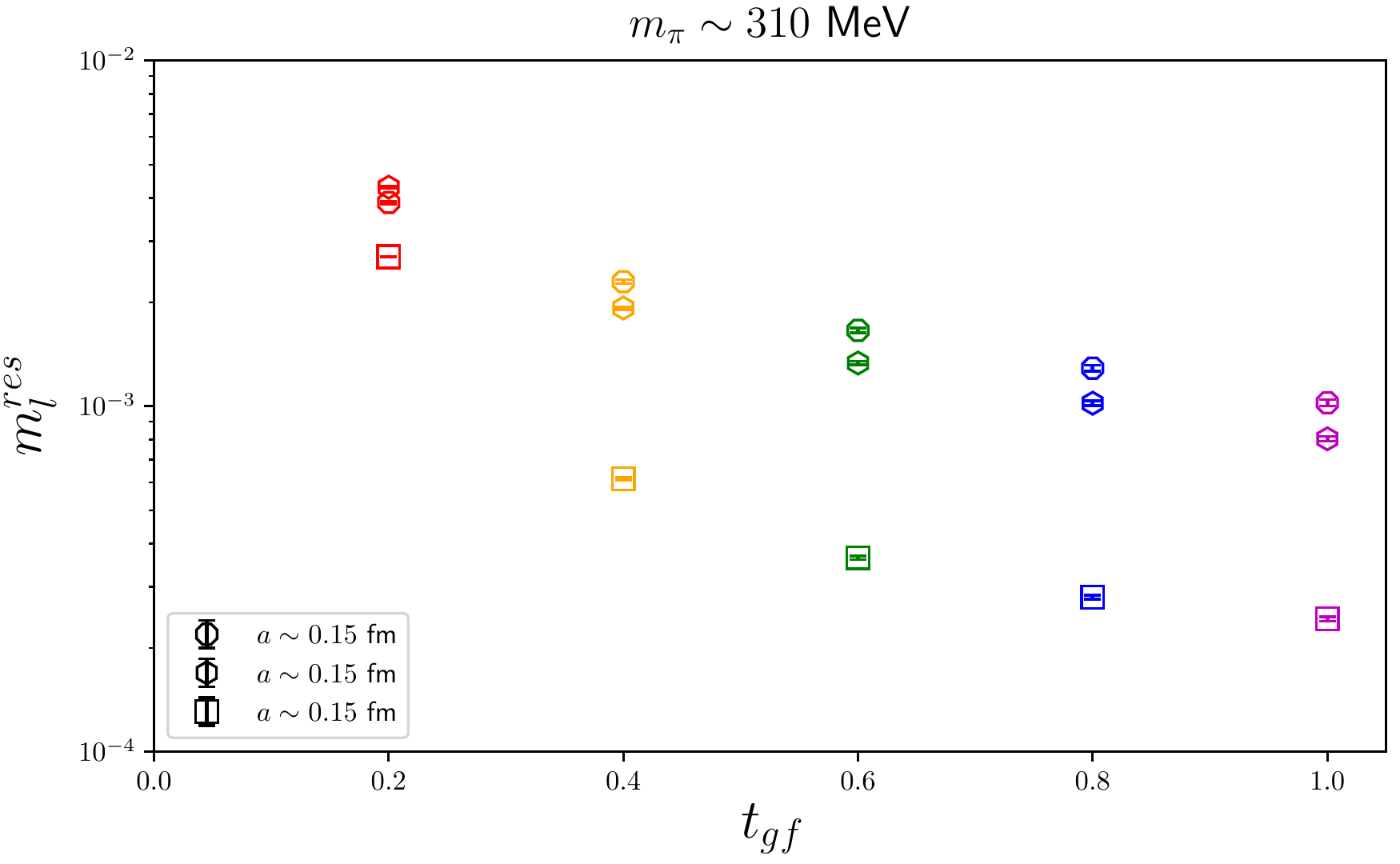}
\caption{\label{fig:za_tgf} $Z_A$ (top) and $m_l^{res}$ (bottom) as a function of flow-time on the $m_\pi\simeq310$~MeV ensembles.  The results of $Z_A$ are slightly shifted horizontally for visual clarity.}
\end{figure}

The first significant benefit observed is that as the flow-time is increased a dramatic reduction of the chiral symmetry breaking properties of the valence MDWF action is achieved.
This can be observed in the significant reduction in $m_{res}$ at fixed pion mass or similarly, the values of $Z_A$ approaching 1 for all gauge couplings, both of which are depicted in Fig.~\ref{fig:za_tgf}.
With the tuning we have chosen, to hold the pion mass, as well as $L_5$, $M_5$, $b_5$, and $c_5$, fixed as we vary the flow-time, we observe an exponential reduction in $m_{res}$ as the flow-time is increased.
Though not depicted in these figures or tables, we also studied the dependence of $m_{res}$ on $L_5$ as the flow-time was varied.
We find that for small flow-time, the reduction in $m_{res}$ as $L_5$ increases is power law, indicating the 5D zero-mode contributions are dominating the residual chiral symmetry breaking.
As we increase the flow-time, $m_{res}$ begins to fall off exponentially in $L_5$, indicating the gradient flow smearing suppresses these zero-mode contributions.

Another significant benefit we observe is that stochastic fluctuations become smaller for increasing flow-time because the gradient flow smearing procedure suppresses the ultraviolet noise.
This is observed from the sample effective mass plots of the nucleon and pion in Fig.~\ref{fig:mpi_tgf}.
The gradient flow is applied in all four spacetime directions, so the neighboring time slices become more correlated, rendering a direct comparison of the effective mass plots more complicated.
However, the list of fitted quantities in Table~\ref{tab:flow_hadron} demonstrates the correlated stochastic uncertainties are reduced for increasing flow-time.
Comparing the $t_{gf}=1$ to $t_{gf}=0.2$ results, we observe approximately a factor of $\sqrt{2}$ reduction of the stochastic uncertainty for equal computing cost for all quantities other than the pseudoscalar meson masses.

%
\section{Flow-time independence of continuum limit \label{sec:cont_lim}}

\begin{figure}
\includegraphics[width=\columnwidth]{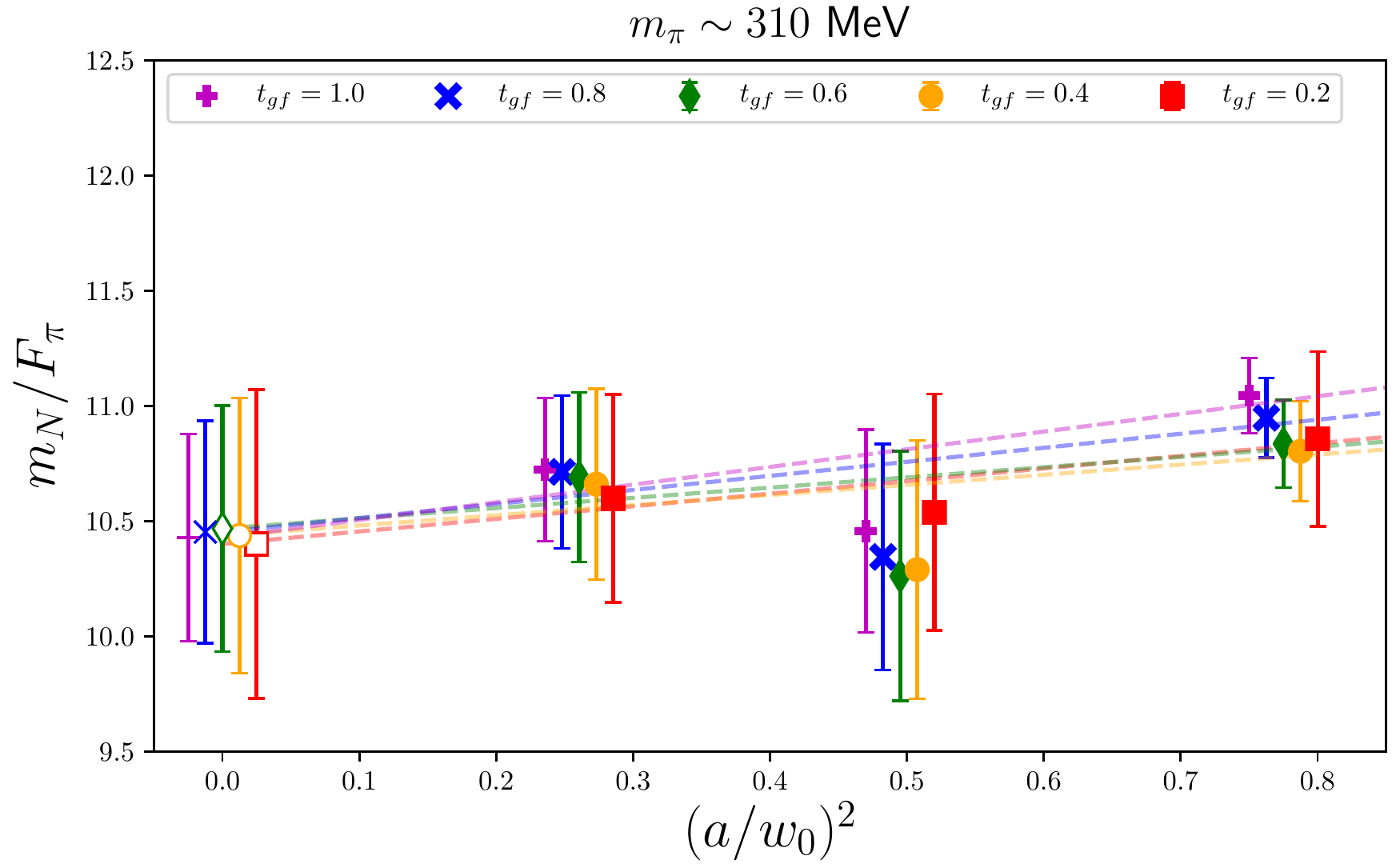}\vspace{5pt}
\includegraphics[width=\columnwidth]{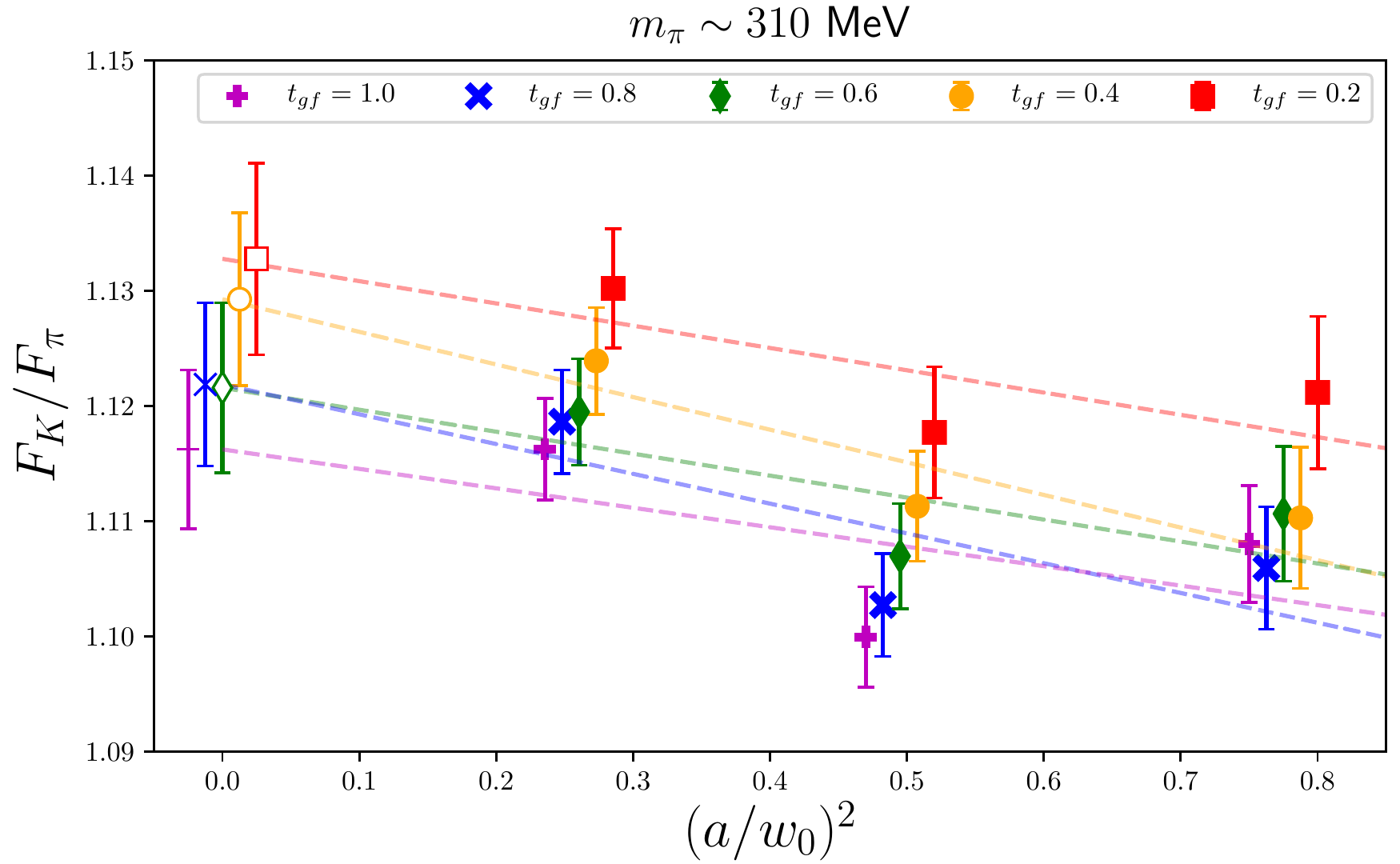}
\caption{\label{fig:mNF_tgf} Flow-time (in)dependence of $m_N/F_\pi$ and $F_K/F_\pi$ on the $m_\pi\sim310$~MeV ensembles.
The filled in symbols are the results of our calculations, and the open symbols clustered at $a/w_0=0$ are the continuum extrapolated results using the simple Ansatz of a constant plus an $(a/w_0)^2$ term.
The results are slightly shifted horizontally for visual clarity.}
\end{figure}

In Fig.~\ref{fig:mNF_tgf}, we show a continuum study of $m_N/F_\pi$ and $F_K/F_\pi$ on the $m_\pi\sim310$~MeV ensembles, for all flow-times used. We explore four different continuum extrapolation Ans\"{a}tze for a quantity $f$:
\begin{equation}\label{eq:continuum_fit}
f(a/w_0) = \left\{
\begin{array}{ll}
    f_0\, ,& \textrm{constant\, ,}
\\
    f_0 + f_2 \frac{a^2}{w_0^2}\, ,
    & \textrm{linear in $a^2$\, ,}
\\
    f_0 + \alpha_s f_2^\prime \frac{a^2}{w_0^2}\, ,
    & \textrm{linear in $\a_s a^2$\, ,}
\\
    f_0 + f_4 \frac{a^4}{w_0^4}\, ,
    & \textrm{quadratic in $a^2$\, .}
\end{array}\right.
\end{equation}
The gradient flow scale $w_0$ was first defined in Ref.~\cite{Borsanyi:2012zs}, and a value of $w_0\text{\cite{Borsanyi:2012zs}}=0.1755(18)(04)$~fm was determined.  The value determined in Ref.~\cite{Bazavov:2015yea} is similar with a slight discrepancy, $w_0\text{\cite{Bazavov:2015yea}}=0.1714({}_{12}^{15})$~fm.  We use this value as we are using the same ensembles on which it was determined.
With only three lattice spacings, we choose not to perform an extrapolation in both $a^2$ and either $\a_s a^2$ or $a^4$ simultaneously.
However, we observe the value of $f_2$ for both $m_N/F_\pi$ and $F_K/F_\pi$ to be small and often consistent with zero.  This motivates exploring the linear in $\a_s a^2$ and $a^4$ fits as estimates of systematic uncertainties in the continuum extrapolation.
We find all four continuum extrapolations show consistency at the 1-sigma level, both between all four different fit Ans\"{a}tze and also between the various flow-time extrapolations.
In Fig.~\ref{fig:mNF_tgf}, we display the continuum extrapolation using the Ansatz linear in $(a/w_0)^2$.
The quark-mass-independent values of $a/w_0$ and $\a_s$ are taken from Ref.~\cite{Bazavov:2015yea}.

For $m_N/F_\pi$, we observe minimal discretization corrections with a very small slope in $(a/w_0)^2$.
For $F_K/F_\pi$, a quantity which is determined much more precisely for equal stochastic sampling, we observe mild, though still quite small, discretization corrections.
While the discretization corrections are basically flow-time independent for $m_N/F_\pi$, they seem to become more pronounced for $F_K/F_\pi$ as the flow-time is increased.
There is an indication of the presence of higher order quartic in $a/w_0$ corrections, but we are not able to resolve these with the numerical results in this work.
Previous studies of the heavy-light decay constants observed that large amounts of APE smearing~\cite{Albanese:1987ds} could induce significant higher order discretization effects~\cite{Bernard:2002pc}.
It is possible that the larger $t_{gf}$ smearings are having a similar effect on the strange quark, and thus the value of $F_K$, at the sub percent level.
These potential systematic uncertainties should be explored in more detail for a sub percent calculation of $F_K/F_\pi$ using this action.

%
\subsection{Mixed-meson mass corrections \label{sec:mixed}}

In order to use the MAEFT extrapolation formulae, there are a few additional quantities which must be determined from the MALQCD calculations.
At NLO in the MAEFT expansion, one needs to know the masses of the mixed valence-sea mesons which propagate in virtual loops and the value of the partial quenching parameter which controls the unitarity violating contributions~\cite{Chen:2006wf,Chen:2007ug}.
In a general MALQCD calculation with a chirally symmetric valence action, one has
\begin{align}
m_{vs}^2 &= \frac{1}{2}\left(m_{vv}^2 + m_{ss}^2\right)
    + a^2 \tilde{\D}_\textrm{Mix}\, ,
\nonumber\\
\D_\textrm{PQ}^2 &= m_{ss}^2 - m_{vv}^2\, ,
\end{align}
where $m_{vv}$ is the mass of the pseudoscalar valence-valence meson,
$m_{ss}$ is the mass of the pseudoscalar sea-sea meson including possible additive discretization corrections,
and $a^2 \tilde{\D}_\textrm{Mix}$ is an additional additive discretization correction to the mass of a meson composed of one valence and one sea quark.
For our MALQCD calculations, these two quantities are given by~\cite{Chen:2006wf,Chen:2007ug,Chen:2009su}
\begin{align}
m_{vs}^2 &= \frac{1}{2}\left(m_{vv}^2 + m_{ss,5}^2\right)
    + a^2 \tilde{\D}_\textrm{Mix}\, ,
\nonumber\\
a^2 \tilde{\D}_\textrm{Mix} &= a^2\D_\textrm{Mix}
    +\frac{a^2}{8}\D_A + \frac{3a^2}{16}\D_T
    +\frac{a^2}{8}\D_V + \frac{a^2}{32}\D_I\, ,
\nonumber\\
a^2 \D_\textrm{Mix} &= \frac{8 a^2 C_\textrm{Mix}}{F^2}\, ,
\nonumber\\
\D_\textrm{PQ}^2 &= m_{ss,5}^2 + a^2\D_I - m_{vv}^2\, ,
\end{align}
where $m_{ss,5}$ is the mass of the taste-5 pseudoscalar meson,
$a^2\D_B$ are the taste splittings between the other taste-meson and the taste-5 meson, $a^2\D_B = m_B^2 - m_5^2$,
$F$ is the leading order pion decay constant,
and $C_\textrm{Mix}$, is the LEC of a new operator present in the MAEFT Lagrangian at $\mc{O}(a^2)$.
The mixed-meson mass splitting, $a^2 \D_\textrm{Mix}$ is universal at LO in the MAEFT expansion~\cite{Bar:2005tu}, regardless of the taste of the staggered sea-quark partnered with the DW quark.
In Ref.~\cite{Orginos:2007tw}, it was observed that there is a noticeable quark-mass dependence of the mixed-meson splitting, as defined, e.g., for the pion
\begin{equation}\label{eq:mixed_split}
\D m_{vs}^2 \equiv
    m_{\pi,vs}^2 - \frac{1}{2} \left(
    m_{\pi,DW}^2 + m_{\pi,5}^2 \right)\, .
\end{equation}
There are three common methods of incorporating these discretization corrections:
\begin{enumerate}
\item Power-series expand the discretization corrections about $a=0$, and use a continuum EFT extrapolation enhanced by general corrections of the form $a^2$, $a^2 \alpha_S$, etc..

\item Extrapolate these mixed-meson discretization corrections to the chiral limit, and use a uniform correction for all mixed mesons with the full MAEFT expressions.

\item Use the on-shell renormalized mixed-meson masses as they are on each ensemble with the full MAEFT expressions.
\end{enumerate}
Provided the discretization corrections are under control, all three methods should agree in the continuum limit.
It is useful, therefore, to determine the mixed-meson masses for all combinations of valence and sea quarks used in the MALQCD calculations.

In order to compute the mixed-meson spectrum, we need to construct pseudoscalar mesons composed of one MDWF and one HISQ fermion propagator.
To compute the MDWF propagators, we have used the QUDA library interfaced from Chroma with solutions generated with gauge-covariant Gaussian smeared sources~\cite{Frommer:1995ik}.
To compute the HISQ propagators, we utilized the MILC code.
To minimize the gauge noise, we similarly used a gauge-covariant source for the staggered fermions.
This source was created in Chroma, with routines added to the \texttt{devel} branch to support writing a source file readable as a \texttt{vector\_field} source by the MILC code.
The MDWF fermions were converted to the \texttt{DD\_PAIRS} format to be read by MILC, which was used to compute the mixed-meson and HISQ-HISQ pseudoscalar spectrum.
To further reduce the gauge noise, the mixed-meson correlation functions were constructed with interpolating operators
\begin{equation}
\mc{O}_{vs} = \bar{q}_{val} \g_5 q_{sea}
\end{equation}
as well as their Hermitian conjugates.  The real part of the averaged conjugate pairs of correlation functions were then used to determine the spectrum, which were computed with all possible pairings of light and strange quarks with one MDWF- and one HISQ-type quark propagator.

In Table~\ref{tab:mixed}, we list the masses of mixed mesons computed in this work, using only flow-time $t_{gf}=1$ ensembles.  In Table~\ref{tab:mixed_relative}, we list the values of the splittings $\D m_{vs}^2$, defined as in Eq.~\eqref{eq:mixed_split}, and $m_{vv}$ and $m_{ss}$ are the pseudoscalar masses of the valence-valence and sea-sea mesons, respectively.
The values are listed in $w_0$ units where the quark-mass-independent values $w_0/a$ are taken from Ref.~\cite{Bazavov:2015yea}.
We use the notation of Ref.~\cite{Chen:2001yi} and denote the various mixed mesons as
\begin{align}
\phi_{uj} &= \textrm{pion: val. light = $u$, sea light = $j$,}
\nonumber\\
\phi_{ur} &= \textrm{kaon: val. light = $u$, sea strange = $r$,}
\nonumber\\
\phi_{sj} &= \textrm{kaon: val. strange = $s$, sea light = $j$,}
\nonumber\\
\phi_{sr} &= \textrm{$\bar{s}\g_5 s$: val. strange = $s$, sea strange = $r$.}
\end{align}

\begin{table}
\caption{\label{tab:mixed} The mixed-meson mass spectrum determined on ensembles used in this work, with flow-time $t_{gf}=1$.}
\begin{ruledtabular}
\begin{tabular}{ccccc}
Ensemble& $am_{uj}$& $am_{sj}$& $am_{ur}$& $am_{sr}$\\
\hline
a15m310& 0.300(6)& 0.432(4)& 0.444(5)& 0.549(2)\\
a12m310& 0.216(2)& 0.334(2)& 0.339(2)& 0.430(1)\\
a09m310& 0.150(1)& 0.243(1)& 0.247(1)& 0.315(1)\\
a15m220& 0.255(3) & 0.416(3)& 0.430(3)& 0.543(1)\\
a12m220& 0.178(2)& 0.321(2)& 0.335(2)& 0.428(1)
\end{tabular}
\end{ruledtabular}
\end{table}

\begin{table}
\caption{\label{tab:mixed_relative} The mixed-meson mass splittings Eq.~\eqref{eq:mixed_split} determined on ensembles used in this work, with flow-time $t_{gf}=1$.
The values of $w_0/a$ are determined from Ref.~\cite{Bazavov:2015yea}.}
\begin{ruledtabular}
\begin{tabular}{ccccc}
Ensemble& $w_0^2 \D m_{uj}^2$& $w_0^2 \D m_{sj}^2$& $w_0^2\D m_{ur}^2$& $w_0^2\D m_{sr}^2$\\
\hline
a15m310& 0.0439(41)& 0.0298(40)& 0.0440(59)& 0.0422(28) \\
a12m310& 0.0214(17)& 0.0123(29)& 0.0199(30)& 0.0206(22) \\
a09m310& 0.0102(09)& 0.0038(18)& 0.0102(19)& 0.0085(14) \\
a15m220& 0.0488(38)& 0.0341(58)& 0.0488(60)& 0.0410(36) \\
a12m220& 0.0279(13)& 0.0142(20)& 0.0334(30)& 0.0212(20)
\end{tabular}
\end{ruledtabular}
\end{table}

%
\section{Benchmark calculation of $F_{K^\pm} / F_{\pi^\pm}$ \label{sec:benchmark}}

After demonstrating the flow-time independence of $m_N/F_\pi$ and $F_K/F_\pi$ in the continuum limit and observing the advantages of larger smearing flow-times $t_{gf}$, we provide a benchmark computation with all systematic errors estimated.
In particular we assess the effects of the extrapolation to the physical pion mass as well as to the continuum and infinite volume limit of $F_K/F_\pi$.
At NLO in the three-flavors chiral expansion, this quantity depends upon only a single LEC, $L_5$~\cite{Gasser:1984gg}. Therefore, with the limited number of ensembles used in this work, we can perform a full extrapolation to the physical point.
Further, $F_K/F_\pi$ is obtained with great precision from many different LQCD calculations and it is one of the quantities reviewed in depth by the FLAG Working Group~\cite{Aoki:2016frl}.
A comparison serves as an important benchmark calculation of our lattice action.

\subsection{$\chi$PT extrapolation at different gradient flow-times.}
We have three lattice spacings and two pion masses with different values of $m_{\pi}L$. Following our findings for the continuum extrapolation at $m_\pi\sim310$~MeV, our chiral-continuum extrapolation is performed with the form
\begin{align}\label{eq:fkfpi_chpt}
\frac{F_K}{F_\pi} &= 1
	+\frac{5}{8} \frac{m_\pi^2}{\L_{\chi}^2} \ell_\pi
	-\frac{1}{4} \frac{m_K^2}{\L_\chi^2} \ell_K
	-\frac{3}{8} \frac{m_\eta^2}{\L_\chi^2} \ell_\eta
\nonumber\\&\phantom{=}
	+\frac{4(m_K^2 -m_\pi^2)}{\L_\chi^2} (4\pi)^2 \left[
		L_5(\L_{\chi}) +\frac{a^2}{w_0^2} L_{a}
	\right]\, .
\end{align}
In this expression, we have used the relation valid at NLO in the $SU(3)$ chiral expansion, $m_\eta^2 = 4m_K^2/3 - m_\pi^2 / 3$, and the definitions $\ell_\phi = \ln(m_\phi^2 / \L_\chi^2)$ ($\phi \in \{ \pi, K, \eta$ \}) and $\L_\chi^2 = (4\pi)^2 F_K F_\pi$.
We have also included the finite volume corrections from the radiative pion loops predicted at one loop in $\chi$PT~\cite{COLANGELO2005136,Durr:2010hr}, but we find they have an irrelevant effect on the fit with the precision we have.
The discretization corrections are flavor independent and so they must vanish in the $SU(3)$ flavor limit where $F_K/F_\pi = 1$ exactly.  Therefore, we parametrize the discretization correction through an unknown LEC that accompanies a term proportional to $(m_K^2 - m_\pi^2) (a/w_0)^2$.

\begin{figure}
\includegraphics[width=\columnwidth]{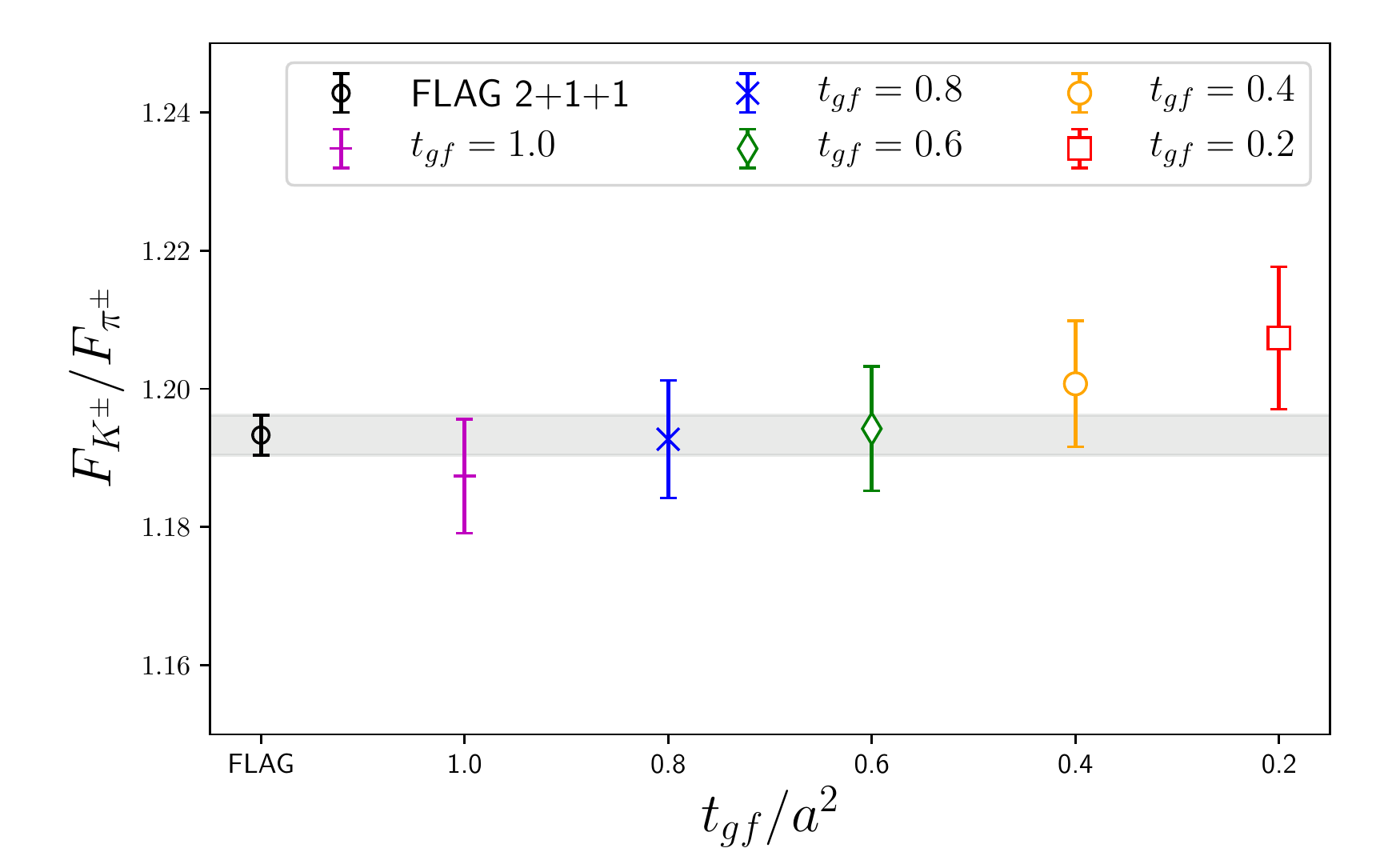}
\caption{\label{fig:FkFpi_chpt_tgf} Flow-time (in)dependence of $F_{K^\pm}/F_{\pi^\pm}$ at the physical point $m_{\pi} \approx 135$ MeV in the continuum limit.
The colored symbols are the results of our calculations extrapolated to the continuum limit and to the physical point using Eq.~\eqref{eq:fkfpi_chpt}.
The benchmark FLAG result is the leftmost black point and it is consistent with our results at all flow-times within 1-sigma (horizontal gray band).
The linear trend in flow-time observed is not present in the full continuum, chiral extrapolation analysis of different, but consistent, analysis of pseudoscalar correlation functions, so we believe this observed trend is not statistically significant.
}
\end{figure}

Using the expression in Eq.~\eqref{eq:fkfpi_chpt}, we fit the five ensembles used in this work for each flow-time independently.
We then extrapolate these results to the isospin symmetric physical point, as determined by FLAG~\cite{Aoki:2016frl} with $m_\pi = 134.8(3)$~MeV and $m_K=494.2(3)$~MeV.
In order to compare with the FLAG determination, we must correct these results from the isospin symmetric point to the ratio of the charged decay constants, as prescribed in Eqs.~(62) and (63) of the most recent FLAG review.
In Fig.~\ref{fig:FkFpi_chpt_tgf}, we display our resulting values of $F_{K^\pm} / F_{\pi^\pm}$ for each flow-time.
We observe good quality in all our fits, as defined by the $Q$-value, which is the Bayesian analog to the $p$-value defined in Eq.~(B4) of Ref.~\cite{Bazavov:2016nty}. For comparison, we plot the FLAG determination of $F_{K^\pm} / F_{\pi^\pm}$ from the average of results using $N_f = 2+1+1$ ensembles.  At the 1-sigma level, our results are self-consistent (flow-time independent) and also consistent with the FLAG average value.
There is a trend of $F_K / F_\pi$ with $t_{gf}$ observed in Fig.~\ref{fig:FkFpi_chpt_tgf}.
However, we do not believe this is statistically significant because the continuum, chiral analysis using different, but consistent, correlation function analysis results as input, results in values of $F_K/F_\pi$ which do not have a trend.

\subsection{MA EFT extrapolation at $t_{gf}=1$}
While the numerical results are sufficient to constrain the unknown LECs, we note that for larger flow-times the quality of the fit decreases, hinting at missing dependence upon the input parameters.
For $t_{gf}=1$, we have also computed the mixed-meson masses, and so we can perform the full MA EFT extrapolation.
The NLO MA EFT expressions for $f_\pi = \sqrt{2} F_\pi$ and $f_K$ are provided in Eqs.~(C1) and (C2) of Ref.~\cite{Chen:2006wf}, respectively.
In our case, we have tuned the valence quark masses such that the pion mass matches the taste-5 HISQ pion mass, which implies $\D_{ju} = \D_{rs} = 0$ in the reference expressions.
Further, the mixed-meson mass splitting is independent of quark mass at LO, allowing us to simplify the extrapolation formula.
To simplify transcribing the expression, we define
\begin{align}\label{eq:params}
&\e_\pi^2 = \frac{m_\pi^2}{\L_\chi^2},&
&\e_{ju}^2 = \frac{m_\pi^2 + a^2\tilde{\D}_\textrm{Mix}}{\L_\chi^2},&
\nonumber\\
&\e_{K}^2 = \frac{m_K^2}{\L_\chi^2},&
&\e_{ru}^2 = \e_{sj}^2 = \frac{m_K^2 + a^2\tilde{\D}_\textrm{Mix}}{\L_\chi^2},&
\nonumber\\
&\e_{ss}^2 = \frac{m_{ss}^2}{\L_\chi^2},&
&\e_{rs}^2 = \frac{m_{ss}^2 + a^2\tilde{\D}_\textrm{Mix}}{\L_\chi^2},&
\nonumber\\
&\d_{PQ}^2 = \frac{a^2 \D_{\rm I}}{\L_\chi^2},&
&\e_X^2 = \frac{4}{3}\e_K^2 -\frac{1}{3}\e_\pi^2 + \d_{PQ}^2,&
\nonumber\\
&\textrm{and}&
&\L_\chi^2 = 16\pi^2 F_\pi F_K.&
\end{align}

The resulting MA EFT expression is
\begin{align}\label{eq:fkfpi_ma}
\frac{F_K}{F_\pi} &= 1
	+\frac{1}{2}\e_{ju}^2 \ell_{ju}
	+\frac{1}{8} \ell_\pi \left\{
		\e_\pi^2 -\frac{\d_{PQ}^2(\e_X^2 + \e_\pi^2)}{\e_X^2 - \e_\pi^2}
		\right.
\nonumber\\ &\phantom{=}\left.
		+\frac{\d_{PQ}^4 \e_X^2}{3(\e_X^2 - \e_\pi^2)^2}
		-\frac{4\d_{PQ}^4 \e_\pi^2}{3(\e_X^2 - \e_\pi^2)(\e_{ss}^2 - \e_\pi^2)}
	\right\}
\nonumber\\ &\phantom{=}
	-\frac{1}{2}\e_{sj}^2 \ell_{sj}
	+\frac{1}{4}\e_{ru}^2 \ell_{ru}
	-\frac{1}{4}\e_{rs}^2 \ell_{rs}
	+\frac{\ell_{ss}}{4} \Bigg\{
		\e_{ss}^2
\nonumber\\ &\phantom{=}
		+\frac{\d_{PQ}^2 (3\e_{ss}^4 +2(\e_K^2-\e_\pi^2)\e_X^2 -3\e_{ss}^2\e_X^2)}
			{3(\e_X^2 -\e_{ss}^2)^2}
\nonumber\\ &\phantom{=}\left.
		-\frac{\d_{PQ}^4(2\e_{ss}^4 -\e_X^2(\e_{ss}^2+\e_\pi^2))}
			{3(\e_X^2 -\e_{ss}^2)^2 (\e_{ss}^2 -\e_\pi^2)} \right\}
	-\frac{3}{8}\e_X^2\ell_X \Bigg\{ 1 
\nonumber\\ &\phantom{=}
		-\frac{2\d_{PQ}^2/3}{(\e_X^2 - \e_\pi^2)}
+\frac{\d_{PQ}^2 [4(\e_K^2-\e_\pi^2) +6(\e_{ss}^2 -\e_X^2)]}
	{9(\e_X^2 -\e_{ss}^2)^2} 
\nonumber\\ &\phantom{=}
		+\frac{\d_{PQ}^4/9}{(\e_X^2 - \e_\pi^2)^2}
		-\frac{2\d_{PQ}^4 (2\e_{ss}^2-\e_\pi^2-\e_{X}^2)}{9(\e_X^2 -\e_{ss}^2)^2 (\e_X^2-\e_\pi^2)}
	\Bigg\} 
\nonumber\\ &\phantom{=}
	+\frac{\d_{PQ}^2 (\e_K^2 -\e_\pi^2)}{6(\e_X^2 -\e_{ss}^2)}
	+\frac{\d_{PQ}^4/24}{(\e_X^2 - \e_\pi^2)}
	-\frac{\d_{PQ}^4/12}{(\e_X^2 -\e_{ss}^2)} 
\nonumber\\ &\phantom{=} -\frac{\d_{PQ}^2}{8}
	+4(\e_K^2 - \e_\pi^2) (4\pi)^2 L_5(\L_{\chi})\, .
\end{align}
In this expression, we have only included the NLO counterterm, which is the same as in $SU(3)$ $\chi$PT, $L_5$.
We observe that with this MA expression, the $a^2 (m_K^2 - m_\pi^2)$ term is no longer needed to fit the data.
When it is included, the fit returns a value of this LEC 2 orders of magnitude smaller than when using Eq.~\eqref{eq:fkfpi_chpt}.
For this analysis, we have taken the values of $w_0^2 \D m_{ju}^2$ from Table~\ref{tab:mixed_relative}, combined with the values of $a/w_0$ from Ref.~\cite{Bazavov:2015yea} to determine the values of $a^2 \tilde{\D}_\textrm{Mix}$.
We have used the values of $r_1^2 a^2 \D_{\rm I}$ and $r_1/a$ from Ref.~\cite{Bazavov:2012xda} to convert them to lattice units and combine them to form the necessary quantities in Eq.~\eqref{eq:params}.
We observe that the MA expression is approximately 150 times more likely to reproduce the observed data when compared to $SU(3)$ $\chi$PT, as determined by the Bayes factors given in Table~\ref{tab:FKFpi_fit}, providing very strong evidence that the MA expression provides the more correct physical point extrapolation. We leave further investigation of $F_K/F_\pi$ with more statistics and more ensembles to future work.

\begin{table}
\caption{\label{tab:FKFpi_fit}
Physical extrapolation from the $F_K/F_\pi$ analysis.
The $Q$-value is the Bayesian analog of the $p$-value defined in Eq.~(B4) of Ref.~\cite{Bazavov:2016nty}. The logGBF denotes the log of the Gaussian Bayes factor and is used to select models under the Bayesian framework. The Bayes factors are suppressed for $t_{gf}$ less than 1.0 since model comparisons are only sensible within the same data set.
}
\begin{ruledtabular}
\begin{tabular}{ccllcc}
$t_{gf}$& Function& $10^3\times L_5$& $F_K/F_\pi$& $Q$-value& logGBF\\
\hline
0.2& Eq.~\eqref{eq:fkfpi_chpt}& 5.55(1.17)& 1.2102(105)& 0.836& --- \\
0.4& Eq.~\eqref{eq:fkfpi_chpt}& 4.79(1.03)& 1.2034(93)& 0.808& --- \\
0.6& Eq.~\eqref{eq:fkfpi_chpt}& 4.05(1.02)& 1.1968(92)& 0.686& --- \\
0.8& Eq.~\eqref{eq:fkfpi_chpt}& 3.88(96)&    1.1952(87)& 0.448& --- \\
1.0& Eq.~\eqref{eq:fkfpi_chpt}& 3.27(93)&    1.1898(84)& 0.278& 6.915 \\
\hline
1.0& Eq.~\eqref{eq:fkfpi_ma}& 3.35(33)& 1.1905(32)& 0.296& 11.947 
\end{tabular}
\end{ruledtabular}
\end{table}

%
\section{MDWF in QUDA: optimizations and performance \label{sec:quda}}

In order to efficiently perform the MDWF solves, we utilize the GPU implementation of the MDWF operator and solver~\cite{Kim:2014mpa} from the highly optimized  QUDA library~\cite{Clark:2009wm,Babich:2011np}.
We added the API for accessing this solver to the Chroma~\cite{Edwards:2004sx} package, which is  publicly available in the most recent version.

The MDWF calculations were performed on three different GPU-enabled machines, Surface and RZHasGPU at LLNL and Titan at OLCF.%
\footnote{Some of the early tuning and flow-time dependence studies were performed at the JLab High Performance Computing Center and at the Fermilab Lattice Gauge Theory Computational Facility.}
The Surface cluster is composed of dual NVIDIA Tesla K40 cards with Intel Xeon E5-2670 CPU nodes.
The RZHasGPU cluster is composed of dual NVIDIA Tesla K80 cards with Intel Xeon E5-2667 v3 CPU nodes.
The Titan supercomputer is composed of single NVIDIA Tesla K20X cards with AMD Opteron CPU nodes.
An interesting feature of the Titan nodes is the use of two 8-core NUMA nodes per node.
We have found that we can provide 2 MPI ranks per GPU, by using both NUMA nodes, and achieve an approximately 69\% performance boost with otherwise identical parameters.
In Table~\ref{tab:quda}, we list the sustained performance on the three machines achieved with the present implementation of the double-half mixed-precision MDWF solver.
The single node performance is notable, and we are at present working on improving the strong scaling of the MDWF solver in QUDA through better overlapping of communication and computation.
Additionally, a significant reduction of the condition number for the symmetric implementation of the MDWF operator has been observed~\cite{jung:2015latt}.
QUDA supports both the symmetric and asymmetric implementations of the MDWF operator.
Currently, Chroma only supports the asymmetric operator, but we plan to investigate possible reduction in the time to solution from switching to the symmetric implementation.

\begin{table}
\caption{\label{tab:quda} Performance of the double-half mixed precision MDWF solver in QUDA on the various compute nodes used with 2, 4 and 1 GPU per node on the Surface, RZHasGPU and Titan computers.
The \% of peak performance is obtained by comparing our sustained to the theoretical single-node single-precision performance.
On Titan, we oversubscribe the GPUs by using 1 MPI rank per NUMA node, which amounts to 2 MPI ranks per GPU, resulting in a $\sim69\%$ performance boost.
}
\begin{ruledtabular}
\begin{tabular}{ccccccc}
Computer& GPUs& MPI& Geometry& \multicolumn{3}{c}{Performance [GFlops]}\\
& & ranks& & Total& per node& \% peak\\
\hline
Surface& 2& 2& 1 1 1 2& 1250& 1250& 44\% \\
RZHasGPU& 4& 4& 1 1 1 4& 1785& 1785& 48\% \\
Titan& 8& 16& 1 1 2 8& 2885& \phantom{0}361& 25\% \\
Titan& 16& 32& 1 2 2 8& 4720& \phantom{0}295& 20\% \\
Titan& 32& 64& 1 2 4 8& 8500& \phantom{0}266& 18\% \\
\end{tabular}
\end{ruledtabular}
\end{table}

%
\section{Conclusions \label{sec:conclusions}}
In this work, we have motivated a new mixed lattice QCD action: M\"{o}bius domain-wall valence fermions solved with the dynamical $N_f=2+1+1$ HISQ sea fermions after a gradient smearing algorithm is used to filter out UV modes of the gluons.
To retain the correct continuum limit, the gradient flow-time is held fixed in lattice units, such that any dependence upon this new scale also vanishes in the continuum limit.
We demonstrate the flow-time independence of the continuum limit by computing two sample quantities, $F_K / F_\pi$ and $m_N / F_\pi$.
An extrapolation of $F_K / F_\pi$ to the continuum, infinite volume and physical pion and kaon mass point is consistent with the FLAG average of the $N_f=2+1+1$ LQCD results for all flow-times explored in this work.

For flow-time of $t_{gf}=1$, we estimate the total systematic error from different chiral and continuum fits to be smaller than our current statistical uncertainty.
Of particular note, we also demonstrate that the gradient flow smearing highly suppresses sources of residual chiral symmetry breaking in the action for moderate values of the flow-time:
the axial renormalization constant becomes effectively lattice spacing independent and close to 1 for all ensembles at a flow-time of $t_{gf}=1$;
the residual chiral symmetry breaking, measured by the quantity $m_{res}$, is exponentially damped with increasing flow-time and less than 10\% of the input light quark mass for all ensembles, including the physical quark-mass ensembles, with $t_{gf}=1$ and moderate values of $L_5$.

This action, coupled with the use of the highly optimized QUDA library, provides an economical method of performing LQCD calculations with an action that respects chiral symmetry to a high degree.
The MILC Collaboration has a long history of making their configurations freely available to all interested parties.
The breadth of parameters used in the generation of the HISQ ensembles allows users to fully control all LQCD systematics: notably the continuum, and infinite volume extrapolations, as well as a physical quark-mass interpolation.

We have plans to use this action for computing various quantities relevant to fundamental nuclear and high-energy physics research, detailed, for example, in the NSAC Long Range Plan for Nuclear Science and the HEPAP P5 Strategic Plan for U.S. Particle Physics.
So far, we have used this mixed action to demonstrate the benefits of a new method for computing hadronic matrix elements~\cite{Bouchard:2016heu},
applied this method to a precise determination of $g_A$~\cite{Berkowitz:2017gql},
and we have computed the $\pi^-\rightarrow\pi^+$ transition matrix elements relevant for the scenario in which heavy lepton-number violating physics beyond the Standard Model contributes to the hypothesized neutrinoless double beta decay of large nuclei~\cite{Nicholson:2016byl}.

\bigskip
\acknowledgments
We gratefully acknowledge the MILC Collaboration for use of the dynamical HISQ ensembles~\cite{Bazavov:2010ru,Bazavov:2012xda}.
The two new ensembles we generated can be made available to any interested person or group.
We thank Carleton DeTar and Doug Toussaint for help compiling and using the MILC code at LLNL and understanding how to write source fields from Chroma that can be read by MILC for the construction of the mixed-meson correlation functions.
We also thank Claude Bernard for useful correspondence regarding scale setting and taste violations with the HISQ action.
Part of this work was performed at the Kavli Institute for Theoretical Physics supported by NSF Grant No. PHY-1125915.

The software used for this work was built on top of the Chroma software suite~\cite{Edwards:2004sx} and the highly optimized QCD GPU library QUDA~\cite{Clark:2009wm,Babich:2011np}.
We also utilized the highly efficient HDF5 I/O Library~\cite{hdf5} with an interface to HDF5 in the USQCD QDP++ package that was added with SciDAC 3 support (CalLat)~\cite{Kurth:2015mqa}, as well as the MILC software for solving for HISQ propagators.
Finally, the HPC jobs were efficiently managed with a \texttt{bash} job manager, \texttt{METAQ}~\cite{berkowitz.metaq}, capable of intelligently backfilling idle nodes in sets of nodes bundled into larger jobs submitted to HPC systems.  \texttt{METAQ} was developed with SciDAC 3 support (CalLat) and is available on \texttt{github}.
The numerical calculations in this work were performed at
the Jefferson Lab High Performance Computing Center and the Fermilab Lattice Gauge Theory Computational Facility on facilities of the USQCD Collaboration, which are funded by the Office of Science of the U.S. Department of Energy;
Lawrence Livermore National Laboratory on the Surface and RZhasGPU GPU clusters as well as the Cab CPU and Vulcan BG/Q clusters;
and the Oak Ridge Leadership Computing Facility at the Oak Ridge National Laboratory, which is supported by the Office of Science of the U.S. Department of Energy under Contract No. DE-AC05-00OR22725, on the Titan machine through a DOE INCITE award (CalLat).
We thank the Lawrence Livermore National Laboratory (LLNL) Institutional Computing Grand Challenge program for the computing allocation.

This work was performed with support from LDRD funding from LLNL 13-ERD-023 (EB, ER, PV);
and by the RIKEN Special Postdoctoral Researcher program (ER).
This work is supported in part by the DFG and the NSFC through funds provided to the Sino-German CRC 110 ``Symmetries and the Emergence of Structure in QCD'' (E. B.).
This work was also performed under the auspices of the U.S. Department of Energy by Lawrence Livermore National Laboratory under Contract DE-AC52-07NA27344 (EB, ER, PV);
under contract DE-AC05-06OR23177, under which Jefferson Science Associates, LLC, manages and operates the Jefferson Lab (BJ, KO) which includes funding from the DOE Office Of Science, Offices of Nuclear Physics, High Energy Physics and Advanced Scientific Computing Research under the SciDAC program (USQCD) (B. J.);
under contract DE-AC02-05CH11231, which the Regents of the University of California manage and operate Lawrence Berkeley National Laboratory and the National Energy Research Scientific Computing Center (CCC, TK, AWL);
This work was further performed under the auspices of the U.S. Department of Energy, Office of Science, Office of Nuclear Physics under contracts:
DE-FG02-04ER41302 (CMB, KNO);
DE-SC00046548 (AN);
DE-SC0015376, Double-Beta Decay Topical Collaboration (AWL);
by the Office of Advanced Scientific Computing Research, Scientific Discovery through Advanced Computing (SciDAC) program under Award Number KB0301052 (EB, TK, AWL);
and by the DOE Early Career Research Program, Office of Nuclear Physics under FWP NQCDAWL (CCC, AWL).

\appendix

\section{Tables of flow-time dependence}
Here, we provide tables of the various quantities computed in this work on the different flow-times used. Tuned quark masses and measured renormalization constants are reported in Table~\ref{tab:flowtune}, while hadron masses and meson decay constants are summarized in Table~\ref{tab:flow_hadron}.

\begin{table*}
\caption{\label{tab:flowtune} The tuned values of the MDWF light and strange quark masses on various ensembles for various flow-times.  We also list the values of the average plaquette after applying the gradient flow as well as $m_{res}$ and the renormalization constants.}
\begin{ruledtabular}
\begin{tabular}{ccccccccccccc}
Ensemble& $M_5$& $L_5$& $b_5$& $c_5$&
    $t_{gf}$& Plaquette&
    $am_l^{mdwf}$& $am_l^{res}$& $Z_A^{ll}$&
    $am_s^{mdwf}$& $am_s^{res}$& $Z_A^{ls}$ \\
\hline
a15m310& 1.3& 12& 1.5& 0.5&
      0.2& 0.87701(2)& 0.00970& 0.003882(38)& 0.8668(36)& 0.06810& 0.003022(31)& 0.8740(13) \\
&&&&& 0.4& 0.95521(1)& 0.01160& 0.002290(29)& 0.8993(34)& 0.07380& 0.001668(22)& 0.9074(12) \\
&&&&& 0.6& 0.97723(1)& 0.01250& 0.001656(26)& 0.9274(26)& 0.08000& 0.001163(19)& 0.9389(12) \\
&&&&& 0.8& 0.98560(1)& 0.01480& 0.001287(24)& 0.9498(24)& 0.08520& 0.000880(17)& 0.9608(11) \\
&&&&& 1.0& 0.98964(1)& 0.01580& 0.001022(23)& 0.9645(21)& 0.09020& 0.000685(15)& 0.9760(09) \\
\hline
a12m310& 1.2& 8& 1.25& 0.25&
      0.2& 0.89320(1)& 0.00680& 0.004298(22)& 0.9007(23)& 0.05300& 0.003416(18)& 0.9034(10) \\
&&&&& 0.4& 0.96401(1)& 0.00960& 0.001922(18)& 0.9201(20)& 0.05830& 0.001352(15)& 0.9243(07) \\
&&&&& 0.6& 0.98251(1)& 0.01086& 0.001332(17)& 0.9418(18)& 0.06280& 0.000860(13)& 0.9464(07) \\
&&&&& 0.8& 0.98925(0)& 0.01176& 0.001019(15)& 0.9565(18)& 0.06650& 0.000615(11)& 0.9608(07) \\
&&&&& 1.0& 0.99242(0)& 0.01260& 0.000804(14)& 0.9660(17)& 0.06930& 0.000467(09)& 0.9705(06) \\
\hline
a09m310& 1.1& 6& 1.25 &0.25&
      0.2& 0.91073(0)& 0.00543& 0.002704(07)& 0.9319(18)& 0.03880& 0.002359(05)& 0.9343(05) \\
&&&&& 0.4& 0.97236(0)& 0.00798& 0.000616(05)& 0.9444(16)& 0.04330& 0.000459(04)& 0.9452(06) \\
&&&&& 0.6& 0.98721(0)& 0.00850& 0.000364(04)& 0.9577(15)& 0.04500& 0.000251(03)& 0.9590(05) \\
&&&&& 0.8& 0.99239(0)& 0.00921& 0.000280(04)& 0.9659(13)& 0.04780& 0.000189(02)& 0.9679(04) \\
&&&&& 1.0& 0.99478(0)& 0.00951& 0.000242(04)& 0.9719(13)& 0.04910& 0.000169(02)& 0.9739(04) \\
\hline
a15m220& 1.3& 16& 1.75 &0.75&
      0.2& 0.87718(1)& 0.00425& 0.002254(18)& 0.8634(38)& 0.06810& 0.001699(17)& 0.8713(12) \\
&&&&& 0.4& 0.95535(1)& 0.00532& 0.001356(16)& 0.8892(33)& 0.07380& 0.000953(14)& 0.9064(12) \\
&&&&& 0.6& 0.97735(1)& 0.00615& 0.000966(14)& 0.9221(31)& 0.08000& 0.000658(11)& 0.9398(13) \\
&&&&& 0.8& 0.98570(1)& 0.00668& 0.000733(11)& 0.9456(27)& 0.08520& 0.000492(10)& 0.9617(11) \\
&&&&& 1.0& 0.98973(1)& 0.00712& 0.000567(10)& 0.9610(26)& 0.09020& 0.000374(09)& 0.9765(09) \\
\hline
a12m220& 1.2& 12& 1.5& 0.5&
      0.2& 0.89332(1)& 0.00365& 0.001562(11)& 0.8923(25)& 0.05480& 0.001085(10)& 0.9026(21) \\
&&&&& 0.4& 0.96410(0)& 0.00456& 0.000935(09)& 0.9132(22)& 0.05880& 0.000582(07)& 0.9240(17) \\
&&&&& 0.6& 0.98259(0)& 0.00522& 0.000673(08)& 0.9409(37)& 0.06280& 0.000391(06)& 0.9466(14) \\
&&&&& 0.8& 0.98931(0)& 0.00575& 0.000511(07)& 0.9546(28)& 0.06660& 0.000286(05)& 0.9621(12) \\
&&&&& 1.0& 0.99248(0)& 0.00600& 0.000390(05)& 0.9615(22)& 0.06930& 0.000216(04)& 0.9718(11) \\
\end{tabular}
\end{ruledtabular}
\end{table*}

\begin{table*}
\caption{\label{tab:flow_hadron}
Various hadronic quantities determined at different flow-times. The posterior distributions related to meson and baryon correlation functions are extracted using a $2+1$-state fit Ansatz for mesons and $2$ states for the nucleons, as described in Secs.~\ref{sec:fitfunc} and~\ref{sec:analysis}. The meson two-point correlation functions are fit simultaneously with the 4D axial-vector current, and then a chained fit~\cite{Bouchard:2014ypa} is used to propagate all remaining correlations. The entire fit strategy is implemented under the Bayesian framework with \texttt{lsqfit}~\cite{lsqfit}. }
\begin{ruledtabular}
\begin{tabular}{cccccccccc}
Ensemble& $t_{gf}$ &
	$am_\pi$ & $am_K$& $am_{ss}$ &
	$aF_\pi$ & $aF_K$ & $am_N$ & $F_K / F_\pi$ & $m_N / F_\pi$ \\
\hline
a15m310
&    0.2& 0.2352(13)\phantom{0}& 0.4025(12)& 0.51904(87)& 0.07781(85)& 0.08724(85)& 0.845(28)& 1.1212(66)& 10.86(38) \\
&    0.4& 0.2327(13)\phantom{0}& 0.4014(12)& 0.51710(89)& 0.07720(69)& 0.08572(64)& 0.834(15)& 1.1103(61)& 10.80(22) \\
&    0.6& 0.2286(11)\phantom{0}& 0.4004(12)& 0.51673(91)& 0.07599(54)& 0.08439(53)& 0.823(13)& 1.1107(58)& 10.84(19) \\
&    0.8& 0.2363(11)\phantom{0}& 0.4028(12)& 0.51673(92)& 0.07543(53)& 0.08343(49)& 0.826(11)& 1.1059(53)& 10.95(17) \\
&    1.0& 0.2367(12)\phantom{0}& 0.4046(12)& 0.51858(94)& 0.07436(51)& 0.08239(46)& 0.821(10)& 1.1080(51)& 11.05(16) \\\hline
a12m310
&    0.2& 0.18876(60)& 0.3233(07)& 0.41835(61)& 0.06385(65)& 0.07137(63)& 0.673(32)& 1.1177(57)& 10.54(51) \\
&    0.4& 0.18842(62)& 0.3233(07)& 0.41773(60)& 0.06306(53)& 0.07008(47)& 0.649(35)& 1.1113(48)& 10.29(56) \\
&    0.6& 0.18837(64)& 0.3232(07)& 0.41754(59)& 0.06243(48)& 0.06911(40)& 0.641(34)& 1.1070(46)& 10.26(54) \\
&    0.8& 0.18833(65)& 0.3234(07)& 0.41776(58)& 0.06196(44)& 0.06832(36)& 0.641(30)& 1.1027(45)& 10.35(49) \\
&    1.0& 0.18911(65)& 0.3232(07)& 0.41721(58)& 0.06142(41)& 0.06755(34)& 0.642(27)& 1.0999(44)& 10.46(44) \\
\hline
a09m310
&    0.2& 0.13982(42)& 0.2411(04)& 0.31227(36)& 0.04578(45)& 0.05174(47)& 0.485(20)& 1.1302(52)& 10.60(45) \\
&    0.4& 0.14017(39)& 0.2423(04)& 0.31392(36)& 0.04590(36)& 0.05159(35)& 0.489(18)& 1.1239(46)& 10.66(41) \\
&    0.6& 0.13860(38)& 0.2396(04)& 0.31041(37)& 0.04568(33)& 0.05113(31)& 0.488(16)& 1.1195(46)& 10.69(37) \\
&    0.8& 0.14026(38)& 0.2416(04)& 0.31280(37)& 0.04550(31)& 0.05090(29)& 0.488(14)& 1.1186(45)& 10.71(33) \\
&    1.0& 0.13978(38)& 0.2405(04)& 0.31129(38)& 0.04521(30)& 0.05047(28)& 0.485(13)& 1.1163(44)& 10.72(31) \\
\hline
a15m220
&    0.2& 0.16707(94)& 0.3838(09)& 0.51227(74)& 0.07616(82)& 0.08794(70)& 0.788(37)& 1.1546(74)& 10.35(49) \\
&    0.4& 0.16668(82)& 0.3848(09)& 0.51195(72)& 0.07521(74)& 0.08638(58)& 0.794(34)& 1.1485(74)& 10.56(47) \\
&    0.6& 0.16683(79)& 0.3852(08)& 0.51184(70)& 0.07425(67)& 0.08481(51)& 0.787(18)& 1.1422(72)& 10.60(27) \\
&    0.8& 0.16647(76)& 0.3853(09)& 0.51195(70)& 0.07343(63)& 0.08338(45)& 0.776(29)& 1.1355(71)& 10.57(41) \\
&    1.0& 0.16629(85)& 0.3866(09)& 0.51388(70)& 0.07231(61)& 0.08205(42)& 0.766(28)& 1.1348(72)& 10.59(40) \\
\hline
a12m220
&    0.2& 0.13305(58)& 0.3080(12)& 0.41732(56)& 0.05732(63)& 0.06618(84)& 0.629(28)& 1.154(12)\phantom{0}& 10.97(50) \\
&    0.4& 0.13370(54)& 0.3086(11)& 0.41636(72)& 0.05773(53)& 0.06610(76)& 0.581(48)& 1.145(12)\phantom{0}& 10.06(85) \\
&    0.6& 0.13354(96)& 0.3088(10)& 0.41583(55)& 0.05784(51)& 0.06582(63)& 0.620(27)& 1.138(10)\phantom{0}& 10.73(48) \\
&    0.8& 0.13491(75)& 0.3103(07)& 0.41690(53)& 0.05778(47)& 0.06572(38)& 0.621(23)& 1.1374(69)& 10.74(41) \\
&    1.0& 0.13424(66)& 0.3097(07)& 0.41618(52)& 0.05731(45)& 0.06514(35)& 0.619(19)& 1.1367(69)& 10.80(36) \\
\end{tabular}
\end{ruledtabular}
\end{table*}

%
\section{Priors for correlator fits}
In Table~\ref{tab:priors} we summarize the Bayesian priors used in the analysis of the mesonic two-point functions, together with the ones for the nucleon correlator and $m_{res}$. Notice that the priors are chosen to be independent of the gradient flow-time.

\begin{table*}
\caption{Priors for correlator fits in lattice units. The priors are all Gaussian distributed and listed as mean(standard deviation). The oscillating and first excited-state energies are defined as splitting from the ground state, where $\Delta_i\equiv \ln(E_i-E_0)$. This leads to a log-normal distributed energy splitting which is positive definite, and as a result enforces a strict hierarchy of states. The priors are chosen to be flow-time independent.
\label{tab:priors}}
\begin{ruledtabular}
\begin{tabular}{cccccccccc}
& $E_0^\pi$ & $z_{0,p}^\pi$ & $z_{0,s}^\pi$ & $E_0^K$ & $z_{0,p}^K$ & $z_{0,s}^K$ & $E_0^{ss}$ & $z_{0,p}^{ss}$ & $z_{0,s}^{ss}$ \\
\hline
a15m310 & 0.2360(236) & 0.255(255) & 0.025(25) & 0.4050(405) & 0.198(198) & 0.0198(198) & 0.520(52) & 0.182(182) & 0.0185(185)\\
a12m310 & 0.190(19) & 0.19(19) & 0.02(2) & 0.3220(322) & 0.148(148) & 0.0159(159) & 0.4180(418) & 0.142(142) & 0.0152(152)\\
a09m310 & 0.140(14) & 0.122(122) & 0.0047(47) & 0.2420(242) & 0.1(1) & 0.0039(39) & 0.3120(312) & 0.1(1) & 0.0037(37)\\
a15m220 & 0.1660(166) & 0.325(325) & 0.031(31) & 0.3850(385) & 0.2(2) & 0.02(2) & 0.5150(515) & 0.18(18) & 0.0184(184)\\
a12m220 & 0.1340(134) & 0.224(224) & 0.0115(115) & 0.310(31) & 0.15(15) & 0.0079(79) & 0.4150(415) & 0.137(137) & 0.0073(73)\\
\hline
& $\Delta_{\textrm{osc.}}^\pi$ & $z_{\textrm{osc.},p}^\pi$ & $z_{\textrm{osc.},s}^\pi$ & $\Delta_{\textrm{osc.}}^K$ & $z_{\textrm{osc.},p}^K$ & $z_{\textrm{osc.},s}^K$ & $\Delta_{\textrm{osc.}}^{ss}$ & $z_{\textrm{osc.},p}^{ss}$ & $z_{\textrm{osc.},s}^{ss}$ \\
\hline
a15m310 & 0(1.45) & 0(0.255) & 0(0.0125) & 0(1.45) & 0(0.198) & 0(0.01) & 0(1.45) & 0(0.182) & 0(0.009)\\
a12m310 & 0(1.67) & 0(0.19) & 0(0.01) & 0(1.67) & 0(0.148) & 0(0.008) & 0(1.67) & 0(0.142) & 0(0.008)\\
a09m310 & 0(1.96) & 0(0.122) & 0(0.00235) & 0(1.96) & 0(0.1) & 0(0.0018) & 0(1.96) & 0(0.1) & 0(0.0018)\\
a15m220 & 0(1.8) & 0(0.325) & 0(0.015) & 0(1.8) & 0(0.2) & 0(0.01) & 0(1.8) & 0(0.18) & 0(0.009)\\
a12m220 & 0(2) & 0(0.224) & 0(0.0057) & 0(2) & 0(0.15) & 0(0.004) & 0(2) & 0(0.137) & 0(0.004)\\
\hline
& $\Delta_1^\pi$ & $z_{1,p}^\pi$ & $z_{1,s}^\pi$ & $\Delta_1^K$ & $z_{1,p}^K$ & $z_{1,s}^K$ & $\Delta_1^{ss}$ & $z_{1,p}^{ss}$ & $z_{1,s}^{ss}$ \\
\hline
a15m310 & -0.75(70) & 0(0.255) & 0(0.0125) &  -0.75(70) & 0(0.198) & 0(0.01) &  -0.75(70) & 0(0.182) & 0(0.009)\\
a12m310 & -0.97(70) & 0(0.19) & 0(0.01) & -0.97(70) & 0(0.148) & 0(0.008) & -0.97(70) & 0(0.142) & 0(0.008)\\
a09m310 & -1.26(70) & 0(0.122) & 0(0.00235) & -1.26(70) & 0(0.1) & 0(0.0018) & -1.26(70) & 0(0.1) & 0(0.0018)\\
a15m220 & -1.1(7) & 0(0.325) & 0(0.015) & -1.1(7) & 0(0.2) & 0(0.01) & -1.1(7) & 0(0.18) & 0(0.009)\\
a12m220 & -1.3(7) & 0(0.224) & 0(0.0057) & -1.3(7) & 0(0.15) & 0(0.004) & -1.3(7) & 0(0.137) & 0(0.004)\\
\hline
& $f_0^\pi$ & $f_{\textrm{osc.}}^\pi$ & $f_1^\pi$ &  $f_0^K$ & $f_{\textrm{osc.}}^K$ & $f_1^K$ &  $f_0^{ss}$ & $f_{\textrm{osc.}}^{ss}$ & $f_1^{ss}$ \\
\hline
a15m310 & 0.0387(387) & 0(0.0387) & 0(0.0387) & 0.054(54) & 0(0.054) & 0(0.054) & 0.0648(648) & 0(0.0648) & 0(0.0648)\\
a12m310 & 0.028(20) & 0(0.028) & 0(0.028) & 0.04(4) & 0(0.04) & 0(0.04)& 0.0485(485) & 0(0.0485) & 0(0.0485)\\
a09m310 & 0.0175(175) & 0(0.0175) & 0(0.0175) & 0.0256(256) & 0(0.0256) & 0(0.0256) & 0.0318(318) & 0(0.0318) & 0(0.0318)\\
a15m220 & 0.0309(309) & 0(0.0309) & 0(0.0309) & 0.0522(522) & 0(0.0522) & 0(0.0522) & 0.0636(636) & 0(0.0636) & 0(0.0636)\\
a12m220 & 0.0221(221) & 0(0.0221) & 0(0.0221) & 0.0375(375) & 0(0.0375) & 0(0.0375) & 0.047(47) & 0(0.047) & 0(0.047)\\
\hline
& $E^{N}_0$ & $z_{0,p}^N$ & $z_{0,s}^N$ & $\D^{N}_1$ & $z_{1,p}^N$ & $z_{1,s}^N$ & $m_{res}^l$ & $m_{res}^s$ &\\
\hline
a15m310 & 0.820(82) & 0.0112(55) & 4.1(4.1)E-4 & -0.75(70) & 0(0.112) & 0(0.0021) & 0(1) & 0(1)\\
a12m310 & 0.670(67) & 0.006(3) & 2.6(2.6)E-4 & -1.0(7) & 0(0.06) & 0(0.0013) & 0(1) & 0(1)\\
a09m310 & 0.50(5) & 0.0024(12) & 2.2(2.2)E-5 & -1.27(68) & 0(0.024) & 0(1.1)E-4 & 0(1) & 0(1)\\
a15m220 & 0.760(76) & 0.011(5) & 4.2(4.2)E-4 & -1.1(7) & 0(0.11) & 0(0.0021) & 0(1) & 0(1)\\
a12m220 & 0.610(61) & 0.0054(27) & 7.9(7.0)E-5 & -1.3(7) & 0(0.054) & 0(4)E-4 & 0(1) & 0(1)\\
\end{tabular}
\end{ruledtabular}
\end{table*}

\section{Correlator analysis fit regions}
A summary of the fit regions for the two-point function analysis is shown in Table~\ref{tab:fit_region} for the three different ensembles used in this work. $q_1q_2$ superscripts identify mesonic states ($\pi$, $\bar{s}\g_5 s$, and $K$.)

\begin{table}
\caption{Fit range in lattice units. The fit region is chosen to be approximately the same in physical units for all pseudoscalar correlator fits, as well as among the nucleon correlator fits. The nucleon correlation functions are fit closer to the origin because of the poorer signal-to-noise ratio as compared to pseudoscalar observables.\label{tab:fit_region}}
\begin{ruledtabular}
\begin{tabular}{ccccc}
$a$& $C^{q_1q_2}$ t$_{\textrm{min}}$ & $C^{q_1q_2}$ t$_{\textrm{max}}$ & $C^N$ t$_{\textrm{min}}$& $C^N$ t$_{\textrm{max}}$\\
\hline
0.15 fm & 7 & 15 & 4 & 10 \\
0.12 fm & 8 & 19 & 5 & 12 \\
0.09 fm & 12 & 25 & 7 & 16
\end{tabular}
\end{ruledtabular}
\end{table}

%
\section{Topological charge evolution on HISQ ensembles}
\begin{figure}
\includegraphics[width=\columnwidth]{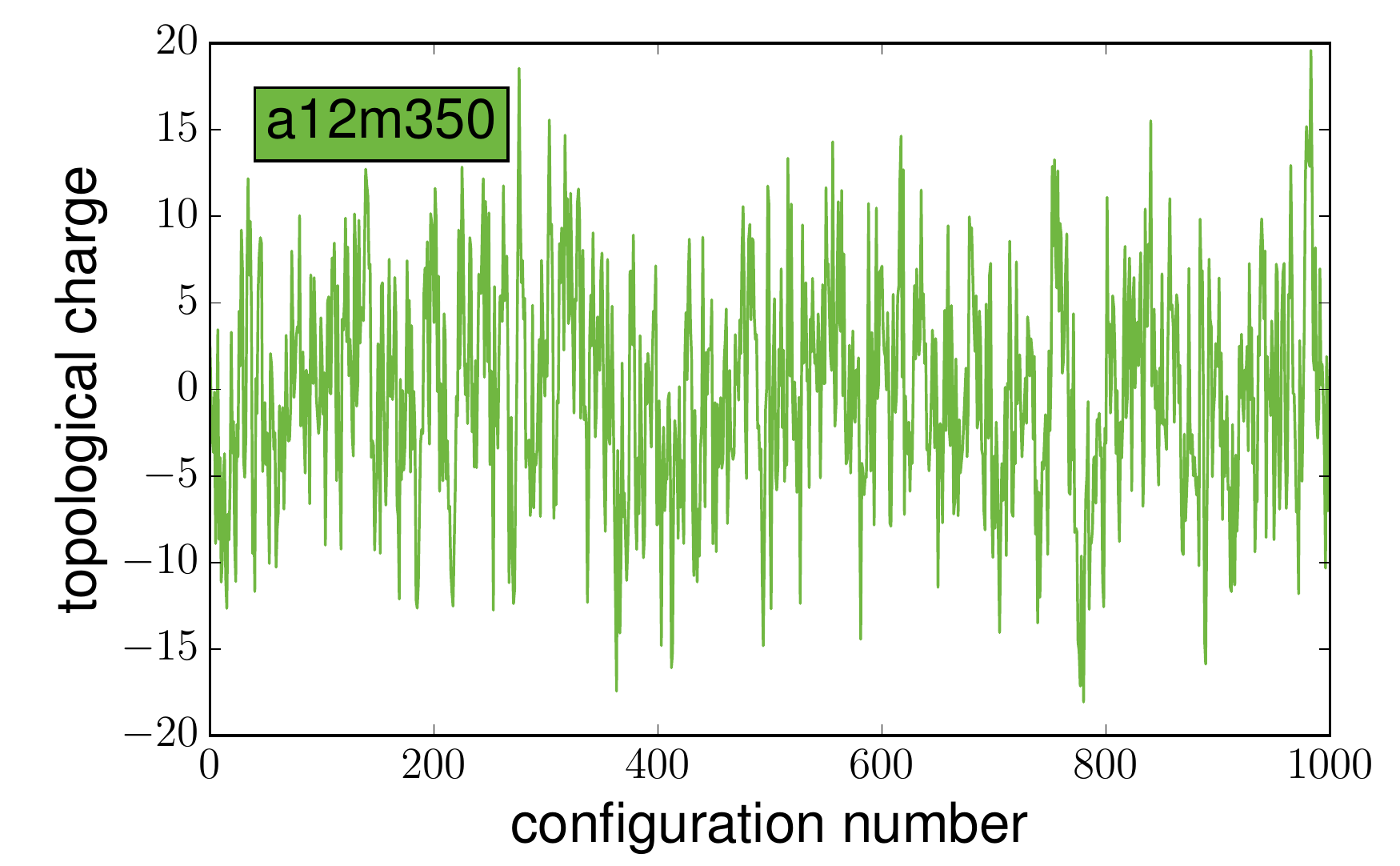}
\includegraphics[width=\columnwidth]{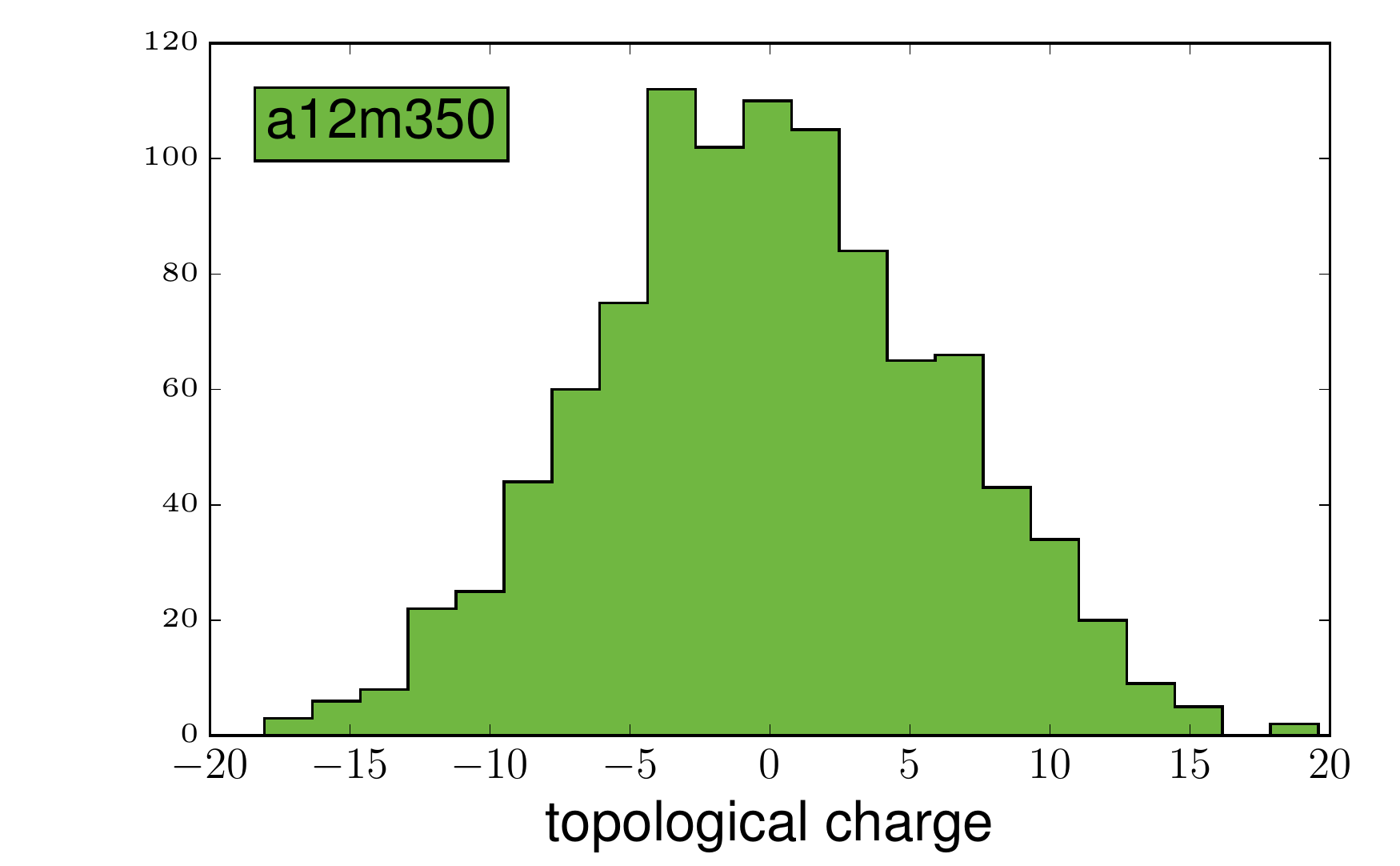}
\includegraphics[width=\columnwidth]{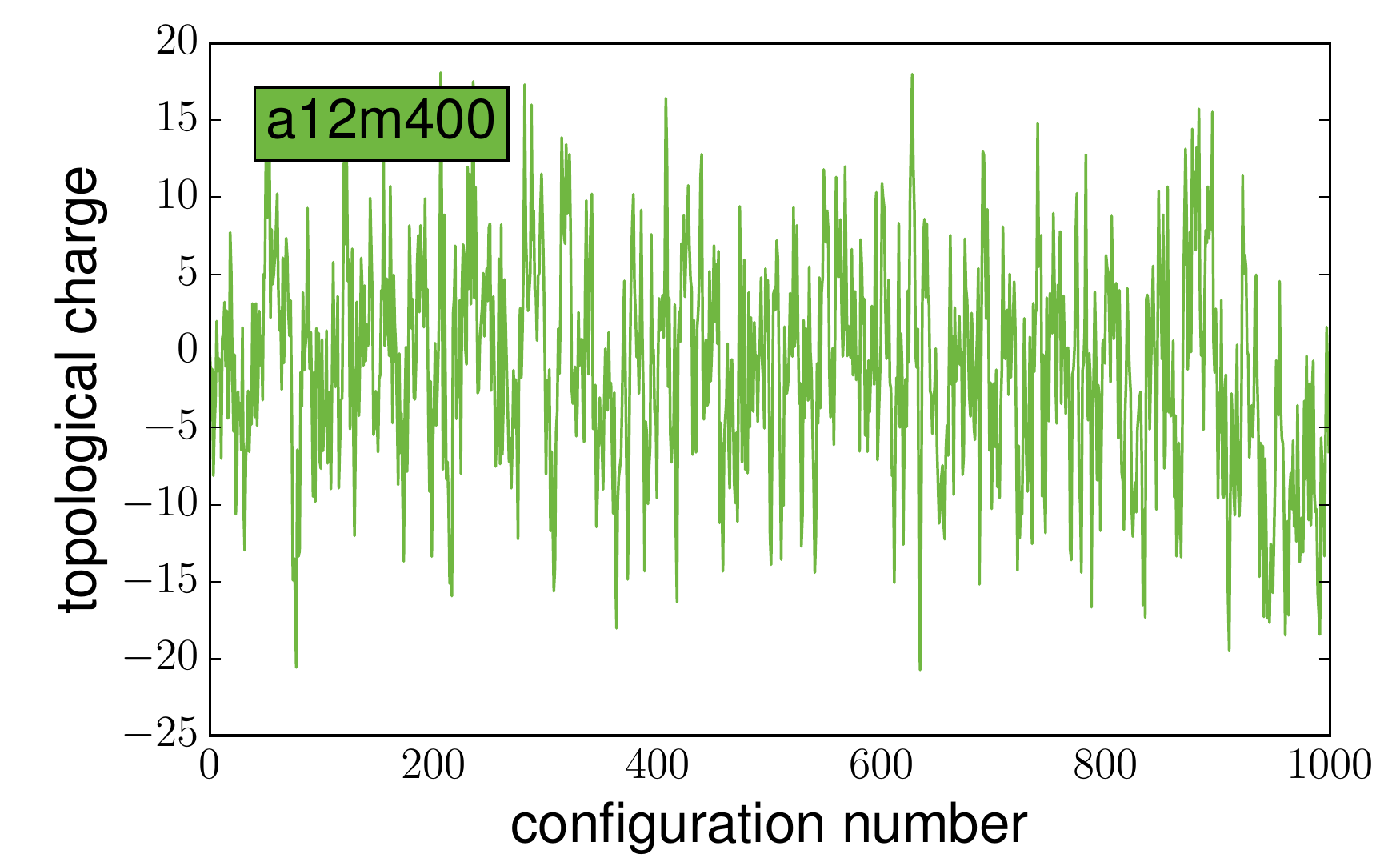}
\includegraphics[width=\columnwidth]{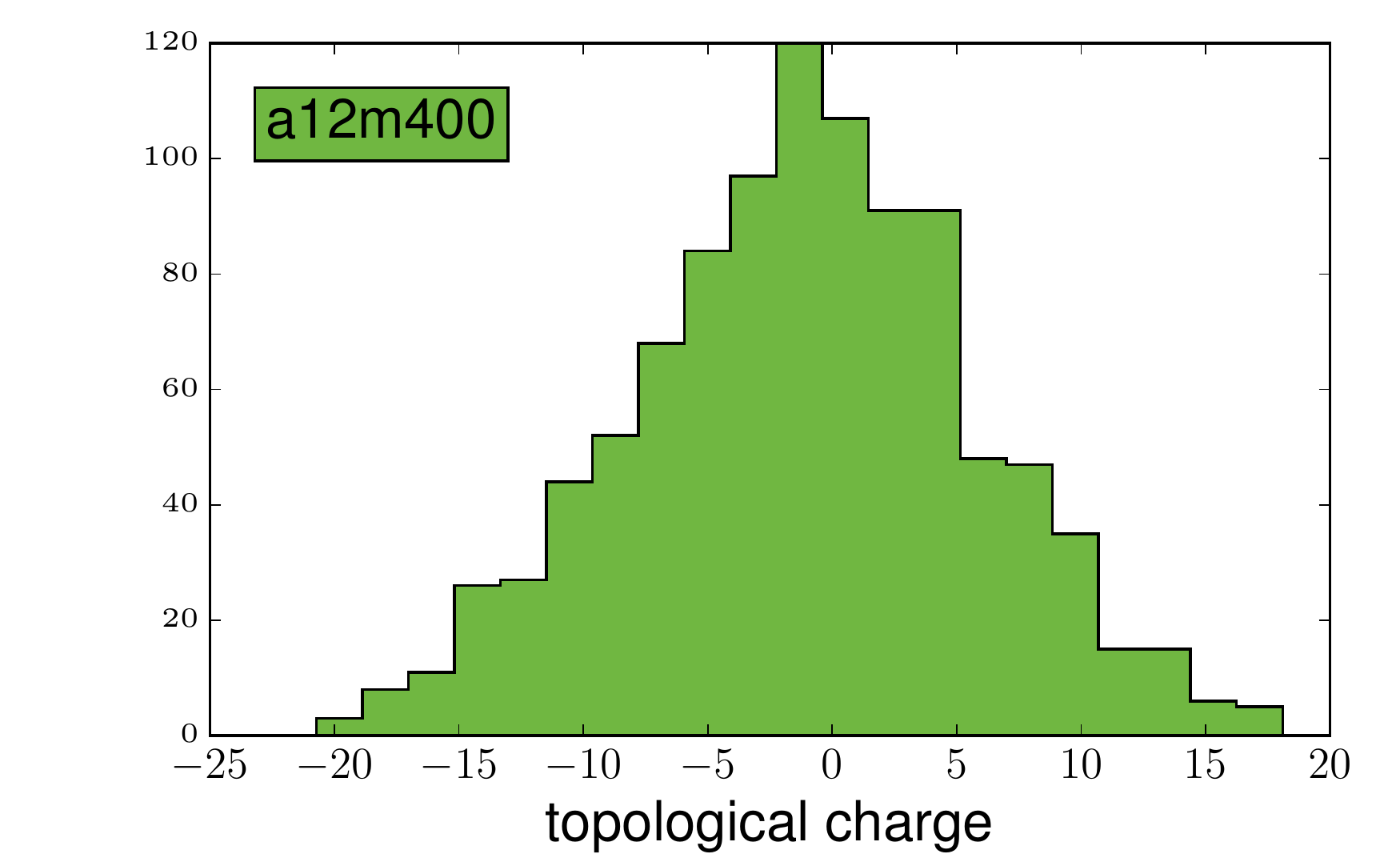}
\caption{\label{fig:topological_charge}
{Topological charge of the a12m350 and a12m400 ensembles at flow-time $t_{gf} = 0.99$. The topological charge randomly fluctuates and shows no long correlation as a function of configuration number (Monte Carlo time) for both ensembles.
The histograms show that the fluctuations are centered around zero, indicating the absence of CP (charge-parity) violation, and are Gaussian distributed, indicating that the volumes are sufficiently large.}
}
\end{figure}

In this Appendix, we provide additional details for the $N_f=2+1+1$ HISQ ensembles at heavy pion masses ($m_\pi \approx$ 350 and 400 MeV).
The ensembles have a lattice spacing of $\approx 0.12$ fm, and we expect the topological charge to fluctuate along the molecular dynamics trajectory and be Gaussian distributed.
This behavior is plotted in Fig.~\ref{fig:topological_charge} for both ensembles.
Each of the new ensembles is obtained by combining configurations from eight independent streams (collected after each stream has thermalized), and they are plotted together in Fig.~\ref{fig:topological_charge}.
We solve the gradient flow equations with the Symanzik action to smooth out the HISQ gauge fields, with a step size of $\epsilon=0.03$ and up to $n=166$ iterations.
We use the symmetric Clover discretization of the bosonic topological charge density operator $G_{\mu\nu}\tilde{G}_{\mu\nu}$.

\bibliography{c51_bib}

\begin{thebibliography}{165}%
\makeatletter
\providecommand \@ifxundefined [1]{%
 \@ifx{#1\undefined}
}%
\providecommand \@ifnum [1]{%
 \ifnum #1\expandafter \@firstoftwo
 \else \expandafter \@secondoftwo
 \fi
}%
\providecommand \@ifx [1]{%
 \ifx #1\expandafter \@firstoftwo
 \else \expandafter \@secondoftwo
 \fi
}%
\providecommand \natexlab [1]{#1}%
\providecommand \enquote  [1]{``#1''}%
\providecommand \bibnamefont  [1]{#1}%
\providecommand \bibfnamefont [1]{#1}%
\providecommand \citenamefont [1]{#1}%
\providecommand \href@noop [0]{\@secondoftwo}%
\providecommand \href [0]{\begingroup \@sanitize@url \@href}%
\providecommand \@href[1]{\@@startlink{#1}\@@href}%
\providecommand \@@href[1]{\endgroup#1\@@endlink}%
\providecommand \@sanitize@url [0]{\catcode `\\12\catcode `\$12\catcode
  `\&12\catcode `\#12\catcode `\^12\catcode `\_12\catcode `\%12\relax}%
\providecommand \@@startlink[1]{}%
\providecommand \@@endlink[0]{}%
\providecommand \url  [0]{\begingroup\@sanitize@url \@url }%
\providecommand \@url [1]{\endgroup\@href {#1}{\urlprefix }}%
\providecommand \urlprefix  [0]{URL }%
\providecommand \Eprint [0]{\href }%
\providecommand \doibase [0]{http://dx.doi.org/}%
\providecommand \selectlanguage [0]{\@gobble}%
\providecommand \bibinfo  [0]{\@secondoftwo}%
\providecommand \bibfield  [0]{\@secondoftwo}%
\providecommand \translation [1]{[#1]}%
\providecommand \BibitemOpen [0]{}%
\providecommand \bibitemStop [0]{}%
\providecommand \bibitemNoStop [0]{.\EOS\space}%
\providecommand \EOS [0]{\spacefactor3000\relax}%
\providecommand \BibitemShut  [1]{\csname bibitem#1\endcsname}%
\let\auto@bib@innerbib\@empty
\bibitem [{\citenamefont {Fritzsch}\ and\ \citenamefont
  {Gell-Mann}(1972)}]{Fritzsch:1972jv}%
  \BibitemOpen
  \bibfield  {author} {\bibinfo {author} {\bibfnamefont {Harald}\ \bibnamefont
  {Fritzsch}}\ and\ \bibinfo {author} {\bibfnamefont {Murray}\ \bibnamefont
  {Gell-Mann}},\ }\bibfield  {title} {\enquote {\bibinfo {title} {{Current
  algebra: Quarks and what else?}}}\ }\bibfield  {booktitle} {\emph {\bibinfo
  {booktitle} {{Proceedings, 16th International Conference on High-Energy
  Physics, ICHEP, Batavia, Illinois, 6-13 Sep 1972}}},\ }\href@noop {}
  {\bibfield  {journal} {\bibinfo  {journal} {eConf}\ }\textbf {\bibinfo
  {volume} {C720906V2}},\ \bibinfo {pages} {135--165} (\bibinfo {year}
  {1972})},\ \Eprint {http://arxiv.org/abs/hep-ph/0208010}
  {arXiv:hep-ph/0208010 [hep-ph]} \BibitemShut {NoStop}%
\bibitem [{\citenamefont {Fritzsch}\ \emph {et~al.}(1973)\citenamefont
  {Fritzsch}, \citenamefont {Gell-Mann},\ and\ \citenamefont
  {Leutwyler}}]{Fritzsch:1973pi}%
  \BibitemOpen
  \bibfield  {author} {\bibinfo {author} {\bibfnamefont {H.}~\bibnamefont
  {Fritzsch}}, \bibinfo {author} {\bibfnamefont {Murray}\ \bibnamefont
  {Gell-Mann}}, \ and\ \bibinfo {author} {\bibfnamefont {H.}~\bibnamefont
  {Leutwyler}},\ }\bibfield  {title} {\enquote {\bibinfo {title} {{Advantages
  of the Color Octet Gluon Picture}},}\ }\href {\doibase
  10.1016/0370-2693(73)90625-4} {\bibfield  {journal} {\bibinfo  {journal}
  {Phys. Lett.}\ }\textbf {\bibinfo {volume} {B47}},\ \bibinfo {pages}
  {365--368} (\bibinfo {year} {1973})}\BibitemShut {NoStop}%
\bibitem [{\citenamefont {Gross}\ and\ \citenamefont
  {Wilczek}(1973)}]{Gross:1973id}%
  \BibitemOpen
  \bibfield  {author} {\bibinfo {author} {\bibfnamefont {David~J.}\
  \bibnamefont {Gross}}\ and\ \bibinfo {author} {\bibfnamefont {Frank}\
  \bibnamefont {Wilczek}},\ }\bibfield  {title} {\enquote {\bibinfo {title}
  {{Ultraviolet Behavior of Nonabelian Gauge Theories}},}\ }\href {\doibase
  10.1103/PhysRevLett.30.1343} {\bibfield  {journal} {\bibinfo  {journal}
  {Phys. Rev. Lett.}\ }\textbf {\bibinfo {volume} {30}},\ \bibinfo {pages}
  {1343--1346} (\bibinfo {year} {1973})}\BibitemShut {NoStop}%
\bibitem [{\citenamefont {Politzer}(1973)}]{Politzer:1973fx}%
  \BibitemOpen
  \bibfield  {author} {\bibinfo {author} {\bibfnamefont {H.~David}\
  \bibnamefont {Politzer}},\ }\bibfield  {title} {\enquote {\bibinfo {title}
  {{Reliable Perturbative Results for Strong Interactions?}}}\ }\href {\doibase
  10.1103/PhysRevLett.30.1346} {\bibfield  {journal} {\bibinfo  {journal}
  {Phys. Rev. Lett.}\ }\textbf {\bibinfo {volume} {30}},\ \bibinfo {pages}
  {1346--1349} (\bibinfo {year} {1973})}\BibitemShut {NoStop}%
\bibitem [{\citenamefont {Weinberg}(1979)}]{Weinberg:1978kz}%
  \BibitemOpen
  \bibfield  {author} {\bibinfo {author} {\bibfnamefont {Steven}\ \bibnamefont
  {Weinberg}},\ }\bibfield  {title} {\enquote {\bibinfo {title}
  {{Phenomenological Lagrangians}},}\ }\href {\doibase
  10.1016/0378-4371(79)90223-1} {\bibfield  {journal} {\bibinfo  {journal}
  {Physica}\ }\textbf {\bibinfo {volume} {A96}},\ \bibinfo {pages} {327--340}
  (\bibinfo {year} {1979})}\BibitemShut {NoStop}%
\bibitem [{\citenamefont {Symanzik}(1983{\natexlab{a}})}]{Symanzik:1983dc}%
  \BibitemOpen
  \bibfield  {author} {\bibinfo {author} {\bibfnamefont {K.}~\bibnamefont
  {Symanzik}},\ }\bibfield  {title} {\enquote {\bibinfo {title} {{Continuum
  Limit and Improved Action in Lattice Theories. 1. Principles and phi**4
  Theory}},}\ }\href {\doibase 10.1016/0550-3213(83)90468-6} {\bibfield
  {journal} {\bibinfo  {journal} {Nucl. Phys.}\ }\textbf {\bibinfo {volume}
  {B226}},\ \bibinfo {pages} {187--204} (\bibinfo {year}
  {1983}{\natexlab{a}})}\BibitemShut {NoStop}%
\bibitem [{\citenamefont {Symanzik}(1983{\natexlab{b}})}]{Symanzik:1983gh}%
  \BibitemOpen
  \bibfield  {author} {\bibinfo {author} {\bibfnamefont {K.}~\bibnamefont
  {Symanzik}},\ }\bibfield  {title} {\enquote {\bibinfo {title} {{Continuum
  Limit and Improved Action in Lattice Theories. 2. O(N) Nonlinear Sigma Model
  in Perturbation Theory}},}\ }\href {\doibase 10.1016/0550-3213(83)90469-8}
  {\bibfield  {journal} {\bibinfo  {journal} {Nucl. Phys.}\ }\textbf {\bibinfo
  {volume} {B226}},\ \bibinfo {pages} {205--227} (\bibinfo {year}
  {1983}{\natexlab{b}})}\BibitemShut {NoStop}%
\bibitem [{\citenamefont {Aoki}\ \emph {et~al.}(2017)\citenamefont {Aoki} \emph
  {et~al.}}]{Aoki:2016frl}%
  \BibitemOpen
  \bibfield  {author} {\bibinfo {author} {\bibfnamefont {S.}~\bibnamefont
  {Aoki}} \emph {et~al.},\ }\bibfield  {title} {\enquote {\bibinfo {title}
  {{Review of lattice results concerning low-energy particle physics}},}\
  }\href {\doibase 10.1140/epjc/s10052-016-4509-7} {\bibfield  {journal}
  {\bibinfo  {journal} {Eur. Phys. J.}\ }\textbf {\bibinfo {volume} {C77}},\
  \bibinfo {pages} {112} (\bibinfo {year} {2017})},\ \Eprint
  {http://arxiv.org/abs/1607.00299} {arXiv:1607.00299 [hep-lat]} \BibitemShut
  {NoStop}%
\bibitem [{\citenamefont {Wilson}(1974)}]{Wilson:1974sk}%
  \BibitemOpen
  \bibfield  {author} {\bibinfo {author} {\bibfnamefont {Kenneth~G.}\
  \bibnamefont {Wilson}},\ }\bibfield  {title} {\enquote {\bibinfo {title}
  {{Confinement of Quarks}},}\ }\href {\doibase 10.1103/PhysRevD.10.2445}
  {\bibfield  {journal} {\bibinfo  {journal} {Phys. Rev.}\ }\textbf {\bibinfo
  {volume} {D10}},\ \bibinfo {pages} {2445--2459} (\bibinfo {year}
  {1974})}\BibitemShut {NoStop}%
\bibitem [{\citenamefont {Sheikholeslami}\ and\ \citenamefont
  {Wohlert}(1985)}]{Sheikholeslami:1985ij}%
  \BibitemOpen
  \bibfield  {author} {\bibinfo {author} {\bibfnamefont {B.}~\bibnamefont
  {Sheikholeslami}}\ and\ \bibinfo {author} {\bibfnamefont {R.}~\bibnamefont
  {Wohlert}},\ }\bibfield  {title} {\enquote {\bibinfo {title} {{Improved
  Continuum Limit Lattice Action for QCD with Wilson Fermions}},}\ }\href
  {\doibase 10.1016/0550-3213(85)90002-1} {\bibfield  {journal} {\bibinfo
  {journal} {Nucl. Phys.}\ }\textbf {\bibinfo {volume} {B259}},\ \bibinfo
  {pages} {572} (\bibinfo {year} {1985})}\BibitemShut {NoStop}%
\bibitem [{\citenamefont {Frezzotti}\ \emph {et~al.}(2001)\citenamefont
  {Frezzotti}, \citenamefont {Grassi}, \citenamefont {Sint},\ and\
  \citenamefont {Weisz}}]{Frezzotti:2000nk}%
  \BibitemOpen
  \bibfield  {author} {\bibinfo {author} {\bibfnamefont {Roberto}\ \bibnamefont
  {Frezzotti}}, \bibinfo {author} {\bibfnamefont {Pietro~Antonio}\ \bibnamefont
  {Grassi}}, \bibinfo {author} {\bibfnamefont {Stefan}\ \bibnamefont {Sint}}, \
  and\ \bibinfo {author} {\bibfnamefont {Peter}\ \bibnamefont {Weisz}}
  (\bibinfo {collaboration} {Alpha}),\ }\bibfield  {title} {\enquote {\bibinfo
  {title} {{Lattice QCD with a chirally twisted mass term}},}\ }\href@noop {}
  {\bibfield  {journal} {\bibinfo  {journal} {JHEP}\ }\textbf {\bibinfo
  {volume} {08}},\ \bibinfo {pages} {058} (\bibinfo {year} {2001})},\ \Eprint
  {http://arxiv.org/abs/hep-lat/0101001} {arXiv:hep-lat/0101001 [hep-lat]}
  \BibitemShut {NoStop}%
\bibitem [{\citenamefont {Frezzotti}\ and\ \citenamefont
  {Rossi}(2004)}]{Frezzotti:2003ni}%
  \BibitemOpen
  \bibfield  {author} {\bibinfo {author} {\bibfnamefont {R.}~\bibnamefont
  {Frezzotti}}\ and\ \bibinfo {author} {\bibfnamefont {G.~C.}\ \bibnamefont
  {Rossi}},\ }\bibfield  {title} {\enquote {\bibinfo {title} {{Chirally
  improving Wilson fermions. 1. O(a) improvement}},}\ }\href {\doibase
  10.1088/1126-6708/2004/08/007} {\bibfield  {journal} {\bibinfo  {journal}
  {JHEP}\ }\textbf {\bibinfo {volume} {08}},\ \bibinfo {pages} {007} (\bibinfo
  {year} {2004})},\ \Eprint {http://arxiv.org/abs/hep-lat/0306014}
  {arXiv:hep-lat/0306014 [hep-lat]} \BibitemShut {NoStop}%
\bibitem [{\citenamefont {Kogut}\ and\ \citenamefont
  {Susskind}(1975)}]{Kogut:1974ag}%
  \BibitemOpen
  \bibfield  {author} {\bibinfo {author} {\bibfnamefont {John~B.}\ \bibnamefont
  {Kogut}}\ and\ \bibinfo {author} {\bibfnamefont {Leonard}\ \bibnamefont
  {Susskind}},\ }\bibfield  {title} {\enquote {\bibinfo {title} {{Hamiltonian
  Formulation of Wilson's Lattice Gauge Theories}},}\ }\href {\doibase
  10.1103/PhysRevD.11.395} {\bibfield  {journal} {\bibinfo  {journal} {Phys.
  Rev.}\ }\textbf {\bibinfo {volume} {D11}},\ \bibinfo {pages} {395--408}
  (\bibinfo {year} {1975})}\BibitemShut {NoStop}%
\bibitem [{\citenamefont {Susskind}(1977)}]{Susskind:1976jm}%
  \BibitemOpen
  \bibfield  {author} {\bibinfo {author} {\bibfnamefont {Leonard}\ \bibnamefont
  {Susskind}},\ }\bibfield  {title} {\enquote {\bibinfo {title} {{Lattice
  Fermions}},}\ }\href {\doibase 10.1103/PhysRevD.16.3031} {\bibfield
  {journal} {\bibinfo  {journal} {Phys. Rev.}\ }\textbf {\bibinfo {volume}
  {D16}},\ \bibinfo {pages} {3031--3039} (\bibinfo {year} {1977})}\BibitemShut
  {NoStop}%
\bibitem [{\citenamefont {Marinari}\ \emph {et~al.}(1981)\citenamefont
  {Marinari}, \citenamefont {Parisi},\ and\ \citenamefont
  {Rebbi}}]{Marinari:1981qf}%
  \BibitemOpen
  \bibfield  {author} {\bibinfo {author} {\bibfnamefont {E.}~\bibnamefont
  {Marinari}}, \bibinfo {author} {\bibfnamefont {G.}~\bibnamefont {Parisi}}, \
  and\ \bibinfo {author} {\bibfnamefont {C.}~\bibnamefont {Rebbi}},\ }\bibfield
   {title} {\enquote {\bibinfo {title} {{Monte Carlo Simulation of the Massive
  Schwinger Model}},}\ }\href {\doibase 10.1016/0550-3213(81)90048-1}
  {\bibfield  {journal} {\bibinfo  {journal} {Nucl. Phys.}\ }\textbf {\bibinfo
  {volume} {B190}},\ \bibinfo {pages} {734} (\bibinfo {year} {1981})},\
  \bibinfo {note} {[,595(1981)]}\BibitemShut {NoStop}%
\bibitem [{\citenamefont {Bernard}\ \emph {et~al.}(2007)\citenamefont
  {Bernard}, \citenamefont {Golterman}, \citenamefont {Shamir},\ and\
  \citenamefont {Sharpe}}]{Bernard:2006vv}%
  \BibitemOpen
  \bibfield  {author} {\bibinfo {author} {\bibfnamefont {Claude}\ \bibnamefont
  {Bernard}}, \bibinfo {author} {\bibfnamefont {Maarten}\ \bibnamefont
  {Golterman}}, \bibinfo {author} {\bibfnamefont {Yigal}\ \bibnamefont
  {Shamir}}, \ and\ \bibinfo {author} {\bibfnamefont {Stephen~R.}\ \bibnamefont
  {Sharpe}},\ }\bibfield  {title} {\enquote {\bibinfo {title} {{Comment on
  `Chiral anomalies and rooted staggered fermions'}},}\ }\href {\doibase
  10.1016/j.physletb.2007.04.018} {\bibfield  {journal} {\bibinfo  {journal}
  {Phys. Lett.}\ }\textbf {\bibinfo {volume} {B649}},\ \bibinfo {pages}
  {235--240} (\bibinfo {year} {2007})},\ \Eprint
  {http://arxiv.org/abs/hep-lat/0603027} {arXiv:hep-lat/0603027 [hep-lat]}
  \BibitemShut {NoStop}%
\bibitem [{\citenamefont {Bernard}\ \emph {et~al.}(2006)\citenamefont
  {Bernard}, \citenamefont {Golterman},\ and\ \citenamefont
  {Shamir}}]{Bernard:2006ee}%
  \BibitemOpen
  \bibfield  {author} {\bibinfo {author} {\bibfnamefont {Claude}\ \bibnamefont
  {Bernard}}, \bibinfo {author} {\bibfnamefont {Maarten}\ \bibnamefont
  {Golterman}}, \ and\ \bibinfo {author} {\bibfnamefont {Yigal}\ \bibnamefont
  {Shamir}},\ }\bibfield  {title} {\enquote {\bibinfo {title} {{Observations on
  staggered fermions at non-zero lattice spacing}},}\ }\href {\doibase
  10.1103/PhysRevD.73.114511} {\bibfield  {journal} {\bibinfo  {journal} {Phys.
  Rev.}\ }\textbf {\bibinfo {volume} {D73}},\ \bibinfo {pages} {114511}
  (\bibinfo {year} {2006})},\ \Eprint {http://arxiv.org/abs/hep-lat/0604017}
  {arXiv:hep-lat/0604017 [hep-lat]} \BibitemShut {NoStop}%
\bibitem [{\citenamefont {Creutz}(2007)}]{Creutz:2007yg}%
  \BibitemOpen
  \bibfield  {author} {\bibinfo {author} {\bibfnamefont {Michael}\ \bibnamefont
  {Creutz}},\ }\bibfield  {title} {\enquote {\bibinfo {title} {{Chiral
  anomalies and rooted staggered fermions}},}\ }\href {\doibase
  10.1016/j.physletb.2007.03.065} {\bibfield  {journal} {\bibinfo  {journal}
  {Phys. Lett.}\ }\textbf {\bibinfo {volume} {B649}},\ \bibinfo {pages}
  {230--234} (\bibinfo {year} {2007})},\ \Eprint
  {http://arxiv.org/abs/hep-lat/0701018} {arXiv:hep-lat/0701018 [hep-lat]}
  \BibitemShut {NoStop}%
\bibitem [{\citenamefont {Bernard}(2005)}]{Bernard:2004ab}%
  \BibitemOpen
  \bibfield  {author} {\bibinfo {author} {\bibfnamefont {C.}~\bibnamefont
  {Bernard}},\ }\bibfield  {title} {\enquote {\bibinfo {title} {{Order of the
  chiral and continuum limits in staggered chiral perturbation theory}},}\
  }\href {\doibase 10.1103/PhysRevD.71.094020} {\bibfield  {journal} {\bibinfo
  {journal} {Phys. Rev.}\ }\textbf {\bibinfo {volume} {D71}},\ \bibinfo {pages}
  {094020} (\bibinfo {year} {2005})},\ \Eprint
  {http://arxiv.org/abs/hep-lat/0412030} {arXiv:hep-lat/0412030 [hep-lat]}
  \BibitemShut {NoStop}%
\bibitem [{\citenamefont {Bernard}(2006)}]{Bernard:2006zw}%
  \BibitemOpen
  \bibfield  {author} {\bibinfo {author} {\bibfnamefont {C.}~\bibnamefont
  {Bernard}},\ }\bibfield  {title} {\enquote {\bibinfo {title} {{Staggered
  chiral perturbation theory and the fourth-root trick}},}\ }\href {\doibase
  10.1103/PhysRevD.73.114503} {\bibfield  {journal} {\bibinfo  {journal} {Phys.
  Rev.}\ }\textbf {\bibinfo {volume} {D73}},\ \bibinfo {pages} {114503}
  (\bibinfo {year} {2006})},\ \Eprint {http://arxiv.org/abs/hep-lat/0603011}
  {arXiv:hep-lat/0603011 [hep-lat]} \BibitemShut {NoStop}%
\bibitem [{\citenamefont {Shamir}(2005)}]{Shamir:2004zc}%
  \BibitemOpen
  \bibfield  {author} {\bibinfo {author} {\bibfnamefont {Yigal}\ \bibnamefont
  {Shamir}},\ }\bibfield  {title} {\enquote {\bibinfo {title} {{Locality of the
  fourth root of the staggered-fermion determinant: Renormalization-group
  approach}},}\ }\href {\doibase 10.1103/PhysRevD.71.034509} {\bibfield
  {journal} {\bibinfo  {journal} {Phys. Rev.}\ }\textbf {\bibinfo {volume}
  {D71}},\ \bibinfo {pages} {034509} (\bibinfo {year} {2005})},\ \Eprint
  {http://arxiv.org/abs/hep-lat/0412014} {arXiv:hep-lat/0412014 [hep-lat]}
  \BibitemShut {NoStop}%
\bibitem [{\citenamefont {Shamir}(2007)}]{Shamir:2006nj}%
  \BibitemOpen
  \bibfield  {author} {\bibinfo {author} {\bibfnamefont {Yigal}\ \bibnamefont
  {Shamir}},\ }\bibfield  {title} {\enquote {\bibinfo {title}
  {{Renormalization-group analysis of the validity of staggered-fermion QCD
  with the fourth-root recipe}},}\ }\href {\doibase 10.1103/PhysRevD.75.054503}
  {\bibfield  {journal} {\bibinfo  {journal} {Phys. Rev.}\ }\textbf {\bibinfo
  {volume} {D75}},\ \bibinfo {pages} {054503} (\bibinfo {year} {2007})},\
  \Eprint {http://arxiv.org/abs/hep-lat/0607007} {arXiv:hep-lat/0607007
  [hep-lat]} \BibitemShut {NoStop}%
\bibitem [{\citenamefont {Bernard}\ \emph
  {et~al.}(2008{\natexlab{a}})\citenamefont {Bernard}, \citenamefont
  {Golterman},\ and\ \citenamefont {Shamir}}]{Bernard:2007ma}%
  \BibitemOpen
  \bibfield  {author} {\bibinfo {author} {\bibfnamefont {Claude}\ \bibnamefont
  {Bernard}}, \bibinfo {author} {\bibfnamefont {Maarten}\ \bibnamefont
  {Golterman}}, \ and\ \bibinfo {author} {\bibfnamefont {Yigal}\ \bibnamefont
  {Shamir}},\ }\bibfield  {title} {\enquote {\bibinfo {title} {{Effective field
  theories for QCD with rooted staggered fermions}},}\ }\href {\doibase
  10.1103/PhysRevD.77.074505} {\bibfield  {journal} {\bibinfo  {journal} {Phys.
  Rev.}\ }\textbf {\bibinfo {volume} {D77}},\ \bibinfo {pages} {074505}
  (\bibinfo {year} {2008}{\natexlab{a}})},\ \Eprint
  {http://arxiv.org/abs/0712.2560} {arXiv:0712.2560 [hep-lat]} \BibitemShut
  {NoStop}%
\bibitem [{\citenamefont {Durr}\ and\ \citenamefont
  {Hoelbling}(2005)}]{Durr:2004ta}%
  \BibitemOpen
  \bibfield  {author} {\bibinfo {author} {\bibfnamefont {Stephan}\ \bibnamefont
  {Durr}}\ and\ \bibinfo {author} {\bibfnamefont {Christian}\ \bibnamefont
  {Hoelbling}},\ }\bibfield  {title} {\enquote {\bibinfo {title} {{Scaling
  tests with dynamical overlap and rooted staggered fermions}},}\ }\href
  {\doibase 10.1103/PhysRevD.71.054501} {\bibfield  {journal} {\bibinfo
  {journal} {Phys. Rev.}\ }\textbf {\bibinfo {volume} {D71}},\ \bibinfo {pages}
  {054501} (\bibinfo {year} {2005})},\ \Eprint
  {http://arxiv.org/abs/hep-lat/0411022} {arXiv:hep-lat/0411022 [hep-lat]}
  \BibitemShut {NoStop}%
\bibitem [{\citenamefont {Durr}\ and\ \citenamefont
  {Hoelbling}(2006)}]{Durr:2006ze}%
  \BibitemOpen
  \bibfield  {author} {\bibinfo {author} {\bibfnamefont {Stephan}\ \bibnamefont
  {Durr}}\ and\ \bibinfo {author} {\bibfnamefont {Christian}\ \bibnamefont
  {Hoelbling}},\ }\bibfield  {title} {\enquote {\bibinfo {title} {{Lattice
  fermions with complex mass}},}\ }\href {\doibase 10.1103/PhysRevD.74.014513}
  {\bibfield  {journal} {\bibinfo  {journal} {Phys. Rev.}\ }\textbf {\bibinfo
  {volume} {D74}},\ \bibinfo {pages} {014513} (\bibinfo {year} {2006})},\
  \Eprint {http://arxiv.org/abs/hep-lat/0604005} {arXiv:hep-lat/0604005
  [hep-lat]} \BibitemShut {NoStop}%
\bibitem [{\citenamefont {Hasenfratz}\ and\ \citenamefont
  {Hoffmann}(2006)}]{Hasenfratz:2006nw}%
  \BibitemOpen
  \bibfield  {author} {\bibinfo {author} {\bibfnamefont {Anna}\ \bibnamefont
  {Hasenfratz}}\ and\ \bibinfo {author} {\bibfnamefont {Roland}\ \bibnamefont
  {Hoffmann}},\ }\bibfield  {title} {\enquote {\bibinfo {title} {{Validity of
  the rooted staggered determinant in the continuum limit}},}\ }\href {\doibase
  10.1103/PhysRevD.74.014511} {\bibfield  {journal} {\bibinfo  {journal} {Phys.
  Rev.}\ }\textbf {\bibinfo {volume} {D74}},\ \bibinfo {pages} {014511}
  (\bibinfo {year} {2006})},\ \Eprint {http://arxiv.org/abs/hep-lat/0604010}
  {arXiv:hep-lat/0604010 [hep-lat]} \BibitemShut {NoStop}%
\bibitem [{\citenamefont {Bernard}\ \emph
  {et~al.}(2008{\natexlab{b}})\citenamefont {Bernard}, \citenamefont
  {Golterman}, \citenamefont {Shamir},\ and\ \citenamefont
  {Sharpe}}]{Bernard:2007eh}%
  \BibitemOpen
  \bibfield  {author} {\bibinfo {author} {\bibfnamefont {Claude}\ \bibnamefont
  {Bernard}}, \bibinfo {author} {\bibfnamefont {Maarten}\ \bibnamefont
  {Golterman}}, \bibinfo {author} {\bibfnamefont {Yigal}\ \bibnamefont
  {Shamir}}, \ and\ \bibinfo {author} {\bibfnamefont {Stephen~R.}\ \bibnamefont
  {Sharpe}},\ }\bibfield  {title} {\enquote {\bibinfo {title} {{'t Hooft
  vertices, partial quenching, and rooted staggered QCD}},}\ }\href {\doibase
  10.1103/PhysRevD.77.114504} {\bibfield  {journal} {\bibinfo  {journal} {Phys.
  Rev.}\ }\textbf {\bibinfo {volume} {D77}},\ \bibinfo {pages} {114504}
  (\bibinfo {year} {2008}{\natexlab{b}})},\ \Eprint
  {http://arxiv.org/abs/0711.0696} {arXiv:0711.0696 [hep-lat]} \BibitemShut
  {NoStop}%
\bibitem [{\citenamefont {Sharpe}(2006)}]{Sharpe:2006re}%
  \BibitemOpen
  \bibfield  {author} {\bibinfo {author} {\bibfnamefont {Stephen~R.}\
  \bibnamefont {Sharpe}},\ }\bibfield  {title} {\enquote {\bibinfo {title}
  {{Rooted staggered fermions: Good, bad or ugly?}}}\ }\bibfield  {booktitle}
  {\emph {\bibinfo {booktitle} {{Proceedings, 24th International Symposium on
  Lattice Field Theory (Lattice 2006): Tucson, USA, July 23-28, 2006}}},\
  }\href@noop {} {\bibfield  {journal} {\bibinfo  {journal} {PoS}\ }\textbf
  {\bibinfo {volume} {LAT2006}},\ \bibinfo {pages} {022} (\bibinfo {year}
  {2006})},\ \Eprint {http://arxiv.org/abs/hep-lat/0610094}
  {arXiv:hep-lat/0610094 [hep-lat]} \BibitemShut {NoStop}%
\bibitem [{\citenamefont {Kronfeld}(2007)}]{Kronfeld:2007ek}%
  \BibitemOpen
  \bibfield  {author} {\bibinfo {author} {\bibfnamefont {Andreas~S.}\
  \bibnamefont {Kronfeld}},\ }\bibfield  {title} {\enquote {\bibinfo {title}
  {{Lattice Gauge Theory with Staggered Fermions: How, Where, and Why
  (Not)}},}\ }\bibfield  {booktitle} {\emph {\bibinfo {booktitle}
  {{Proceedings, 25th International Symposium on Lattice field theory (Lattice
  2007): Regensburg, Germany, July 30-August 4, 2007}}},\ }\href@noop {}
  {\bibfield  {journal} {\bibinfo  {journal} {PoS}\ }\textbf {\bibinfo {volume}
  {LAT2007}},\ \bibinfo {pages} {016} (\bibinfo {year} {2007})},\ \Eprint
  {http://arxiv.org/abs/0711.0699} {arXiv:0711.0699 [hep-lat]} \BibitemShut
  {NoStop}%
\bibitem [{\citenamefont {Bazavov}\ \emph
  {et~al.}(2010{\natexlab{a}})\citenamefont {Bazavov} \emph
  {et~al.}}]{Bazavov:2009bb}%
  \BibitemOpen
  \bibfield  {author} {\bibinfo {author} {\bibfnamefont {A.}~\bibnamefont
  {Bazavov}} \emph {et~al.} (\bibinfo {collaboration} {MILC}),\ }\bibfield
  {title} {\enquote {\bibinfo {title} {{Nonperturbative QCD Simulations with
  2+1 Flavors of Improved Staggered Quarks}},}\ }\href {\doibase
  10.1103/RevModPhys.82.1349} {\bibfield  {journal} {\bibinfo  {journal} {Rev.
  Mod. Phys.}\ }\textbf {\bibinfo {volume} {82}},\ \bibinfo {pages}
  {1349--1417} (\bibinfo {year} {2010}{\natexlab{a}})},\ \Eprint
  {http://arxiv.org/abs/0903.3598} {arXiv:0903.3598 [hep-lat]} \BibitemShut
  {NoStop}%
\bibitem [{\citenamefont {Nielsen}\ and\ \citenamefont
  {Ninomiya}(1981{\natexlab{a}})}]{Nielsen:1980rz}%
  \BibitemOpen
  \bibfield  {author} {\bibinfo {author} {\bibfnamefont {Holger~Bech}\
  \bibnamefont {Nielsen}}\ and\ \bibinfo {author} {\bibfnamefont
  {M.}~\bibnamefont {Ninomiya}},\ }\bibfield  {title} {\enquote {\bibinfo
  {title} {{Absence of Neutrinos on a Lattice. 1. Proof by Homotopy Theory}},}\
  }\href {\doibase 10.1016/0550-3213(81)90361-8, 10.1016/0550-3213(82)90011-6}
  {\bibfield  {journal} {\bibinfo  {journal} {Nucl. Phys.}\ }\textbf {\bibinfo
  {volume} {B185}},\ \bibinfo {pages} {20} (\bibinfo {year}
  {1981}{\natexlab{a}})},\ \bibinfo {note} {[,533(1980)]}\BibitemShut {NoStop}%
\bibitem [{\citenamefont {Nielsen}\ and\ \citenamefont
  {Ninomiya}(1981{\natexlab{b}})}]{Nielsen:1981xu}%
  \BibitemOpen
  \bibfield  {author} {\bibinfo {author} {\bibfnamefont {Holger~Bech}\
  \bibnamefont {Nielsen}}\ and\ \bibinfo {author} {\bibfnamefont
  {M.}~\bibnamefont {Ninomiya}},\ }\bibfield  {title} {\enquote {\bibinfo
  {title} {{Absence of Neutrinos on a Lattice. 2. Intuitive Topological
  Proof}},}\ }\href {\doibase 10.1016/0550-3213(81)90524-1} {\bibfield
  {journal} {\bibinfo  {journal} {Nucl. Phys.}\ }\textbf {\bibinfo {volume}
  {B193}},\ \bibinfo {pages} {173--194} (\bibinfo {year}
  {1981}{\natexlab{b}})}\BibitemShut {NoStop}%
\bibitem [{\citenamefont {Nielsen}\ and\ \citenamefont
  {Ninomiya}(1981{\natexlab{c}})}]{Nielsen:1981hk}%
  \BibitemOpen
  \bibfield  {author} {\bibinfo {author} {\bibfnamefont {Holger~Bech}\
  \bibnamefont {Nielsen}}\ and\ \bibinfo {author} {\bibfnamefont
  {M.}~\bibnamefont {Ninomiya}},\ }\bibfield  {title} {\enquote {\bibinfo
  {title} {{No Go Theorem for Regularizing Chiral Fermions}},}\ }\href
  {\doibase 10.1016/0370-2693(81)91026-1} {\bibfield  {journal} {\bibinfo
  {journal} {Phys. Lett.}\ }\textbf {\bibinfo {volume} {B105}},\ \bibinfo
  {pages} {219--223} (\bibinfo {year} {1981}{\natexlab{c}})}\BibitemShut
  {NoStop}%
\bibitem [{\citenamefont {Ginsparg}\ and\ \citenamefont
  {Wilson}(1982)}]{Ginsparg:1981bj}%
  \BibitemOpen
  \bibfield  {author} {\bibinfo {author} {\bibfnamefont {Paul~H.}\ \bibnamefont
  {Ginsparg}}\ and\ \bibinfo {author} {\bibfnamefont {Kenneth~G.}\ \bibnamefont
  {Wilson}},\ }\bibfield  {title} {\enquote {\bibinfo {title} {{A Remnant of
  Chiral Symmetry on the Lattice}},}\ }\href {\doibase
  10.1103/PhysRevD.25.2649} {\bibfield  {journal} {\bibinfo  {journal} {Phys.
  Rev.}\ }\textbf {\bibinfo {volume} {D25}},\ \bibinfo {pages} {2649} (\bibinfo
  {year} {1982})}\BibitemShut {NoStop}%
\bibitem [{\citenamefont {{L\"uscher}}(1998)}]{Luscher:1998pqa}%
  \BibitemOpen
  \bibfield  {author} {\bibinfo {author} {\bibfnamefont {Martin}\ \bibnamefont
  {{L\"uscher}}},\ }\bibfield  {title} {\enquote {\bibinfo {title} {{Exact
  chiral symmetry on the lattice and the Ginsparg-Wilson relation}},}\ }\href
  {\doibase 10.1016/S0370-2693(98)00423-7} {\bibfield  {journal} {\bibinfo
  {journal} {Phys. Lett.}\ }\textbf {\bibinfo {volume} {B428}},\ \bibinfo
  {pages} {342--345} (\bibinfo {year} {1998})},\ \Eprint
  {http://arxiv.org/abs/hep-lat/9802011} {arXiv:hep-lat/9802011 [hep-lat]}
  \BibitemShut {NoStop}%
\bibitem [{\citenamefont {Kaplan}(1992)}]{Kaplan:1992bt}%
  \BibitemOpen
  \bibfield  {author} {\bibinfo {author} {\bibfnamefont {David~B.}\
  \bibnamefont {Kaplan}},\ }\bibfield  {title} {\enquote {\bibinfo {title} {{A
  Method for simulating chiral fermions on the lattice}},}\ }\href {\doibase
  10.1016/0370-2693(92)91112-M} {\bibfield  {journal} {\bibinfo  {journal}
  {Phys. Lett.}\ }\textbf {\bibinfo {volume} {B288}},\ \bibinfo {pages}
  {342--347} (\bibinfo {year} {1992})},\ \Eprint
  {http://arxiv.org/abs/hep-lat/9206013} {arXiv:hep-lat/9206013 [hep-lat]}
  \BibitemShut {NoStop}%
\bibitem [{\citenamefont {Shamir}(1993)}]{Shamir:1993zy}%
  \BibitemOpen
  \bibfield  {author} {\bibinfo {author} {\bibfnamefont {Yigal}\ \bibnamefont
  {Shamir}},\ }\bibfield  {title} {\enquote {\bibinfo {title} {{Chiral fermions
  from lattice boundaries}},}\ }\href {\doibase 10.1016/0550-3213(93)90162-I}
  {\bibfield  {journal} {\bibinfo  {journal} {Nucl. Phys.}\ }\textbf {\bibinfo
  {volume} {B406}},\ \bibinfo {pages} {90--106} (\bibinfo {year} {1993})},\
  \Eprint {http://arxiv.org/abs/hep-lat/9303005} {arXiv:hep-lat/9303005
  [hep-lat]} \BibitemShut {NoStop}%
\bibitem [{\citenamefont {Furman}\ and\ \citenamefont
  {Shamir}(1995)}]{Furman:1994ky}%
  \BibitemOpen
  \bibfield  {author} {\bibinfo {author} {\bibfnamefont {Vadim}\ \bibnamefont
  {Furman}}\ and\ \bibinfo {author} {\bibfnamefont {Yigal}\ \bibnamefont
  {Shamir}},\ }\bibfield  {title} {\enquote {\bibinfo {title} {{Axial
  symmetries in lattice QCD with Kaplan fermions}},}\ }\href {\doibase
  10.1016/0550-3213(95)00031-M} {\bibfield  {journal} {\bibinfo  {journal}
  {Nucl. Phys.}\ }\textbf {\bibinfo {volume} {B439}},\ \bibinfo {pages}
  {54--78} (\bibinfo {year} {1995})},\ \Eprint
  {http://arxiv.org/abs/hep-lat/9405004} {arXiv:hep-lat/9405004 [hep-lat]}
  \BibitemShut {NoStop}%
\bibitem [{\citenamefont {Narayanan}\ and\ \citenamefont
  {Neuberger}(1994)}]{Narayanan:1993sk}%
  \BibitemOpen
  \bibfield  {author} {\bibinfo {author} {\bibfnamefont {Rajamani}\
  \bibnamefont {Narayanan}}\ and\ \bibinfo {author} {\bibfnamefont {Herbert}\
  \bibnamefont {Neuberger}},\ }\bibfield  {title} {\enquote {\bibinfo {title}
  {{Chiral determinant as an overlap of two vacua}},}\ }\href {\doibase
  10.1016/0550-3213(94)90393-X} {\bibfield  {journal} {\bibinfo  {journal}
  {Nucl. Phys.}\ }\textbf {\bibinfo {volume} {B412}},\ \bibinfo {pages}
  {574--606} (\bibinfo {year} {1994})},\ \Eprint
  {http://arxiv.org/abs/hep-lat/9307006} {arXiv:hep-lat/9307006 [hep-lat]}
  \BibitemShut {NoStop}%
\bibitem [{\citenamefont {Narayanan}\ and\ \citenamefont
  {Neuberger}(1993)}]{Narayanan:1993ss}%
  \BibitemOpen
  \bibfield  {author} {\bibinfo {author} {\bibfnamefont {Rajamani}\
  \bibnamefont {Narayanan}}\ and\ \bibinfo {author} {\bibfnamefont {Herbert}\
  \bibnamefont {Neuberger}},\ }\bibfield  {title} {\enquote {\bibinfo {title}
  {{Chiral fermions on the lattice}},}\ }\href {\doibase
  10.1103/PhysRevLett.71.3251} {\bibfield  {journal} {\bibinfo  {journal}
  {Phys. Rev. Lett.}\ }\textbf {\bibinfo {volume} {71}},\ \bibinfo {pages}
  {3251} (\bibinfo {year} {1993})},\ \Eprint
  {http://arxiv.org/abs/hep-lat/9308011} {arXiv:hep-lat/9308011 [hep-lat]}
  \BibitemShut {NoStop}%
\bibitem [{\citenamefont {Narayanan}\ and\ \citenamefont
  {Neuberger}(1995)}]{Narayanan:1994gw}%
  \BibitemOpen
  \bibfield  {author} {\bibinfo {author} {\bibfnamefont {Rajamani}\
  \bibnamefont {Narayanan}}\ and\ \bibinfo {author} {\bibfnamefont {Herbert}\
  \bibnamefont {Neuberger}},\ }\bibfield  {title} {\enquote {\bibinfo {title}
  {{A Construction of lattice chiral gauge theories}},}\ }\href {\doibase
  10.1016/0550-3213(95)00111-5} {\bibfield  {journal} {\bibinfo  {journal}
  {Nucl. Phys.}\ }\textbf {\bibinfo {volume} {B443}},\ \bibinfo {pages}
  {305--385} (\bibinfo {year} {1995})},\ \Eprint
  {http://arxiv.org/abs/hep-th/9411108} {arXiv:hep-th/9411108 [hep-th]}
  \BibitemShut {NoStop}%
\bibitem [{\citenamefont {Borici}(2000)}]{Borici:1999zw}%
  \BibitemOpen
  \bibfield  {author} {\bibinfo {author} {\bibfnamefont {A.}~\bibnamefont
  {Borici}},\ }\bibfield  {title} {\enquote {\bibinfo {title} {{Truncated
  overlap fermions}},}\ }\bibfield  {booktitle} {\emph {\bibinfo {booktitle}
  {{Lattice field theory. Proceedings, 17th International Symposium,
  Lattice'99, Pisa, Italy, June 29-July 3, 1999}}},\ }\href {\doibase
  10.1016/S0920-5632(00)91802-4} {\bibfield  {journal} {\bibinfo  {journal}
  {Nucl. Phys. Proc. Suppl.}\ }\textbf {\bibinfo {volume} {83}},\ \bibinfo
  {pages} {771--773} (\bibinfo {year} {2000})},\ \Eprint
  {http://arxiv.org/abs/hep-lat/9909057} {arXiv:hep-lat/9909057 [hep-lat]}
  \BibitemShut {NoStop}%
\bibitem [{\citenamefont {Borici}(1999)}]{Borici:1999da}%
  \BibitemOpen
  \bibfield  {author} {\bibinfo {author} {\bibfnamefont {Artan}\ \bibnamefont
  {Borici}},\ }\bibfield  {title} {\enquote {\bibinfo {title} {{Truncated
  overlap fermions: The Link between overlap and domain wall fermions}},}\ }in\
  \href@noop {} {\emph {\bibinfo {booktitle} {{Lattice fermions and structure
  of the vacuum. Proceedings, NATO Advanced Research Workshop, Dubna, Russia,
  October 5-9, 1999}}}}\ (\bibinfo {year} {1999})\ pp.\ \bibinfo {pages}
  {41--52},\ \Eprint {http://arxiv.org/abs/hep-lat/9912040}
  {arXiv:hep-lat/9912040 [hep-lat]} \BibitemShut {NoStop}%
\bibitem [{\citenamefont {Kennedy}(2005)}]{Kennedy:2004ae}%
  \BibitemOpen
  \bibfield  {author} {\bibinfo {author} {\bibfnamefont {A.~D.}\ \bibnamefont
  {Kennedy}},\ }\bibfield  {title} {\enquote {\bibinfo {title} {{Algorithms for
  lattice QCD with dynamical fermions}},}\ }\bibfield  {booktitle} {\emph
  {\bibinfo {booktitle} {{Lattice field theory. Proceedings, 22nd International
  Symposium, Lattice 2004, Batavia, USA, June 21-26, 2004}}},\ }\href {\doibase
  10.1016/j.nuclphysbps.2004.11.243} {\bibfield  {journal} {\bibinfo  {journal}
  {Nucl. Phys. Proc. Suppl.}\ }\textbf {\bibinfo {volume} {140}},\ \bibinfo
  {pages} {190--203} (\bibinfo {year} {2005})},\ \bibinfo {note}
  {[,190(2004)]},\ \Eprint {http://arxiv.org/abs/hep-lat/0409167}
  {arXiv:hep-lat/0409167 [hep-lat]} \BibitemShut {NoStop}%
\bibitem [{\citenamefont {Bar}\ \emph {et~al.}(2003)\citenamefont {Bar},
  \citenamefont {Rupak},\ and\ \citenamefont {Shoresh}}]{Bar:2002nr}%
  \BibitemOpen
  \bibfield  {author} {\bibinfo {author} {\bibfnamefont {Oliver}\ \bibnamefont
  {Bar}}, \bibinfo {author} {\bibfnamefont {Gautam}\ \bibnamefont {Rupak}}, \
  and\ \bibinfo {author} {\bibfnamefont {Noam}\ \bibnamefont {Shoresh}},\
  }\bibfield  {title} {\enquote {\bibinfo {title} {{Simulations with different
  lattice Dirac operators for valence and sea quarks}},}\ }\href {\doibase
  10.1103/PhysRevD.67.114505} {\bibfield  {journal} {\bibinfo  {journal} {Phys.
  Rev.}\ }\textbf {\bibinfo {volume} {D67}},\ \bibinfo {pages} {114505}
  (\bibinfo {year} {2003})},\ \Eprint {http://arxiv.org/abs/hep-lat/0210050}
  {arXiv:hep-lat/0210050 [hep-lat]} \BibitemShut {NoStop}%
\bibitem [{\citenamefont {Renner}\ \emph {et~al.}(2005)\citenamefont {Renner},
  \citenamefont {Schroers}, \citenamefont {Edwards}, \citenamefont {Fleming},
  \citenamefont {Hagler}, \citenamefont {Negele}, \citenamefont {Orginos},
  \citenamefont {Pochinski},\ and\ \citenamefont {Richards}}]{Renner:2004ck}%
  \BibitemOpen
  \bibfield  {author} {\bibinfo {author} {\bibfnamefont {Dru~Bryant}\
  \bibnamefont {Renner}}, \bibinfo {author} {\bibfnamefont {W.}~\bibnamefont
  {Schroers}}, \bibinfo {author} {\bibfnamefont {R.}~\bibnamefont {Edwards}},
  \bibinfo {author} {\bibfnamefont {George~Tamminga}\ \bibnamefont {Fleming}},
  \bibinfo {author} {\bibfnamefont {Ph.}\ \bibnamefont {Hagler}}, \bibinfo
  {author} {\bibfnamefont {John~W.}\ \bibnamefont {Negele}}, \bibinfo {author}
  {\bibfnamefont {K.}~\bibnamefont {Orginos}}, \bibinfo {author} {\bibfnamefont
  {A.~V.}\ \bibnamefont {Pochinski}}, \ and\ \bibinfo {author} {\bibfnamefont
  {D.}~\bibnamefont {Richards}} (\bibinfo {collaboration} {LHP}),\ }\bibfield
  {title} {\enquote {\bibinfo {title} {{Hadronic physics with domain-wall
  valence and improved staggered sea quarks}},}\ }\bibfield  {booktitle} {\emph
  {\bibinfo {booktitle} {{Lattice field theory. Proceedings, 22nd International
  Symposium, Lattice 2004, Batavia, USA, June 21-26, 2004}}},\ }\href {\doibase
  10.1016/j.nuclphysbps.2004.11.357} {\bibfield  {journal} {\bibinfo  {journal}
  {Nucl. Phys. Proc. Suppl.}\ }\textbf {\bibinfo {volume} {140}},\ \bibinfo
  {pages} {255--260} (\bibinfo {year} {2005})},\ \bibinfo {note}
  {[,255(2004)]},\ \Eprint {http://arxiv.org/abs/hep-lat/0409130}
  {arXiv:hep-lat/0409130 [hep-lat]} \BibitemShut {NoStop}%
\bibitem [{\citenamefont {Orginos}\ and\ \citenamefont
  {Toussaint}(1999)}]{Orginos:1998ue}%
  \BibitemOpen
  \bibfield  {author} {\bibinfo {author} {\bibfnamefont {Kostas}\ \bibnamefont
  {Orginos}}\ and\ \bibinfo {author} {\bibfnamefont {Doug}\ \bibnamefont
  {Toussaint}} (\bibinfo {collaboration} {MILC}),\ }\bibfield  {title}
  {\enquote {\bibinfo {title} {{Testing improved actions for dynamical
  Kogut-Susskind quarks}},}\ }\href {\doibase 10.1103/PhysRevD.59.014501}
  {\bibfield  {journal} {\bibinfo  {journal} {Phys. Rev.}\ }\textbf {\bibinfo
  {volume} {D59}},\ \bibinfo {pages} {014501} (\bibinfo {year} {1999})},\
  \Eprint {http://arxiv.org/abs/hep-lat/9805009} {arXiv:hep-lat/9805009
  [hep-lat]} \BibitemShut {NoStop}%
\bibitem [{\citenamefont {Orginos}\ \emph {et~al.}(1999)\citenamefont
  {Orginos}, \citenamefont {Toussaint},\ and\ \citenamefont
  {Sugar}}]{Orginos:1999cr}%
  \BibitemOpen
  \bibfield  {author} {\bibinfo {author} {\bibfnamefont {Kostas}\ \bibnamefont
  {Orginos}}, \bibinfo {author} {\bibfnamefont {Doug}\ \bibnamefont
  {Toussaint}}, \ and\ \bibinfo {author} {\bibfnamefont {R.~L.}\ \bibnamefont
  {Sugar}} (\bibinfo {collaboration} {MILC}),\ }\bibfield  {title} {\enquote
  {\bibinfo {title} {{Variants of fattening and flavor symmetry
  restoration}},}\ }\href {\doibase 10.1103/PhysRevD.60.054503} {\bibfield
  {journal} {\bibinfo  {journal} {Phys. Rev.}\ }\textbf {\bibinfo {volume}
  {D60}},\ \bibinfo {pages} {054503} (\bibinfo {year} {1999})},\ \Eprint
  {http://arxiv.org/abs/hep-lat/9903032} {arXiv:hep-lat/9903032 [hep-lat]}
  \BibitemShut {NoStop}%
\bibitem [{\citenamefont {Bernard}\ \emph {et~al.}(2001)\citenamefont
  {Bernard}, \citenamefont {Burch}, \citenamefont {Orginos}, \citenamefont
  {Toussaint}, \citenamefont {DeGrand}, \citenamefont {Detar}, \citenamefont
  {Datta}, \citenamefont {Gottlieb}, \citenamefont {Heller},\ and\
  \citenamefont {Sugar}}]{Bernard:2001av}%
  \BibitemOpen
  \bibfield  {author} {\bibinfo {author} {\bibfnamefont {Claude~W.}\
  \bibnamefont {Bernard}}, \bibinfo {author} {\bibfnamefont {Tom}\ \bibnamefont
  {Burch}}, \bibinfo {author} {\bibfnamefont {Kostas}\ \bibnamefont {Orginos}},
  \bibinfo {author} {\bibfnamefont {Doug}\ \bibnamefont {Toussaint}}, \bibinfo
  {author} {\bibfnamefont {Thomas~A.}\ \bibnamefont {DeGrand}}, \bibinfo
  {author} {\bibfnamefont {Carleton~E.}\ \bibnamefont {Detar}}, \bibinfo
  {author} {\bibfnamefont {Saumen}\ \bibnamefont {Datta}}, \bibinfo {author}
  {\bibfnamefont {Steven~A.}\ \bibnamefont {Gottlieb}}, \bibinfo {author}
  {\bibfnamefont {Urs~M.}\ \bibnamefont {Heller}}, \ and\ \bibinfo {author}
  {\bibfnamefont {Robert}\ \bibnamefont {Sugar}},\ }\bibfield  {title}
  {\enquote {\bibinfo {title} {{The QCD spectrum with three quark flavors}},}\
  }\href {\doibase 10.1103/PhysRevD.64.054506} {\bibfield  {journal} {\bibinfo
  {journal} {Phys. Rev.}\ }\textbf {\bibinfo {volume} {D64}},\ \bibinfo {pages}
  {054506} (\bibinfo {year} {2001})},\ \Eprint
  {http://arxiv.org/abs/hep-lat/0104002} {arXiv:hep-lat/0104002 [hep-lat]}
  \BibitemShut {NoStop}%
\bibitem [{\citenamefont {Edwards}\ \emph {et~al.}(2006)\citenamefont
  {Edwards}, \citenamefont {Fleming}, \citenamefont {Hagler}, \citenamefont
  {Negele}, \citenamefont {Orginos}, \citenamefont {Pochinsky}, \citenamefont
  {Renner}, \citenamefont {Richards},\ and\ \citenamefont
  {Schroers}}]{Edwards:2005ym}%
  \BibitemOpen
  \bibfield  {author} {\bibinfo {author} {\bibfnamefont {R.~G.}\ \bibnamefont
  {Edwards}}, \bibinfo {author} {\bibfnamefont {G.~T.}\ \bibnamefont
  {Fleming}}, \bibinfo {author} {\bibfnamefont {Ph.}\ \bibnamefont {Hagler}},
  \bibinfo {author} {\bibfnamefont {J.~W.}\ \bibnamefont {Negele}}, \bibinfo
  {author} {\bibfnamefont {K.}~\bibnamefont {Orginos}}, \bibinfo {author}
  {\bibfnamefont {A.~V.}\ \bibnamefont {Pochinsky}}, \bibinfo {author}
  {\bibfnamefont {D.~B.}\ \bibnamefont {Renner}}, \bibinfo {author}
  {\bibfnamefont {D.~G.}\ \bibnamefont {Richards}}, \ and\ \bibinfo {author}
  {\bibfnamefont {W.}~\bibnamefont {Schroers}} (\bibinfo {collaboration}
  {LHPC}),\ }\bibfield  {title} {\enquote {\bibinfo {title} {{The Nucleon axial
  charge in full lattice QCD}},}\ }\href {\doibase
  10.1103/PhysRevLett.96.052001} {\bibfield  {journal} {\bibinfo  {journal}
  {Phys. Rev. Lett.}\ }\textbf {\bibinfo {volume} {96}},\ \bibinfo {pages}
  {052001} (\bibinfo {year} {2006})},\ \Eprint
  {http://arxiv.org/abs/hep-lat/0510062} {arXiv:hep-lat/0510062 [hep-lat]}
  \BibitemShut {NoStop}%
\bibitem [{\citenamefont {Hagler}\ \emph {et~al.}(2008)\citenamefont {Hagler}
  \emph {et~al.}}]{Hagler:2007xi}%
  \BibitemOpen
  \bibfield  {author} {\bibinfo {author} {\bibfnamefont {Ph.}\ \bibnamefont
  {Hagler}} \emph {et~al.} (\bibinfo {collaboration} {LHPC}),\ }\bibfield
  {title} {\enquote {\bibinfo {title} {{Nucleon Generalized Parton
  Distributions from Full Lattice QCD}},}\ }\href {\doibase
  10.1103/PhysRevD.77.094502} {\bibfield  {journal} {\bibinfo  {journal} {Phys.
  Rev.}\ }\textbf {\bibinfo {volume} {D77}},\ \bibinfo {pages} {094502}
  (\bibinfo {year} {2008})},\ \Eprint {http://arxiv.org/abs/0705.4295}
  {arXiv:0705.4295 [hep-lat]} \BibitemShut {NoStop}%
\bibitem [{\citenamefont {Bratt}\ \emph {et~al.}(2010)\citenamefont {Bratt}
  \emph {et~al.}}]{Bratt:2010jn}%
  \BibitemOpen
  \bibfield  {author} {\bibinfo {author} {\bibfnamefont {J.~D.}\ \bibnamefont
  {Bratt}} \emph {et~al.} (\bibinfo {collaboration} {LHPC}),\ }\bibfield
  {title} {\enquote {\bibinfo {title} {{Nucleon structure from mixed action
  calculations using 2+1 flavors of asqtad sea and domain wall valence
  fermions}},}\ }\href {\doibase 10.1103/PhysRevD.82.094502} {\bibfield
  {journal} {\bibinfo  {journal} {Phys. Rev.}\ }\textbf {\bibinfo {volume}
  {D82}},\ \bibinfo {pages} {094502} (\bibinfo {year} {2010})},\ \Eprint
  {http://arxiv.org/abs/1001.3620} {arXiv:1001.3620 [hep-lat]} \BibitemShut
  {NoStop}%
\bibitem [{\citenamefont {Beane}\ \emph {et~al.}(2006)\citenamefont {Beane},
  \citenamefont {Bedaque}, \citenamefont {Orginos},\ and\ \citenamefont
  {Savage}}]{Beane:2006mx}%
  \BibitemOpen
  \bibfield  {author} {\bibinfo {author} {\bibfnamefont {S.~R.}\ \bibnamefont
  {Beane}}, \bibinfo {author} {\bibfnamefont {P.~F.}\ \bibnamefont {Bedaque}},
  \bibinfo {author} {\bibfnamefont {K.}~\bibnamefont {Orginos}}, \ and\
  \bibinfo {author} {\bibfnamefont {M.~J.}\ \bibnamefont {Savage}} (\bibinfo
  {collaboration} {NPLQCD}),\ }\bibfield  {title} {\enquote {\bibinfo {title}
  {{Nucleon nucleon scattering from fully-dynamical lattice QCD}},}\ }\href
  {\doibase 10.1103/PhysRevLett.97.012001} {\bibfield  {journal} {\bibinfo
  {journal} {Phys. Rev. Lett.}\ }\textbf {\bibinfo {volume} {97}},\ \bibinfo
  {pages} {012001} (\bibinfo {year} {2006})},\ \Eprint
  {http://arxiv.org/abs/hep-lat/0602010} {arXiv:hep-lat/0602010} \BibitemShut
  {NoStop}%
\bibitem [{\citenamefont {Beane}\ \emph {et~al.}(2008)\citenamefont {Beane}
  \emph {et~al.}}]{Beane:2007xs}%
  \BibitemOpen
  \bibfield  {author} {\bibinfo {author} {\bibfnamefont {Silas~R.}\
  \bibnamefont {Beane}} \emph {et~al.} (\bibinfo {collaboration} {NPLQCD}),\
  }\bibfield  {title} {\enquote {\bibinfo {title} {{Precise Determination of
  the I=2 pipi Scattering Length from Mixed-Action Lattice QCD}},}\ }\href
  {\doibase 10.1103/PhysRevD.77.014505} {\bibfield  {journal} {\bibinfo
  {journal} {Phys. Rev.}\ }\textbf {\bibinfo {volume} {D77}},\ \bibinfo {pages}
  {014505} (\bibinfo {year} {2008})},\ \Eprint {http://arxiv.org/abs/0706.3026}
  {arXiv:0706.3026 [hep-lat]} \BibitemShut {NoStop}%
\bibitem [{\citenamefont {Walker-Loud}\ \emph {et~al.}(2009)\citenamefont
  {Walker-Loud} \emph {et~al.}}]{WalkerLoud:2008bp}%
  \BibitemOpen
  \bibfield  {author} {\bibinfo {author} {\bibfnamefont {A.}~\bibnamefont
  {Walker-Loud}} \emph {et~al.},\ }\bibfield  {title} {\enquote {\bibinfo
  {title} {{Light hadron spectroscopy using domain wall valence quarks on an
  Asqtad sea}},}\ }\href {\doibase 10.1103/PhysRevD.79.054502} {\bibfield
  {journal} {\bibinfo  {journal} {Phys. Rev.}\ }\textbf {\bibinfo {volume}
  {D79}},\ \bibinfo {pages} {054502} (\bibinfo {year} {2009})},\ \Eprint
  {http://arxiv.org/abs/0806.4549} {arXiv:0806.4549 [hep-lat]} \BibitemShut
  {NoStop}%
\bibitem [{\citenamefont {Aubin}\ \emph {et~al.}(2010)\citenamefont {Aubin},
  \citenamefont {Laiho},\ and\ \citenamefont {Van~de Water}}]{Aubin:2009jh}%
  \BibitemOpen
  \bibfield  {author} {\bibinfo {author} {\bibfnamefont {C.}~\bibnamefont
  {Aubin}}, \bibinfo {author} {\bibfnamefont {Jack}\ \bibnamefont {Laiho}}, \
  and\ \bibinfo {author} {\bibfnamefont {Ruth~S.}\ \bibnamefont {Van~de
  Water}},\ }\bibfield  {title} {\enquote {\bibinfo {title} {{The Neutral kaon
  mixing parameter B(K) from unquenched mixed-action lattice QCD}},}\ }\href
  {\doibase 10.1103/PhysRevD.81.014507} {\bibfield  {journal} {\bibinfo
  {journal} {Phys. Rev.}\ }\textbf {\bibinfo {volume} {D81}},\ \bibinfo {pages}
  {014507} (\bibinfo {year} {2010})},\ \Eprint {http://arxiv.org/abs/0905.3947}
  {arXiv:0905.3947 [hep-lat]} \BibitemShut {NoStop}%
\bibitem [{\citenamefont {Langacker}\ and\ \citenamefont
  {Pagels}(1973)}]{Langacker:1973hh}%
  \BibitemOpen
  \bibfield  {author} {\bibinfo {author} {\bibfnamefont {Paul}\ \bibnamefont
  {Langacker}}\ and\ \bibinfo {author} {\bibfnamefont {Heinz}\ \bibnamefont
  {Pagels}},\ }\bibfield  {title} {\enquote {\bibinfo {title} {{Chiral
  perturbation theory}},}\ }\href {\doibase 10.1103/PhysRevD.8.4595} {\bibfield
   {journal} {\bibinfo  {journal} {Phys. Rev.}\ }\textbf {\bibinfo {volume}
  {D8}},\ \bibinfo {pages} {4595--4619} (\bibinfo {year} {1973})}\BibitemShut
  {NoStop}%
\bibitem [{\citenamefont {Gasser}\ and\ \citenamefont
  {Leutwyler}(1984)}]{Gasser:1983yg}%
  \BibitemOpen
  \bibfield  {author} {\bibinfo {author} {\bibfnamefont {J.}~\bibnamefont
  {Gasser}}\ and\ \bibinfo {author} {\bibfnamefont {H.}~\bibnamefont
  {Leutwyler}},\ }\bibfield  {title} {\enquote {\bibinfo {title} {{Chiral
  Perturbation Theory to One Loop}},}\ }\href {\doibase
  10.1016/0003-4916(84)90242-2} {\bibfield  {journal} {\bibinfo  {journal}
  {Annals Phys.}\ }\textbf {\bibinfo {volume} {158}},\ \bibinfo {pages} {142}
  (\bibinfo {year} {1984})}\BibitemShut {NoStop}%
\bibitem [{\citenamefont {Leutwyler}(1994)}]{Leutwyler:1993iq}%
  \BibitemOpen
  \bibfield  {author} {\bibinfo {author} {\bibfnamefont {H.}~\bibnamefont
  {Leutwyler}},\ }\bibfield  {title} {\enquote {\bibinfo {title} {{On the
  foundations of chiral perturbation theory}},}\ }\href {\doibase
  10.1006/aphy.1994.1094} {\bibfield  {journal} {\bibinfo  {journal} {Annals
  Phys.}\ }\textbf {\bibinfo {volume} {235}},\ \bibinfo {pages} {165--203}
  (\bibinfo {year} {1994})},\ \Eprint {http://arxiv.org/abs/hep-ph/9311274}
  {arXiv:hep-ph/9311274 [hep-ph]} \BibitemShut {NoStop}%
\bibitem [{\citenamefont {Sharpe}\ and\ \citenamefont
  {Singleton}(1998)}]{Sharpe:1998xm}%
  \BibitemOpen
  \bibfield  {author} {\bibinfo {author} {\bibfnamefont {Stephen~R.}\
  \bibnamefont {Sharpe}}\ and\ \bibinfo {author} {\bibfnamefont {Robert~L.}\
  \bibnamefont {Singleton}, \bibfnamefont {Jr}},\ }\bibfield  {title} {\enquote
  {\bibinfo {title} {{Spontaneous flavor and parity breaking with Wilson
  fermions}},}\ }\href {\doibase 10.1103/PhysRevD.58.074501} {\bibfield
  {journal} {\bibinfo  {journal} {Phys. Rev.}\ }\textbf {\bibinfo {volume}
  {D58}},\ \bibinfo {pages} {074501} (\bibinfo {year} {1998})},\ \Eprint
  {http://arxiv.org/abs/hep-lat/9804028} {arXiv:hep-lat/9804028 [hep-lat]}
  \BibitemShut {NoStop}%
\bibitem [{\citenamefont {Bar}\ \emph {et~al.}(2004)\citenamefont {Bar},
  \citenamefont {Rupak},\ and\ \citenamefont {Shoresh}}]{Bar:2003mh}%
  \BibitemOpen
  \bibfield  {author} {\bibinfo {author} {\bibfnamefont {Oliver}\ \bibnamefont
  {Bar}}, \bibinfo {author} {\bibfnamefont {Gautam}\ \bibnamefont {Rupak}}, \
  and\ \bibinfo {author} {\bibfnamefont {Noam}\ \bibnamefont {Shoresh}},\
  }\bibfield  {title} {\enquote {\bibinfo {title} {{Chiral perturbation theory
  at O(a**2) for lattice QCD}},}\ }\href {\doibase 10.1103/PhysRevD.70.034508}
  {\bibfield  {journal} {\bibinfo  {journal} {Phys. Rev.}\ }\textbf {\bibinfo
  {volume} {D70}},\ \bibinfo {pages} {034508} (\bibinfo {year} {2004})},\
  \Eprint {http://arxiv.org/abs/hep-lat/0306021} {arXiv:hep-lat/0306021
  [hep-lat]} \BibitemShut {NoStop}%
\bibitem [{\citenamefont {Bar}\ \emph {et~al.}(2005)\citenamefont {Bar},
  \citenamefont {Bernard}, \citenamefont {Rupak},\ and\ \citenamefont
  {Shoresh}}]{Bar:2005tu}%
  \BibitemOpen
  \bibfield  {author} {\bibinfo {author} {\bibfnamefont {Oliver}\ \bibnamefont
  {Bar}}, \bibinfo {author} {\bibfnamefont {Claude}\ \bibnamefont {Bernard}},
  \bibinfo {author} {\bibfnamefont {Gautam}\ \bibnamefont {Rupak}}, \ and\
  \bibinfo {author} {\bibfnamefont {Noam}\ \bibnamefont {Shoresh}},\ }\bibfield
   {title} {\enquote {\bibinfo {title} {{Chiral perturbation theory for
  staggered sea quarks and Ginsparg-Wilson valence quarks}},}\ }\href {\doibase
  10.1103/PhysRevD.72.054502} {\bibfield  {journal} {\bibinfo  {journal} {Phys.
  Rev.}\ }\textbf {\bibinfo {volume} {D72}},\ \bibinfo {pages} {054502}
  (\bibinfo {year} {2005})},\ \Eprint {http://arxiv.org/abs/hep-lat/0503009}
  {arXiv:hep-lat/0503009 [hep-lat]} \BibitemShut {NoStop}%
\bibitem [{\citenamefont {Tiburzi}(2005{\natexlab{a}})}]{Tiburzi:2005is}%
  \BibitemOpen
  \bibfield  {author} {\bibinfo {author} {\bibfnamefont {Brian~C.}\
  \bibnamefont {Tiburzi}},\ }\bibfield  {title} {\enquote {\bibinfo {title}
  {{Baryons with Ginsparg-Wilson quarks in a staggered sea}},}\ }\href
  {\doibase 10.1103/PhysRevD.72.094501, 10.1103/PhysRevD.79.039904} {\bibfield
  {journal} {\bibinfo  {journal} {Phys. Rev.}\ }\textbf {\bibinfo {volume}
  {D72}},\ \bibinfo {pages} {094501} (\bibinfo {year} {2005}{\natexlab{a}})},\
  \bibinfo {note} {[Erratum: Phys. Rev.D79,039904(2009)]},\ \Eprint
  {http://arxiv.org/abs/hep-lat/0508019} {arXiv:hep-lat/0508019 [hep-lat]}
  \BibitemShut {NoStop}%
\bibitem [{\citenamefont {Chen}\ \emph {et~al.}(2006)\citenamefont {Chen},
  \citenamefont {O'Connell}, \citenamefont {Van~de Water},\ and\ \citenamefont
  {Walker-Loud}}]{Chen:2005ab}%
  \BibitemOpen
  \bibfield  {author} {\bibinfo {author} {\bibfnamefont {Jiunn-Wei}\
  \bibnamefont {Chen}}, \bibinfo {author} {\bibfnamefont {Donal}\ \bibnamefont
  {O'Connell}}, \bibinfo {author} {\bibfnamefont {Ruth~S.}\ \bibnamefont
  {Van~de Water}}, \ and\ \bibinfo {author} {\bibfnamefont {Andre}\
  \bibnamefont {Walker-Loud}},\ }\bibfield  {title} {\enquote {\bibinfo {title}
  {{Ginsparg-Wilson pions scattering on a staggered sea}},}\ }\href {\doibase
  10.1103/PhysRevD.73.074510} {\bibfield  {journal} {\bibinfo  {journal} {Phys.
  Rev.}\ }\textbf {\bibinfo {volume} {D73}},\ \bibinfo {pages} {074510}
  (\bibinfo {year} {2006})},\ \Eprint {http://arxiv.org/abs/hep-lat/0510024}
  {arXiv:hep-lat/0510024 [hep-lat]} \BibitemShut {NoStop}%
\bibitem [{\citenamefont {Chen}\ \emph {et~al.}(2007)\citenamefont {Chen},
  \citenamefont {O'Connell},\ and\ \citenamefont {Walker-Loud}}]{Chen:2006wf}%
  \BibitemOpen
  \bibfield  {author} {\bibinfo {author} {\bibfnamefont {Jiunn-Wei}\
  \bibnamefont {Chen}}, \bibinfo {author} {\bibfnamefont {Donal}\ \bibnamefont
  {O'Connell}}, \ and\ \bibinfo {author} {\bibfnamefont {Andre}\ \bibnamefont
  {Walker-Loud}},\ }\bibfield  {title} {\enquote {\bibinfo {title} {{Two Meson
  Systems with Ginsparg-Wilson Valence Quarks}},}\ }\href {\doibase
  10.1103/PhysRevD.75.054501} {\bibfield  {journal} {\bibinfo  {journal} {Phys.
  Rev.}\ }\textbf {\bibinfo {volume} {D75}},\ \bibinfo {pages} {054501}
  (\bibinfo {year} {2007})},\ \Eprint {http://arxiv.org/abs/hep-lat/0611003}
  {arXiv:hep-lat/0611003 [hep-lat]} \BibitemShut {NoStop}%
\bibitem [{\citenamefont {Orginos}\ and\ \citenamefont
  {Walker-Loud}(2008)}]{Orginos:2007tw}%
  \BibitemOpen
  \bibfield  {author} {\bibinfo {author} {\bibfnamefont {Kostas}\ \bibnamefont
  {Orginos}}\ and\ \bibinfo {author} {\bibfnamefont {Andre}\ \bibnamefont
  {Walker-Loud}},\ }\bibfield  {title} {\enquote {\bibinfo {title} {{Mixed
  meson masses with domain-wall valence and staggered sea fermions}},}\ }\href
  {\doibase 10.1103/PhysRevD.77.094505} {\bibfield  {journal} {\bibinfo
  {journal} {Phys. Rev.}\ }\textbf {\bibinfo {volume} {D77}},\ \bibinfo {pages}
  {094505} (\bibinfo {year} {2008})},\ \Eprint {http://arxiv.org/abs/0705.0572}
  {arXiv:0705.0572 [hep-lat]} \BibitemShut {NoStop}%
\bibitem [{\citenamefont {Jiang}(2007)}]{Jiang:2007sn}%
  \BibitemOpen
  \bibfield  {author} {\bibinfo {author} {\bibfnamefont {Fu-Jiun}\ \bibnamefont
  {Jiang}},\ }\bibfield  {title} {\enquote {\bibinfo {title} {{Mixed Action
  Lattice Spacing Effects on the Nucleon Axial Charge}},}\ }\href@noop {} {\
  (\bibinfo {year} {2007})},\ \Eprint {http://arxiv.org/abs/hep-lat/0703012}
  {arXiv:hep-lat/0703012 [hep-lat]} \BibitemShut {NoStop}%
\bibitem [{\citenamefont {Chen}\ \emph
  {et~al.}(2009{\natexlab{a}})\citenamefont {Chen}, \citenamefont {O'Connell},\
  and\ \citenamefont {Walker-Loud}}]{Chen:2007ug}%
  \BibitemOpen
  \bibfield  {author} {\bibinfo {author} {\bibfnamefont {Jiunn-Wei}\
  \bibnamefont {Chen}}, \bibinfo {author} {\bibfnamefont {Donal}\ \bibnamefont
  {O'Connell}}, \ and\ \bibinfo {author} {\bibfnamefont {Andre}\ \bibnamefont
  {Walker-Loud}},\ }\bibfield  {title} {\enquote {\bibinfo {title}
  {{Universality of mixed action extrapolation formulae}},}\ }\href {\doibase
  10.1088/1126-6708/2009/04/090} {\bibfield  {journal} {\bibinfo  {journal}
  {JHEP}\ }\textbf {\bibinfo {volume} {04}},\ \bibinfo {pages} {090} (\bibinfo
  {year} {2009}{\natexlab{a}})},\ \Eprint {http://arxiv.org/abs/0706.0035}
  {arXiv:0706.0035 [hep-lat]} \BibitemShut {NoStop}%
\bibitem [{\citenamefont {Chen}\ \emph
  {et~al.}(2009{\natexlab{b}})\citenamefont {Chen}, \citenamefont {Golterman},
  \citenamefont {O'Connell},\ and\ \citenamefont {Walker-Loud}}]{Chen:2009su}%
  \BibitemOpen
  \bibfield  {author} {\bibinfo {author} {\bibfnamefont {Jiunn-Wei}\
  \bibnamefont {Chen}}, \bibinfo {author} {\bibfnamefont {Maarten}\
  \bibnamefont {Golterman}}, \bibinfo {author} {\bibfnamefont {Donal}\
  \bibnamefont {O'Connell}}, \ and\ \bibinfo {author} {\bibfnamefont {Andre}\
  \bibnamefont {Walker-Loud}},\ }\bibfield  {title} {\enquote {\bibinfo {title}
  {{Mixed Action Effective Field Theory: An Addendum}},}\ }\href {\doibase
  10.1103/PhysRevD.79.117502} {\bibfield  {journal} {\bibinfo  {journal} {Phys.
  Rev.}\ }\textbf {\bibinfo {volume} {D79}},\ \bibinfo {pages} {117502}
  (\bibinfo {year} {2009}{\natexlab{b}})},\ \Eprint
  {http://arxiv.org/abs/0905.2566} {arXiv:0905.2566 [hep-lat]} \BibitemShut
  {NoStop}%
\bibitem [{\citenamefont {Bernard}\ and\ \citenamefont
  {Golterman}(1994)}]{Bernard:1993sv}%
  \BibitemOpen
  \bibfield  {author} {\bibinfo {author} {\bibfnamefont {Claude~W.}\
  \bibnamefont {Bernard}}\ and\ \bibinfo {author} {\bibfnamefont {Maarten
  F.~L.}\ \bibnamefont {Golterman}},\ }\bibfield  {title} {\enquote {\bibinfo
  {title} {{Partially quenched gauge theories and an application to staggered
  fermions}},}\ }\href {\doibase 10.1103/PhysRevD.49.486} {\bibfield  {journal}
  {\bibinfo  {journal} {Phys. Rev.}\ }\textbf {\bibinfo {volume} {D49}},\
  \bibinfo {pages} {486--494} (\bibinfo {year} {1994})},\ \Eprint
  {http://arxiv.org/abs/hep-lat/9306005} {arXiv:hep-lat/9306005 [hep-lat]}
  \BibitemShut {NoStop}%
\bibitem [{\citenamefont {Sharpe}\ and\ \citenamefont
  {Shoresh}(2000)}]{Sharpe:2000bc}%
  \BibitemOpen
  \bibfield  {author} {\bibinfo {author} {\bibfnamefont {Stephen~R.}\
  \bibnamefont {Sharpe}}\ and\ \bibinfo {author} {\bibfnamefont {Noam}\
  \bibnamefont {Shoresh}},\ }\bibfield  {title} {\enquote {\bibinfo {title}
  {{Physical results from unphysical simulations}},}\ }\href {\doibase
  10.1103/PhysRevD.62.094503} {\bibfield  {journal} {\bibinfo  {journal} {Phys.
  Rev.}\ }\textbf {\bibinfo {volume} {D62}},\ \bibinfo {pages} {094503}
  (\bibinfo {year} {2000})},\ \Eprint {http://arxiv.org/abs/hep-lat/0006017}
  {arXiv:hep-lat/0006017 [hep-lat]} \BibitemShut {NoStop}%
\bibitem [{\citenamefont {Sharpe}\ and\ \citenamefont
  {Shoresh}(2001)}]{Sharpe:2001fh}%
  \BibitemOpen
  \bibfield  {author} {\bibinfo {author} {\bibfnamefont {Stephen~R.}\
  \bibnamefont {Sharpe}}\ and\ \bibinfo {author} {\bibfnamefont {Noam}\
  \bibnamefont {Shoresh}},\ }\bibfield  {title} {\enquote {\bibinfo {title}
  {{Partially quenched chiral perturbation theory without Phi0}},}\ }\href
  {\doibase 10.1103/PhysRevD.64.114510} {\bibfield  {journal} {\bibinfo
  {journal} {Phys. Rev.}\ }\textbf {\bibinfo {volume} {D64}},\ \bibinfo {pages}
  {114510} (\bibinfo {year} {2001})},\ \Eprint
  {http://arxiv.org/abs/hep-lat/0108003} {arXiv:hep-lat/0108003 [hep-lat]}
  \BibitemShut {NoStop}%
\bibitem [{\citenamefont {Chen}\ and\ \citenamefont
  {Savage}(2002)}]{Chen:2001yi}%
  \BibitemOpen
  \bibfield  {author} {\bibinfo {author} {\bibfnamefont {Jiunn-Wei}\
  \bibnamefont {Chen}}\ and\ \bibinfo {author} {\bibfnamefont {Martin~J.}\
  \bibnamefont {Savage}},\ }\bibfield  {title} {\enquote {\bibinfo {title}
  {{Baryons in partially quenched chiral perturbation theory}},}\ }\href
  {\doibase 10.1103/PhysRevD.65.094001} {\bibfield  {journal} {\bibinfo
  {journal} {Phys. Rev.}\ }\textbf {\bibinfo {volume} {D65}},\ \bibinfo {pages}
  {094001} (\bibinfo {year} {2002})},\ \Eprint
  {http://arxiv.org/abs/hep-lat/0111050} {arXiv:hep-lat/0111050 [hep-lat]}
  \BibitemShut {NoStop}%
\bibitem [{\citenamefont {Sharpe}\ and\ \citenamefont {Van~de
  Water}(2004)}]{Sharpe:2003vy}%
  \BibitemOpen
  \bibfield  {author} {\bibinfo {author} {\bibfnamefont {Stephen~R.}\
  \bibnamefont {Sharpe}}\ and\ \bibinfo {author} {\bibfnamefont {Ruth~S.}\
  \bibnamefont {Van~de Water}},\ }\bibfield  {title} {\enquote {\bibinfo
  {title} {{Unphysical operators in partially quenched QCD}},}\ }\href
  {\doibase 10.1103/PhysRevD.69.054027} {\bibfield  {journal} {\bibinfo
  {journal} {Phys. Rev.}\ }\textbf {\bibinfo {volume} {D69}},\ \bibinfo {pages}
  {054027} (\bibinfo {year} {2004})},\ \Eprint
  {http://arxiv.org/abs/hep-lat/0310012} {arXiv:hep-lat/0310012 [hep-lat]}
  \BibitemShut {NoStop}%
\bibitem [{\citenamefont {Arndt}\ and\ \citenamefont
  {Tiburzi}(2003{\natexlab{a}})}]{Arndt:2003ww}%
  \BibitemOpen
  \bibfield  {author} {\bibinfo {author} {\bibfnamefont {Daniel}\ \bibnamefont
  {Arndt}}\ and\ \bibinfo {author} {\bibfnamefont {Brian~C.}\ \bibnamefont
  {Tiburzi}},\ }\bibfield  {title} {\enquote {\bibinfo {title} {{Charge radii
  of the meson and baryon octets in quenched and partially quenched chiral
  perturbation theory}},}\ }\href {\doibase 10.1103/PhysRevD.68.094501}
  {\bibfield  {journal} {\bibinfo  {journal} {Phys. Rev.}\ }\textbf {\bibinfo
  {volume} {D68}},\ \bibinfo {pages} {094501} (\bibinfo {year}
  {2003}{\natexlab{a}})},\ \Eprint {http://arxiv.org/abs/hep-lat/0307003}
  {arXiv:hep-lat/0307003 [hep-lat]} \BibitemShut {NoStop}%
\bibitem [{\citenamefont {Walker-Loud}(2005)}]{WalkerLoud:2004hf}%
  \BibitemOpen
  \bibfield  {author} {\bibinfo {author} {\bibfnamefont {Andre}\ \bibnamefont
  {Walker-Loud}},\ }\bibfield  {title} {\enquote {\bibinfo {title} {{Octet
  baryon masses in partially quenched chiral perturbation theory}},}\ }\href
  {\doibase 10.1016/j.nuclphysa.2004.10.007} {\bibfield  {journal} {\bibinfo
  {journal} {Nucl. Phys.}\ }\textbf {\bibinfo {volume} {A747}},\ \bibinfo
  {pages} {476--507} (\bibinfo {year} {2005})},\ \Eprint
  {http://arxiv.org/abs/hep-lat/0405007} {arXiv:hep-lat/0405007 [hep-lat]}
  \BibitemShut {NoStop}%
\bibitem [{\citenamefont {Bernard}\ and\ \citenamefont
  {Golterman}(2010)}]{Bernard:2010qc}%
  \BibitemOpen
  \bibfield  {author} {\bibinfo {author} {\bibfnamefont {Claude}\ \bibnamefont
  {Bernard}}\ and\ \bibinfo {author} {\bibfnamefont {Maarten}\ \bibnamefont
  {Golterman}},\ }\bibfield  {title} {\enquote {\bibinfo {title} {{Transfer
  Matrix for Partially Quenched QCD}},}\ }\bibfield  {booktitle} {\emph
  {\bibinfo {booktitle} {{Proceedings, 28th International Symposium on Lattice
  field theory (Lattice 2010): Villasimius, Italy, June 14-19, 2010}}},\
  }\href@noop {} {\bibfield  {journal} {\bibinfo  {journal} {PoS}\ }\textbf
  {\bibinfo {volume} {LATTICE2010}},\ \bibinfo {pages} {252} (\bibinfo {year}
  {2010})},\ \Eprint {http://arxiv.org/abs/1011.0184} {arXiv:1011.0184
  [hep-lat]} \BibitemShut {NoStop}%
\bibitem [{\citenamefont {Bernard}\ and\ \citenamefont
  {Golterman}(2013)}]{Bernard:2013kwa}%
  \BibitemOpen
  \bibfield  {author} {\bibinfo {author} {\bibfnamefont {Claude}\ \bibnamefont
  {Bernard}}\ and\ \bibinfo {author} {\bibfnamefont {Maarten}\ \bibnamefont
  {Golterman}},\ }\bibfield  {title} {\enquote {\bibinfo {title} {{On the
  foundations of partially quenched chiral perturbation theory}},}\ }\href
  {\doibase 10.1103/PhysRevD.88.014004} {\bibfield  {journal} {\bibinfo
  {journal} {Phys. Rev.}\ }\textbf {\bibinfo {volume} {D88}},\ \bibinfo {pages}
  {014004} (\bibinfo {year} {2013})},\ \Eprint {http://arxiv.org/abs/1304.1948}
  {arXiv:1304.1948 [hep-lat]} \BibitemShut {NoStop}%
\bibitem [{\citenamefont {Beane}\ and\ \citenamefont
  {Savage}(2002)}]{Beane:2002vq}%
  \BibitemOpen
  \bibfield  {author} {\bibinfo {author} {\bibfnamefont {Silas~R.}\
  \bibnamefont {Beane}}\ and\ \bibinfo {author} {\bibfnamefont {Martin~J.}\
  \bibnamefont {Savage}},\ }\bibfield  {title} {\enquote {\bibinfo {title}
  {{Nucleons in two flavor partially quenched chiral perturbation theory}},}\
  }\href {\doibase 10.1016/S0375-9474(02)01086-2} {\bibfield  {journal}
  {\bibinfo  {journal} {Nucl. Phys.}\ }\textbf {\bibinfo {volume} {A709}},\
  \bibinfo {pages} {319--344} (\bibinfo {year} {2002})},\ \Eprint
  {http://arxiv.org/abs/hep-lat/0203003} {arXiv:hep-lat/0203003 [hep-lat]}
  \BibitemShut {NoStop}%
\bibitem [{\citenamefont {Beane}\ and\ \citenamefont
  {Savage}(2003)}]{Beane:2002np}%
  \BibitemOpen
  \bibfield  {author} {\bibinfo {author} {\bibfnamefont {Silas~R.}\
  \bibnamefont {Beane}}\ and\ \bibinfo {author} {\bibfnamefont {Martin~J.}\
  \bibnamefont {Savage}},\ }\bibfield  {title} {\enquote {\bibinfo {title}
  {{Partially quenched nucleon nucleon scattering}},}\ }\href {\doibase
  10.1103/PhysRevD.67.054502} {\bibfield  {journal} {\bibinfo  {journal} {Phys.
  Rev.}\ }\textbf {\bibinfo {volume} {D67}},\ \bibinfo {pages} {054502}
  (\bibinfo {year} {2003})},\ \Eprint {http://arxiv.org/abs/hep-lat/0210046}
  {arXiv:hep-lat/0210046 [hep-lat]} \BibitemShut {NoStop}%
\bibitem [{\citenamefont {Arndt}\ and\ \citenamefont
  {Tiburzi}(2003{\natexlab{b}})}]{Arndt:2003we}%
  \BibitemOpen
  \bibfield  {author} {\bibinfo {author} {\bibfnamefont {Daniel}\ \bibnamefont
  {Arndt}}\ and\ \bibinfo {author} {\bibfnamefont {Brian~C}\ \bibnamefont
  {Tiburzi}},\ }\bibfield  {title} {\enquote {\bibinfo {title}
  {{Electromagnetic properties of the baryon decuplet in quenched and partially
  quenched chiral perturbation theory}},}\ }\href {\doibase
  10.1103/PhysRevD.69.059904, 10.1103/PhysRevD.68.114503} {\bibfield  {journal}
  {\bibinfo  {journal} {Phys. Rev.}\ }\textbf {\bibinfo {volume} {D68}},\
  \bibinfo {pages} {114503} (\bibinfo {year} {2003}{\natexlab{b}})},\ \bibinfo
  {note} {[Erratum: Phys. Rev.D69,059904(2004)]},\ \Eprint
  {http://arxiv.org/abs/hep-lat/0308001} {arXiv:hep-lat/0308001 [hep-lat]}
  \BibitemShut {NoStop}%
\bibitem [{\citenamefont {Arndt}\ and\ \citenamefont
  {Tiburzi}(2004)}]{Arndt:2003vd}%
  \BibitemOpen
  \bibfield  {author} {\bibinfo {author} {\bibfnamefont {Daniel}\ \bibnamefont
  {Arndt}}\ and\ \bibinfo {author} {\bibfnamefont {Brian~C.}\ \bibnamefont
  {Tiburzi}},\ }\bibfield  {title} {\enquote {\bibinfo {title} {{Baryon
  decuplet to octet electromagnetic transitions in quenched and partially
  quenched chiral perturbation theory}},}\ }\href {\doibase
  10.1103/PhysRevD.69.014501} {\bibfield  {journal} {\bibinfo  {journal} {Phys.
  Rev.}\ }\textbf {\bibinfo {volume} {D69}},\ \bibinfo {pages} {014501}
  (\bibinfo {year} {2004})},\ \Eprint {http://arxiv.org/abs/hep-lat/0309013}
  {arXiv:hep-lat/0309013 [hep-lat]} \BibitemShut {NoStop}%
\bibitem [{\citenamefont {Tiburzi}\ and\ \citenamefont
  {Walker-Loud}(2005)}]{Tiburzi:2004rh}%
  \BibitemOpen
  \bibfield  {author} {\bibinfo {author} {\bibfnamefont {Brian~C.}\
  \bibnamefont {Tiburzi}}\ and\ \bibinfo {author} {\bibfnamefont {Andre}\
  \bibnamefont {Walker-Loud}},\ }\bibfield  {title} {\enquote {\bibinfo {title}
  {{Decuplet baryon masses in partially quenched chiral perturbation
  theory}},}\ }\href {\doibase 10.1016/j.nuclphysa.2004.11.012} {\bibfield
  {journal} {\bibinfo  {journal} {Nucl. Phys.}\ }\textbf {\bibinfo {volume}
  {A748}},\ \bibinfo {pages} {513--536} (\bibinfo {year} {2005})},\ \Eprint
  {http://arxiv.org/abs/hep-lat/0407030} {arXiv:hep-lat/0407030 [hep-lat]}
  \BibitemShut {NoStop}%
\bibitem [{\citenamefont {Tiburzi}(2005{\natexlab{b}})}]{Tiburzi:2004kd}%
  \BibitemOpen
  \bibfield  {author} {\bibinfo {author} {\bibfnamefont {Brian~C.}\
  \bibnamefont {Tiburzi}},\ }\bibfield  {title} {\enquote {\bibinfo {title}
  {{Baryon masses in partially quenched heavy hadron chiral perturbation
  theory}},}\ }\href {\doibase 10.1103/PhysRevD.71.034501} {\bibfield
  {journal} {\bibinfo  {journal} {Phys. Rev.}\ }\textbf {\bibinfo {volume}
  {D71}},\ \bibinfo {pages} {034501} (\bibinfo {year} {2005}{\natexlab{b}})},\
  \Eprint {http://arxiv.org/abs/hep-lat/0410033} {arXiv:hep-lat/0410033
  [hep-lat]} \BibitemShut {NoStop}%
\bibitem [{\citenamefont {Tiburzi}\ and\ \citenamefont
  {Walker-Loud}(2006)}]{Tiburzi:2005na}%
  \BibitemOpen
  \bibfield  {author} {\bibinfo {author} {\bibfnamefont {Brian~C.}\
  \bibnamefont {Tiburzi}}\ and\ \bibinfo {author} {\bibfnamefont {Andre}\
  \bibnamefont {Walker-Loud}},\ }\bibfield  {title} {\enquote {\bibinfo {title}
  {{Strong isospin breaking in the nucleon and Delta masses}},}\ }\href
  {\doibase 10.1016/j.nuclphysa.2005.08.013} {\bibfield  {journal} {\bibinfo
  {journal} {Nucl. Phys.}\ }\textbf {\bibinfo {volume} {A764}},\ \bibinfo
  {pages} {274--302} (\bibinfo {year} {2006})},\ \Eprint
  {http://arxiv.org/abs/hep-lat/0501018} {arXiv:hep-lat/0501018 [hep-lat]}
  \BibitemShut {NoStop}%
\bibitem [{\citenamefont {O'Connell}\ and\ \citenamefont
  {Savage}(2006)}]{OConnell:2005mfp}%
  \BibitemOpen
  \bibfield  {author} {\bibinfo {author} {\bibfnamefont {Donal}\ \bibnamefont
  {O'Connell}}\ and\ \bibinfo {author} {\bibfnamefont {Martin~J.}\ \bibnamefont
  {Savage}},\ }\bibfield  {title} {\enquote {\bibinfo {title} {{Extrapolation
  formulas for neutron EDM calculations in lattice QCD}},}\ }\href {\doibase
  10.1016/j.physletb.2005.11.053} {\bibfield  {journal} {\bibinfo  {journal}
  {Phys. Lett.}\ }\textbf {\bibinfo {volume} {B633}},\ \bibinfo {pages}
  {319--324} (\bibinfo {year} {2006})},\ \Eprint
  {http://arxiv.org/abs/hep-lat/0508009} {arXiv:hep-lat/0508009 [hep-lat]}
  \BibitemShut {NoStop}%
\bibitem [{\citenamefont {Prelovsek}(2006)}]{Prelovsek:2005rf}%
  \BibitemOpen
  \bibfield  {author} {\bibinfo {author} {\bibfnamefont {Sasa}\ \bibnamefont
  {Prelovsek}},\ }\bibfield  {title} {\enquote {\bibinfo {title} {{Effects of
  staggered fermions and mixed actions on the scalar correlator}},}\ }\href
  {\doibase 10.1103/PhysRevD.73.014506} {\bibfield  {journal} {\bibinfo
  {journal} {Phys. Rev.}\ }\textbf {\bibinfo {volume} {D73}},\ \bibinfo {pages}
  {014506} (\bibinfo {year} {2006})},\ \Eprint
  {http://arxiv.org/abs/hep-lat/0510080} {arXiv:hep-lat/0510080 [hep-lat]}
  \BibitemShut {NoStop}%
\bibitem [{\citenamefont {Aubin}\ \emph {et~al.}(2008)\citenamefont {Aubin},
  \citenamefont {Laiho},\ and\ \citenamefont {Van~de Water}}]{Aubin:2008wk}%
  \BibitemOpen
  \bibfield  {author} {\bibinfo {author} {\bibfnamefont {C.}~\bibnamefont
  {Aubin}}, \bibinfo {author} {\bibfnamefont {Jack}\ \bibnamefont {Laiho}}, \
  and\ \bibinfo {author} {\bibfnamefont {Ruth~S.}\ \bibnamefont {Van~de
  Water}},\ }\bibfield  {title} {\enquote {\bibinfo {title} {{Discretization
  effects and the scalar meson correlator in mixed-action lattice
  simulations}},}\ }\href {\doibase 10.1103/PhysRevD.77.114501} {\bibfield
  {journal} {\bibinfo  {journal} {Phys. Rev.}\ }\textbf {\bibinfo {volume}
  {D77}},\ \bibinfo {pages} {114501} (\bibinfo {year} {2008})},\ \Eprint
  {http://arxiv.org/abs/0803.0129} {arXiv:0803.0129 [hep-lat]} \BibitemShut
  {NoStop}%
\bibitem [{\citenamefont {Na}\ \emph {et~al.}(2015)\citenamefont {Na},
  \citenamefont {Bouchard}, \citenamefont {Lepage}, \citenamefont {Monahan},\
  and\ \citenamefont {Shigemitsu}}]{Na:2015kha}%
  \BibitemOpen
  \bibfield  {author} {\bibinfo {author} {\bibfnamefont {Heechang}\
  \bibnamefont {Na}}, \bibinfo {author} {\bibfnamefont {Chris~M.}\ \bibnamefont
  {Bouchard}}, \bibinfo {author} {\bibfnamefont {G.~Peter}\ \bibnamefont
  {Lepage}}, \bibinfo {author} {\bibfnamefont {Chris}\ \bibnamefont {Monahan}},
  \ and\ \bibinfo {author} {\bibfnamefont {Junko}\ \bibnamefont {Shigemitsu}}
  (\bibinfo {collaboration} {HPQCD}),\ }\bibfield  {title} {\enquote {\bibinfo
  {title} {{$B \rightarrow D l \nu$ form factors at nonzero recoil and
  extraction of $|V_{cb}|$}},}\ }\href {\doibase 10.1103/PhysRevD.93.119906,
  10.1103/PhysRevD.92.054510} {\bibfield  {journal} {\bibinfo  {journal} {Phys.
  Rev.}\ }\textbf {\bibinfo {volume} {D92}},\ \bibinfo {pages} {054510}
  (\bibinfo {year} {2015})},\ \bibinfo {note} {[Erratum: Phys.
  Rev.D93,no.11,119906(2016)]},\ \Eprint {http://arxiv.org/abs/1505.03925}
  {arXiv:1505.03925 [hep-lat]} \BibitemShut {NoStop}%
\bibitem [{\citenamefont {Donald}\ \emph {et~al.}(2014)\citenamefont {Donald},
  \citenamefont {Davies}, \citenamefont {Koponen},\ and\ \citenamefont
  {Lepage}}]{Donald:2013sra}%
  \BibitemOpen
  \bibfield  {author} {\bibinfo {author} {\bibfnamefont {G.~C.}\ \bibnamefont
  {Donald}}, \bibinfo {author} {\bibfnamefont {C.~T.~H.}\ \bibnamefont
  {Davies}}, \bibinfo {author} {\bibfnamefont {J.}~\bibnamefont {Koponen}}, \
  and\ \bibinfo {author} {\bibfnamefont {G.~P.}\ \bibnamefont {Lepage}},\
  }\bibfield  {title} {\enquote {\bibinfo {title} {{Prediction of the $D_s^*$
  width from a calculation of its radiative decay in full lattice QCD}},}\
  }\href {\doibase 10.1103/PhysRevLett.112.212002} {\bibfield  {journal}
  {\bibinfo  {journal} {Phys. Rev. Lett.}\ }\textbf {\bibinfo {volume} {112}},\
  \bibinfo {pages} {212002} (\bibinfo {year} {2014})},\ \Eprint
  {http://arxiv.org/abs/1312.5264} {arXiv:1312.5264 [hep-lat]} \BibitemShut
  {NoStop}%
\bibitem [{\citenamefont {Li}\ \emph {et~al.}(2010)\citenamefont {Li} \emph
  {et~al.}}]{Li:2010pw}%
  \BibitemOpen
  \bibfield  {author} {\bibinfo {author} {\bibfnamefont {A.}~\bibnamefont {Li}}
  \emph {et~al.} (\bibinfo {collaboration} {xQCD}),\ }\bibfield  {title}
  {\enquote {\bibinfo {title} {{Overlap Valence on 2+1 Flavor Domain Wall
  Fermion Configurations with Deflation and Low-mode Substitution}},}\ }\href
  {\doibase 10.1103/PhysRevD.82.114501} {\bibfield  {journal} {\bibinfo
  {journal} {Phys. Rev.}\ }\textbf {\bibinfo {volume} {D82}},\ \bibinfo {pages}
  {114501} (\bibinfo {year} {2010})},\ \Eprint {http://arxiv.org/abs/1005.5424}
  {arXiv:1005.5424 [hep-lat]} \BibitemShut {NoStop}%
\bibitem [{\citenamefont {Lujan}\ \emph {et~al.}(2012)\citenamefont {Lujan},
  \citenamefont {Alexandru}, \citenamefont {Chen}, \citenamefont {Draper},
  \citenamefont {Freeman}, \citenamefont {Gong}, \citenamefont {Lee},
  \citenamefont {Li}, \citenamefont {Liu},\ and\ \citenamefont
  {Mathur}}]{Lujan:2012wg}%
  \BibitemOpen
  \bibfield  {author} {\bibinfo {author} {\bibfnamefont {M.}~\bibnamefont
  {Lujan}}, \bibinfo {author} {\bibfnamefont {A.}~\bibnamefont {Alexandru}},
  \bibinfo {author} {\bibfnamefont {Y.}~\bibnamefont {Chen}}, \bibinfo {author}
  {\bibfnamefont {T.}~\bibnamefont {Draper}}, \bibinfo {author} {\bibfnamefont
  {W.}~\bibnamefont {Freeman}}, \bibinfo {author} {\bibfnamefont
  {M.}~\bibnamefont {Gong}}, \bibinfo {author} {\bibfnamefont {F.~X.}\
  \bibnamefont {Lee}}, \bibinfo {author} {\bibfnamefont {A.}~\bibnamefont
  {Li}}, \bibinfo {author} {\bibfnamefont {K.~F.}\ \bibnamefont {Liu}}, \ and\
  \bibinfo {author} {\bibfnamefont {N.}~\bibnamefont {Mathur}},\ }\bibfield
  {title} {\enquote {\bibinfo {title} {{The $\Delta_{mix}$ parameter in the
  overlap on domain-wall mixed action}},}\ }\href {\doibase
  10.1103/PhysRevD.86.014501} {\bibfield  {journal} {\bibinfo  {journal} {Phys.
  Rev.}\ }\textbf {\bibinfo {volume} {D86}},\ \bibinfo {pages} {014501}
  (\bibinfo {year} {2012})},\ \Eprint {http://arxiv.org/abs/1204.6256}
  {arXiv:1204.6256 [hep-lat]} \BibitemShut {NoStop}%
\bibitem [{\citenamefont {Gong}\ \emph {et~al.}(2013)\citenamefont {Gong} \emph
  {et~al.}}]{Gong:2013vja}%
  \BibitemOpen
  \bibfield  {author} {\bibinfo {author} {\bibfnamefont {M.}~\bibnamefont
  {Gong}} \emph {et~al.} (\bibinfo {collaboration} {XQCD}),\ }\bibfield
  {title} {\enquote {\bibinfo {title} {{Strangeness and charmness content of
  the nucleon from overlap fermions on 2+1-flavor domain-wall fermion
  configurations}},}\ }\href {\doibase 10.1103/PhysRevD.88.014503} {\bibfield
  {journal} {\bibinfo  {journal} {Phys. Rev.}\ }\textbf {\bibinfo {volume}
  {D88}},\ \bibinfo {pages} {014503} (\bibinfo {year} {2013})},\ \Eprint
  {http://arxiv.org/abs/1304.1194} {arXiv:1304.1194 [hep-ph]} \BibitemShut
  {NoStop}%
\bibitem [{\citenamefont {Allton}\ \emph {et~al.}(2007)\citenamefont {Allton}
  \emph {et~al.}}]{Allton:2007hx}%
  \BibitemOpen
  \bibfield  {author} {\bibinfo {author} {\bibfnamefont {C.}~\bibnamefont
  {Allton}} \emph {et~al.} (\bibinfo {collaboration} {RBC, UKQCD}),\ }\bibfield
   {title} {\enquote {\bibinfo {title} {{2+1 flavor domain wall QCD on a (2
  fm)*83 lattice: Light meson spectroscopy with L(s) = 16}},}\ }\href {\doibase
  10.1103/PhysRevD.76.014504} {\bibfield  {journal} {\bibinfo  {journal} {Phys.
  Rev.}\ }\textbf {\bibinfo {volume} {D76}},\ \bibinfo {pages} {014504}
  (\bibinfo {year} {2007})},\ \Eprint {http://arxiv.org/abs/hep-lat/0701013}
  {arXiv:hep-lat/0701013 [hep-lat]} \BibitemShut {NoStop}%
\bibitem [{\citenamefont {Aoki}\ \emph {et~al.}(2011)\citenamefont {Aoki} \emph
  {et~al.}}]{Aoki:2010dy}%
  \BibitemOpen
  \bibfield  {author} {\bibinfo {author} {\bibfnamefont {Y.}~\bibnamefont
  {Aoki}} \emph {et~al.} (\bibinfo {collaboration} {RBC, UKQCD}),\ }\bibfield
  {title} {\enquote {\bibinfo {title} {{Continuum Limit Physics from 2+1 Flavor
  Domain Wall QCD}},}\ }\href {\doibase 10.1103/PhysRevD.83.074508} {\bibfield
  {journal} {\bibinfo  {journal} {Phys. Rev.}\ }\textbf {\bibinfo {volume}
  {D83}},\ \bibinfo {pages} {074508} (\bibinfo {year} {2011})},\ \Eprint
  {http://arxiv.org/abs/1011.0892} {arXiv:1011.0892 [hep-lat]} \BibitemShut
  {NoStop}%
\bibitem [{\citenamefont {Basak}\ \emph {et~al.}(2012)\citenamefont {Basak},
  \citenamefont {Datta}, \citenamefont {Padmanath}, \citenamefont {Majumdar},\
  and\ \citenamefont {Mathur}}]{Basak:2012py}%
  \BibitemOpen
  \bibfield  {author} {\bibinfo {author} {\bibfnamefont {S.}~\bibnamefont
  {Basak}}, \bibinfo {author} {\bibfnamefont {S.}~\bibnamefont {Datta}},
  \bibinfo {author} {\bibfnamefont {M.}~\bibnamefont {Padmanath}}, \bibinfo
  {author} {\bibfnamefont {P.}~\bibnamefont {Majumdar}}, \ and\ \bibinfo
  {author} {\bibfnamefont {N.}~\bibnamefont {Mathur}},\ }\bibfield  {title}
  {\enquote {\bibinfo {title} {{Charm and strange hadron spectra from overlap
  fermions on HISQ gauge configurations}},}\ }\bibfield  {booktitle} {\emph
  {\bibinfo {booktitle} {{Proceedings, 30th International Symposium on Lattice
  Field Theory (Lattice 2012): Cairns, Australia, June 24-29, 2012}}},\
  }\href@noop {} {\bibfield  {journal} {\bibinfo  {journal} {PoS}\ }\textbf
  {\bibinfo {volume} {LATTICE2012}},\ \bibinfo {pages} {141} (\bibinfo {year}
  {2012})},\ \Eprint {http://arxiv.org/abs/1211.6277} {arXiv:1211.6277
  [hep-lat]} \BibitemShut {NoStop}%
\bibitem [{\citenamefont {Basak}\ \emph {et~al.}(2014)\citenamefont {Basak},
  \citenamefont {Datta}, \citenamefont {Lytle}, \citenamefont {Padmanath},
  \citenamefont {Majumdar},\ and\ \citenamefont {Mathur}}]{Basak:2013oya}%
  \BibitemOpen
  \bibfield  {author} {\bibinfo {author} {\bibfnamefont {S.}~\bibnamefont
  {Basak}}, \bibinfo {author} {\bibfnamefont {S.}~\bibnamefont {Datta}},
  \bibinfo {author} {\bibfnamefont {A.~T.}\ \bibnamefont {Lytle}}, \bibinfo
  {author} {\bibfnamefont {M.}~\bibnamefont {Padmanath}}, \bibinfo {author}
  {\bibfnamefont {P.}~\bibnamefont {Majumdar}}, \ and\ \bibinfo {author}
  {\bibfnamefont {N.}~\bibnamefont {Mathur}},\ }\bibfield  {title} {\enquote
  {\bibinfo {title} {{Hadron spectra from overlap fermions on HISQ gauge
  configurations}},}\ }\bibfield  {booktitle} {\emph {\bibinfo {booktitle}
  {{Proceedings, 31st International Symposium on Lattice Field Theory (Lattice
  2013): Mainz, Germany, July 29-August 3, 2013}}},\ }\href@noop {} {\bibfield
  {journal} {\bibinfo  {journal} {PoS}\ }\textbf {\bibinfo {volume}
  {LATTICE2013}},\ \bibinfo {pages} {243} (\bibinfo {year} {2014})},\ \Eprint
  {http://arxiv.org/abs/1312.3050} {arXiv:1312.3050 [hep-lat]} \BibitemShut
  {NoStop}%
\bibitem [{\citenamefont {Basak}\ \emph {et~al.}(2015)\citenamefont {Basak},
  \citenamefont {Datta}, \citenamefont {Mathur}, \citenamefont {Lytle},
  \citenamefont {Majumdar},\ and\ \citenamefont {Padmanath}}]{Basak:2014kma}%
  \BibitemOpen
  \bibfield  {author} {\bibinfo {author} {\bibfnamefont {S.}~\bibnamefont
  {Basak}}, \bibinfo {author} {\bibfnamefont {S.}~\bibnamefont {Datta}},
  \bibinfo {author} {\bibfnamefont {Nilmani}\ \bibnamefont {Mathur}}, \bibinfo
  {author} {\bibfnamefont {A.~T.}\ \bibnamefont {Lytle}}, \bibinfo {author}
  {\bibfnamefont {Pushan}\ \bibnamefont {Majumdar}}, \ and\ \bibinfo {author}
  {\bibfnamefont {M.}~\bibnamefont {Padmanath}} (\bibinfo {collaboration}
  {ILGTI}),\ }\bibfield  {title} {\enquote {\bibinfo {title} {{Hadron spectra
  and $\Delta_{mix}$ from overlap quarks on a HISQ sea}},}\ }\bibfield
  {booktitle} {\emph {\bibinfo {booktitle} {{Proceedings, 32nd International
  Symposium on Lattice Field Theory (Lattice 2014): Brookhaven, NY, USA, June
  23-28, 2014}}},\ }\href@noop {} {\bibfield  {journal} {\bibinfo  {journal}
  {PoS}\ }\textbf {\bibinfo {volume} {LATTICE2014}},\ \bibinfo {pages} {083}
  (\bibinfo {year} {2015})},\ \Eprint {http://arxiv.org/abs/1412.7248}
  {arXiv:1412.7248 [hep-lat]} \BibitemShut {NoStop}%
\bibitem [{\citenamefont {Mathur}\ \emph {et~al.}(2016)\citenamefont {Mathur},
  \citenamefont {Padmanath},\ and\ \citenamefont {Lewis}}]{Mathur:2016hsm}%
  \BibitemOpen
  \bibfield  {author} {\bibinfo {author} {\bibfnamefont {Nilmani}\ \bibnamefont
  {Mathur}}, \bibinfo {author} {\bibfnamefont {M.}~\bibnamefont {Padmanath}}, \
  and\ \bibinfo {author} {\bibfnamefont {Randy}\ \bibnamefont {Lewis}},\
  }\bibfield  {title} {\enquote {\bibinfo {title} {{Charmed-Bottom Mesons from
  Lattice QCD}},}\ }\bibfield  {booktitle} {\emph {\bibinfo {booktitle}
  {{Proceedings, 34th International Symposium on Lattice Field Theory (Lattice
  2016): Southampton, UK, July 24-30, 2016}}},\ }\href@noop {} {\bibfield
  {journal} {\bibinfo  {journal} {PoS}\ }\textbf {\bibinfo {volume}
  {LATTICE2016}},\ \bibinfo {pages} {100} (\bibinfo {year} {2016})},\ \Eprint
  {http://arxiv.org/abs/1611.04085} {arXiv:1611.04085 [hep-lat]} \BibitemShut
  {NoStop}%
\bibitem [{\citenamefont {Bhattacharya}\ \emph {et~al.}(2014)\citenamefont
  {Bhattacharya}, \citenamefont {Cohen}, \citenamefont {Gupta}, \citenamefont
  {Joseph}, \citenamefont {Lin},\ and\ \citenamefont
  {Yoon}}]{Bhattacharya:2013ehc}%
  \BibitemOpen
  \bibfield  {author} {\bibinfo {author} {\bibfnamefont {Tanmoy}\ \bibnamefont
  {Bhattacharya}}, \bibinfo {author} {\bibfnamefont {Saul~D.}\ \bibnamefont
  {Cohen}}, \bibinfo {author} {\bibfnamefont {Rajan}\ \bibnamefont {Gupta}},
  \bibinfo {author} {\bibfnamefont {Anosh}\ \bibnamefont {Joseph}}, \bibinfo
  {author} {\bibfnamefont {Huey-Wen}\ \bibnamefont {Lin}}, \ and\ \bibinfo
  {author} {\bibfnamefont {Boram}\ \bibnamefont {Yoon}},\ }\bibfield  {title}
  {\enquote {\bibinfo {title} {{Nucleon Charges and Electromagnetic Form
  Factors from 2+1+1-Flavor Lattice QCD}},}\ }\href {\doibase
  10.1103/PhysRevD.89.094502} {\bibfield  {journal} {\bibinfo  {journal} {Phys.
  Rev.}\ }\textbf {\bibinfo {volume} {D89}},\ \bibinfo {pages} {094502}
  (\bibinfo {year} {2014})},\ \Eprint {http://arxiv.org/abs/1306.5435}
  {arXiv:1306.5435 [hep-lat]} \BibitemShut {NoStop}%
\bibitem [{\citenamefont {Bhattacharya}\ \emph {et~al.}(2015)\citenamefont
  {Bhattacharya}, \citenamefont {Cirigliano}, \citenamefont {Cohen},
  \citenamefont {Gupta}, \citenamefont {Joseph}, \citenamefont {Lin},\ and\
  \citenamefont {Yoon}}]{Bhattacharya:2015wna}%
  \BibitemOpen
  \bibfield  {author} {\bibinfo {author} {\bibfnamefont {Tanmoy}\ \bibnamefont
  {Bhattacharya}}, \bibinfo {author} {\bibfnamefont {Vincenzo}\ \bibnamefont
  {Cirigliano}}, \bibinfo {author} {\bibfnamefont {Saul}\ \bibnamefont
  {Cohen}}, \bibinfo {author} {\bibfnamefont {Rajan}\ \bibnamefont {Gupta}},
  \bibinfo {author} {\bibfnamefont {Anosh}\ \bibnamefont {Joseph}}, \bibinfo
  {author} {\bibfnamefont {Huey-Wen}\ \bibnamefont {Lin}}, \ and\ \bibinfo
  {author} {\bibfnamefont {Boram}\ \bibnamefont {Yoon}} (\bibinfo
  {collaboration} {PNDME}),\ }\bibfield  {title} {\enquote {\bibinfo {title}
  {{Iso-vector and Iso-scalar Tensor Charges of the Nucleon from Lattice
  QCD}},}\ }\href {\doibase 10.1103/PhysRevD.92.094511} {\bibfield  {journal}
  {\bibinfo  {journal} {Phys. Rev.}\ }\textbf {\bibinfo {volume} {D92}},\
  \bibinfo {pages} {094511} (\bibinfo {year} {2015})},\ \Eprint
  {http://arxiv.org/abs/1506.06411} {arXiv:1506.06411 [hep-lat]} \BibitemShut
  {NoStop}%
\bibitem [{\citenamefont {Durr}\ \emph {et~al.}(2012)\citenamefont {Durr} \emph
  {et~al.}}]{Durr:2011mp}%
  \BibitemOpen
  \bibfield  {author} {\bibinfo {author} {\bibfnamefont {S.}~\bibnamefont
  {Durr}} \emph {et~al.},\ }\bibfield  {title} {\enquote {\bibinfo {title}
  {{Sigma term and strangeness content of octet baryons}},}\ }\href {\doibase
  10.1103/PhysRevD.85.014509, 10.1103/PhysRevD.93.039905} {\bibfield  {journal}
  {\bibinfo  {journal} {Phys. Rev.}\ }\textbf {\bibinfo {volume} {D85}},\
  \bibinfo {pages} {014509} (\bibinfo {year} {2012})},\ \bibinfo {note}
  {[Erratum: Phys. Rev.D93,no.3,039905(2016)]},\ \Eprint
  {http://arxiv.org/abs/1109.4265} {arXiv:1109.4265 [hep-lat]} \BibitemShut
  {NoStop}%
\bibitem [{\citenamefont {Durr}\ \emph {et~al.}(2016)\citenamefont {Durr} \emph
  {et~al.}}]{Durr:2015dna}%
  \BibitemOpen
  \bibfield  {author} {\bibinfo {author} {\bibfnamefont {S.}~\bibnamefont
  {Durr}} \emph {et~al.},\ }\bibfield  {title} {\enquote {\bibinfo {title}
  {{Lattice computation of the nucleon scalar quark contents at the physical
  point}},}\ }\href {\doibase 10.1103/PhysRevLett.116.172001} {\bibfield
  {journal} {\bibinfo  {journal} {Phys. Rev. Lett.}\ }\textbf {\bibinfo
  {volume} {116}},\ \bibinfo {pages} {172001} (\bibinfo {year} {2016})},\
  \Eprint {http://arxiv.org/abs/1510.08013} {arXiv:1510.08013 [hep-lat]}
  \BibitemShut {NoStop}%
\bibitem [{\citenamefont {Yang}\ \emph {et~al.}(2016)\citenamefont {Yang},
  \citenamefont {Alexandru}, \citenamefont {Draper}, \citenamefont {Liang},\
  and\ \citenamefont {Liu}}]{Yang:2015uis}%
  \BibitemOpen
  \bibfield  {author} {\bibinfo {author} {\bibfnamefont {Yi-Bo}\ \bibnamefont
  {Yang}}, \bibinfo {author} {\bibfnamefont {Andrei}\ \bibnamefont
  {Alexandru}}, \bibinfo {author} {\bibfnamefont {Terrence}\ \bibnamefont
  {Draper}}, \bibinfo {author} {\bibfnamefont {Jian}\ \bibnamefont {Liang}}, \
  and\ \bibinfo {author} {\bibfnamefont {Keh-Fei}\ \bibnamefont {Liu}}
  (\bibinfo {collaboration} {xQCD}),\ }\bibfield  {title} {\enquote {\bibinfo
  {title} {{$\pi$N and strangeness sigma terms at the physical point with
  chiral fermions}},}\ }\href {\doibase 10.1103/PhysRevD.94.054503} {\bibfield
  {journal} {\bibinfo  {journal} {Phys. Rev.}\ }\textbf {\bibinfo {volume}
  {D94}},\ \bibinfo {pages} {054503} (\bibinfo {year} {2016})},\ \Eprint
  {http://arxiv.org/abs/1511.09089} {arXiv:1511.09089 [hep-lat]} \BibitemShut
  {NoStop}%
\bibitem [{\citenamefont {Sufian}\ \emph {et~al.}(2017)\citenamefont {Sufian},
  \citenamefont {Yang}, \citenamefont {Alexandru}, \citenamefont {Draper},
  \citenamefont {Liang},\ and\ \citenamefont {Liu}}]{Sufian:2016pex}%
  \BibitemOpen
  \bibfield  {author} {\bibinfo {author} {\bibfnamefont {Raza~Sabbir}\
  \bibnamefont {Sufian}}, \bibinfo {author} {\bibfnamefont {Yi-Bo}\
  \bibnamefont {Yang}}, \bibinfo {author} {\bibfnamefont {Andrei}\ \bibnamefont
  {Alexandru}}, \bibinfo {author} {\bibfnamefont {Terrence}\ \bibnamefont
  {Draper}}, \bibinfo {author} {\bibfnamefont {Jian}\ \bibnamefont {Liang}}, \
  and\ \bibinfo {author} {\bibfnamefont {Keh-Fei}\ \bibnamefont {Liu}},\
  }\bibfield  {title} {\enquote {\bibinfo {title} {{Strange Quark Magnetic
  Moment of the Nucleon at the Physical Point}},}\ }\href {\doibase
  10.1103/PhysRevLett.118.042001} {\bibfield  {journal} {\bibinfo  {journal}
  {Phys. Rev. Lett.}\ }\textbf {\bibinfo {volume} {118}},\ \bibinfo {pages}
  {042001} (\bibinfo {year} {2017})},\ \Eprint
  {http://arxiv.org/abs/1606.07075} {arXiv:1606.07075 [hep-ph]} \BibitemShut
  {NoStop}%
\bibitem [{\citenamefont {Bhattacharya}\ \emph {et~al.}(2016)\citenamefont
  {Bhattacharya}, \citenamefont {Cirigliano}, \citenamefont {Cohen},
  \citenamefont {Gupta}, \citenamefont {Lin},\ and\ \citenamefont
  {Yoon}}]{Bhattacharya:2016zcn}%
  \BibitemOpen
  \bibfield  {author} {\bibinfo {author} {\bibfnamefont {Tanmoy}\ \bibnamefont
  {Bhattacharya}}, \bibinfo {author} {\bibfnamefont {Vincenzo}\ \bibnamefont
  {Cirigliano}}, \bibinfo {author} {\bibfnamefont {Saul}\ \bibnamefont
  {Cohen}}, \bibinfo {author} {\bibfnamefont {Rajan}\ \bibnamefont {Gupta}},
  \bibinfo {author} {\bibfnamefont {Huey-Wen}\ \bibnamefont {Lin}}, \ and\
  \bibinfo {author} {\bibfnamefont {Boram}\ \bibnamefont {Yoon}},\ }\bibfield
  {title} {\enquote {\bibinfo {title} {{Axial, Scalar and Tensor Charges of the
  Nucleon from 2+1+1-flavor Lattice QCD}},}\ }\href {\doibase
  10.1103/PhysRevD.94.054508} {\bibfield  {journal} {\bibinfo  {journal} {Phys.
  Rev.}\ }\textbf {\bibinfo {volume} {D94}},\ \bibinfo {pages} {054508}
  (\bibinfo {year} {2016})},\ \Eprint {http://arxiv.org/abs/1606.07049}
  {arXiv:1606.07049 [hep-lat]} \BibitemShut {NoStop}%
\bibitem [{\citenamefont {Borsanyi}\ \emph {et~al.}(2013)\citenamefont
  {Borsanyi} \emph {et~al.}}]{Borsanyi:2013lga}%
  \BibitemOpen
  \bibfield  {author} {\bibinfo {author} {\bibfnamefont {Sz.}\ \bibnamefont
  {Borsanyi}} \emph {et~al.} (\bibinfo {collaboration}
  {Budapest-Marseille-Wuppertal}),\ }\bibfield  {title} {\enquote {\bibinfo
  {title} {{Isospin splittings in the light baryon octet from lattice QCD and
  QED}},}\ }\href {\doibase 10.1103/PhysRevLett.111.252001} {\bibfield
  {journal} {\bibinfo  {journal} {Phys. Rev. Lett.}\ }\textbf {\bibinfo
  {volume} {111}},\ \bibinfo {pages} {252001} (\bibinfo {year} {2013})},\
  \Eprint {http://arxiv.org/abs/1306.2287} {arXiv:1306.2287 [hep-lat]}
  \BibitemShut {NoStop}%
\bibitem [{\citenamefont {Borsanyi}\ \emph {et~al.}(2015)\citenamefont
  {Borsanyi} \emph {et~al.}}]{Borsanyi:2014jba}%
  \BibitemOpen
  \bibfield  {author} {\bibinfo {author} {\bibfnamefont {Sz.}\ \bibnamefont
  {Borsanyi}} \emph {et~al.},\ }\bibfield  {title} {\enquote {\bibinfo {title}
  {{Ab initio calculation of the neutron-proton mass difference}},}\ }\href
  {\doibase 10.1126/science.1257050} {\bibfield  {journal} {\bibinfo  {journal}
  {Science}\ }\textbf {\bibinfo {volume} {347}},\ \bibinfo {pages} {1452--1455}
  (\bibinfo {year} {2015})},\ \Eprint {http://arxiv.org/abs/1406.4088}
  {arXiv:1406.4088 [hep-lat]} \BibitemShut {NoStop}%
\bibitem [{\citenamefont {Brantley}\ \emph {et~al.}(2016)\citenamefont
  {Brantley}, \citenamefont {Joo}, \citenamefont {Mastropas}, \citenamefont
  {Mereghetti}, \citenamefont {Monge-Camacho}, \citenamefont {Tiburzi},\ and\
  \citenamefont {Walker-Loud}}]{Brantley:2016our}%
  \BibitemOpen
  \bibfield  {author} {\bibinfo {author} {\bibfnamefont {David~A.}\
  \bibnamefont {Brantley}}, \bibinfo {author} {\bibfnamefont {Balint}\
  \bibnamefont {Joo}}, \bibinfo {author} {\bibfnamefont {Ekaterina~V.}\
  \bibnamefont {Mastropas}}, \bibinfo {author} {\bibfnamefont {Emanuele}\
  \bibnamefont {Mereghetti}}, \bibinfo {author} {\bibfnamefont {Henry}\
  \bibnamefont {Monge-Camacho}}, \bibinfo {author} {\bibfnamefont {Brian~C.}\
  \bibnamefont {Tiburzi}}, \ and\ \bibinfo {author} {\bibfnamefont {Andre}\
  \bibnamefont {Walker-Loud}},\ }\bibfield  {title} {\enquote {\bibinfo {title}
  {{Strong isospin violation and chiral logarithms in the baryon spectrum}},}\
  }\href@noop {} {\  (\bibinfo {year} {2016})},\ \Eprint
  {http://arxiv.org/abs/1612.07733} {arXiv:1612.07733 [hep-lat]} \BibitemShut
  {NoStop}%
\bibitem [{\citenamefont {Follana}\ \emph {et~al.}(2007)\citenamefont
  {Follana}, \citenamefont {Mason}, \citenamefont {Davies}, \citenamefont
  {Hornbostel}, \citenamefont {Lepage}, \citenamefont {Shigemitsu},
  \citenamefont {Trottier},\ and\ \citenamefont {Wong}}]{Follana:2006rc}%
  \BibitemOpen
  \bibfield  {author} {\bibinfo {author} {\bibfnamefont {E.}~\bibnamefont
  {Follana}}, \bibinfo {author} {\bibfnamefont {Q.}~\bibnamefont {Mason}},
  \bibinfo {author} {\bibfnamefont {C.}~\bibnamefont {Davies}}, \bibinfo
  {author} {\bibfnamefont {K.}~\bibnamefont {Hornbostel}}, \bibinfo {author}
  {\bibfnamefont {G.~P.}\ \bibnamefont {Lepage}}, \bibinfo {author}
  {\bibfnamefont {J.}~\bibnamefont {Shigemitsu}}, \bibinfo {author}
  {\bibfnamefont {H.}~\bibnamefont {Trottier}}, \ and\ \bibinfo {author}
  {\bibfnamefont {K.}~\bibnamefont {Wong}} (\bibinfo {collaboration} {HPQCD,
  UKQCD}),\ }\bibfield  {title} {\enquote {\bibinfo {title} {{Highly improved
  staggered quarks on the lattice, with applications to charm physics}},}\
  }\href {\doibase 10.1103/PhysRevD.75.054502} {\bibfield  {journal} {\bibinfo
  {journal} {Phys. Rev.}\ }\textbf {\bibinfo {volume} {D75}},\ \bibinfo {pages}
  {054502} (\bibinfo {year} {2007})},\ \Eprint
  {http://arxiv.org/abs/hep-lat/0610092} {arXiv:hep-lat/0610092 [hep-lat]}
  \BibitemShut {NoStop}%
\bibitem [{\citenamefont {Bazavov}\ \emph
  {et~al.}(2010{\natexlab{b}})\citenamefont {Bazavov} \emph
  {et~al.}}]{Bazavov:2010ru}%
  \BibitemOpen
  \bibfield  {author} {\bibinfo {author} {\bibfnamefont {A.}~\bibnamefont
  {Bazavov}} \emph {et~al.} (\bibinfo {collaboration} {MILC}),\ }\bibfield
  {title} {\enquote {\bibinfo {title} {{Scaling studies of QCD with the
  dynamical HISQ action}},}\ }\href {\doibase 10.1103/PhysRevD.82.074501}
  {\bibfield  {journal} {\bibinfo  {journal} {Phys. Rev.}\ }\textbf {\bibinfo
  {volume} {D82}},\ \bibinfo {pages} {074501} (\bibinfo {year}
  {2010}{\natexlab{b}})},\ \Eprint {http://arxiv.org/abs/1004.0342}
  {arXiv:1004.0342 [hep-lat]} \BibitemShut {NoStop}%
\bibitem [{\citenamefont {Bazavov}\ \emph {et~al.}(2013)\citenamefont {Bazavov}
  \emph {et~al.}}]{Bazavov:2012xda}%
  \BibitemOpen
  \bibfield  {author} {\bibinfo {author} {\bibfnamefont {A.}~\bibnamefont
  {Bazavov}} \emph {et~al.} (\bibinfo {collaboration} {MILC}),\ }\bibfield
  {title} {\enquote {\bibinfo {title} {{Lattice QCD ensembles with four flavors
  of highly improved staggered quarks}},}\ }\href {\doibase
  10.1103/PhysRevD.87.054505} {\bibfield  {journal} {\bibinfo  {journal} {Phys.
  Rev.}\ }\textbf {\bibinfo {volume} {D87}},\ \bibinfo {pages} {054505}
  (\bibinfo {year} {2013})},\ \Eprint {http://arxiv.org/abs/1212.4768}
  {arXiv:1212.4768 [hep-lat]} \BibitemShut {NoStop}%
\bibitem [{\citenamefont {Brower}\ \emph {et~al.}(2005)\citenamefont {Brower},
  \citenamefont {Neff},\ and\ \citenamefont {Orginos}}]{Brower:2004xi}%
  \BibitemOpen
  \bibfield  {author} {\bibinfo {author} {\bibfnamefont {Richard~C.}\
  \bibnamefont {Brower}}, \bibinfo {author} {\bibfnamefont {Hartmut}\
  \bibnamefont {Neff}}, \ and\ \bibinfo {author} {\bibfnamefont {Kostas}\
  \bibnamefont {Orginos}},\ }\bibfield  {title} {\enquote {\bibinfo {title}
  {{Mobius fermions: Improved domain wall chiral fermions}},}\ }\bibfield
  {booktitle} {\emph {\bibinfo {booktitle} {{Lattice field theory. Proceedings,
  22nd International Symposium, Lattice 2004, Batavia, USA, June 21-26,
  2004}}},\ }\href {\doibase 10.1016/j.nuclphysbps.2004.11.180} {\bibfield
  {journal} {\bibinfo  {journal} {Nucl. Phys. Proc. Suppl.}\ }\textbf {\bibinfo
  {volume} {140}},\ \bibinfo {pages} {686--688} (\bibinfo {year} {2005})},\
  \bibinfo {note} {[,686(2004)]},\ \Eprint
  {http://arxiv.org/abs/hep-lat/0409118} {arXiv:hep-lat/0409118 [hep-lat]}
  \BibitemShut {NoStop}%
\bibitem [{\citenamefont {Brower}\ \emph {et~al.}(2006)\citenamefont {Brower},
  \citenamefont {Neff},\ and\ \citenamefont {Orginos}}]{Brower:2005qw}%
  \BibitemOpen
  \bibfield  {author} {\bibinfo {author} {\bibfnamefont {R.~C.}\ \bibnamefont
  {Brower}}, \bibinfo {author} {\bibfnamefont {H.}~\bibnamefont {Neff}}, \ and\
  \bibinfo {author} {\bibfnamefont {K.}~\bibnamefont {Orginos}},\ }\bibfield
  {title} {\enquote {\bibinfo {title} {{Mobius fermions}},}\ }\bibfield
  {booktitle} {\emph {\bibinfo {booktitle} {{Hadron physics, proceedings of the
  Workshop on Computational Hadron Physics, University of Cyprus, Nicosia,
  Cyprus, 14-17 September 2005}}},\ }\href {\doibase
  10.1016/j.nuclphysbps.2006.01.047} {\bibfield  {journal} {\bibinfo  {journal}
  {Nucl. Phys. Proc. Suppl.}\ }\textbf {\bibinfo {volume} {153}},\ \bibinfo
  {pages} {191--198} (\bibinfo {year} {2006})},\ \Eprint
  {http://arxiv.org/abs/hep-lat/0511031} {arXiv:hep-lat/0511031 [hep-lat]}
  \BibitemShut {NoStop}%
\bibitem [{\citenamefont {Brower}\ \emph {et~al.}(2012)\citenamefont {Brower},
  \citenamefont {Neff},\ and\ \citenamefont {Orginos}}]{Brower:2012vk}%
  \BibitemOpen
  \bibfield  {author} {\bibinfo {author} {\bibfnamefont {Richard~C.}\
  \bibnamefont {Brower}}, \bibinfo {author} {\bibfnamefont {Harmut}\
  \bibnamefont {Neff}}, \ and\ \bibinfo {author} {\bibfnamefont {Kostas}\
  \bibnamefont {Orginos}},\ }\bibfield  {title} {\enquote {\bibinfo {title}
  {{The M\'obius Domain Wall Fermion Algorithm}},}\ }\href@noop {} {\
  (\bibinfo {year} {2012})},\ \Eprint {http://arxiv.org/abs/1206.5214}
  {arXiv:1206.5214 [hep-lat]} \BibitemShut {NoStop}%
\bibitem [{\citenamefont {Neuberger}(1998{\natexlab{a}})}]{Neuberger:1997bg}%
  \BibitemOpen
  \bibfield  {author} {\bibinfo {author} {\bibfnamefont {Herbert}\ \bibnamefont
  {Neuberger}},\ }\bibfield  {title} {\enquote {\bibinfo {title} {{Vector -
  like gauge theories with almost massless fermions on the lattice}},}\ }\href
  {\doibase 10.1103/PhysRevD.57.5417} {\bibfield  {journal} {\bibinfo
  {journal} {Phys. Rev.}\ }\textbf {\bibinfo {volume} {D57}},\ \bibinfo {pages}
  {5417--5433} (\bibinfo {year} {1998}{\natexlab{a}})},\ \Eprint
  {http://arxiv.org/abs/hep-lat/9710089} {arXiv:hep-lat/9710089 [hep-lat]}
  \BibitemShut {NoStop}%
\bibitem [{\citenamefont {Neuberger}(1998{\natexlab{b}})}]{Neuberger:1997fp}%
  \BibitemOpen
  \bibfield  {author} {\bibinfo {author} {\bibfnamefont {Herbert}\ \bibnamefont
  {Neuberger}},\ }\bibfield  {title} {\enquote {\bibinfo {title} {{Exactly
  massless quarks on the lattice}},}\ }\href {\doibase
  10.1016/S0370-2693(97)01368-3} {\bibfield  {journal} {\bibinfo  {journal}
  {Phys. Lett.}\ }\textbf {\bibinfo {volume} {B417}},\ \bibinfo {pages}
  {141--144} (\bibinfo {year} {1998}{\natexlab{b}})},\ \Eprint
  {http://arxiv.org/abs/hep-lat/9707022} {arXiv:hep-lat/9707022 [hep-lat]}
  \BibitemShut {NoStop}%
\bibitem [{\citenamefont {Vranas}(1998)}]{Vranas:1997da}%
  \BibitemOpen
  \bibfield  {author} {\bibinfo {author} {\bibfnamefont {Pavlos~M.}\
  \bibnamefont {Vranas}},\ }\bibfield  {title} {\enquote {\bibinfo {title}
  {{Chiral symmetry restoration in the Schwinger model with domain wall
  fermions}},}\ }\href {\doibase 10.1103/PhysRevD.57.1415} {\bibfield
  {journal} {\bibinfo  {journal} {Phys. Rev.}\ }\textbf {\bibinfo {volume}
  {D57}},\ \bibinfo {pages} {1415--1432} (\bibinfo {year} {1998})},\ \Eprint
  {http://arxiv.org/abs/hep-lat/9705023} {arXiv:hep-lat/9705023 [hep-lat]}
  \BibitemShut {NoStop}%
\bibitem [{\citenamefont {Kikukawa}\ and\ \citenamefont
  {Noguchi}(2000)}]{Kikukawa:1999sy}%
  \BibitemOpen
  \bibfield  {author} {\bibinfo {author} {\bibfnamefont {Yoshio}\ \bibnamefont
  {Kikukawa}}\ and\ \bibinfo {author} {\bibfnamefont {Tatsuya}\ \bibnamefont
  {Noguchi}},\ }\bibfield  {title} {\enquote {\bibinfo {title} {{Low-energy
  effective action of domain wall fermion and the Ginsparg-Wilson relation}},}\
  }\bibfield  {booktitle} {\emph {\bibinfo {booktitle} {{Lattice field theory.
  Proceedings, 17th International Symposium, Lattice'99, Pisa, Italy, June
  29-July 3, 1999}}},\ }\href {\doibase 10.1016/S0920-5632(00)91758-4}
  {\bibfield  {journal} {\bibinfo  {journal} {Nucl. Phys. Proc. Suppl.}\
  }\textbf {\bibinfo {volume} {83}},\ \bibinfo {pages} {630--632} (\bibinfo
  {year} {2000})},\ \Eprint {http://arxiv.org/abs/hep-lat/9902022}
  {arXiv:hep-lat/9902022 [hep-lat]} \BibitemShut {NoStop}%
\bibitem [{\citenamefont {Edwards}\ and\ \citenamefont
  {Heller}(2001)}]{Edwards:2000qv}%
  \BibitemOpen
  \bibfield  {author} {\bibinfo {author} {\bibfnamefont {Robert~G.}\
  \bibnamefont {Edwards}}\ and\ \bibinfo {author} {\bibfnamefont {Urs~M.}\
  \bibnamefont {Heller}},\ }\bibfield  {title} {\enquote {\bibinfo {title}
  {{Domain wall fermions with exact chiral symmetry}},}\ }\href {\doibase
  10.1103/PhysRevD.63.094505} {\bibfield  {journal} {\bibinfo  {journal} {Phys.
  Rev.}\ }\textbf {\bibinfo {volume} {D63}},\ \bibinfo {pages} {094505}
  (\bibinfo {year} {2001})},\ \Eprint {http://arxiv.org/abs/hep-lat/0005002}
  {arXiv:hep-lat/0005002 [hep-lat]} \BibitemShut {NoStop}%
\bibitem [{\citenamefont {Kennedy}(2006)}]{Kennedy:2006ax}%
  \BibitemOpen
  \bibfield  {author} {\bibinfo {author} {\bibfnamefont {A.~D.}\ \bibnamefont
  {Kennedy}},\ }\bibfield  {title} {\enquote {\bibinfo {title} {{Algorithms for
  dynamical fermions}},}\ }\href@noop {} {\  (\bibinfo {year} {2006})},\
  \Eprint {http://arxiv.org/abs/hep-lat/0607038} {arXiv:hep-lat/0607038
  [hep-lat]} \BibitemShut {NoStop}%
\bibitem [{\citenamefont {Edwards}\ and\ \citenamefont
  {Joo}(2005)}]{Edwards:2004sx}%
  \BibitemOpen
  \bibfield  {author} {\bibinfo {author} {\bibfnamefont {Robert~G.}\
  \bibnamefont {Edwards}}\ and\ \bibinfo {author} {\bibfnamefont {Balint}\
  \bibnamefont {Joo}} (\bibinfo {collaboration} {SciDAC Collaboration, LHPC
  Collaboration, UKQCD Collaboration}),\ }\bibfield  {title} {\enquote
  {\bibinfo {title} {{The Chroma software system for lattice QCD}},}\ }\href
  {\doibase 10.1016/j.nuclphysbps.2004.11.254} {\bibfield  {journal} {\bibinfo
  {journal} {Nucl.Phys.Proc.Suppl.}\ }\textbf {\bibinfo {volume} {140}},\
  \bibinfo {pages} {832} (\bibinfo {year} {2005})},\ \Eprint
  {http://arxiv.org/abs/hep-lat/0409003} {arXiv:hep-lat/0409003 [hep-lat]}
  \BibitemShut {NoStop}%
\bibitem [{\citenamefont {Hasenfratz}\ and\ \citenamefont
  {Knechtli}(2001)}]{Hasenfratz:2001hp}%
  \BibitemOpen
  \bibfield  {author} {\bibinfo {author} {\bibfnamefont {Anna}\ \bibnamefont
  {Hasenfratz}}\ and\ \bibinfo {author} {\bibfnamefont {Francesco}\
  \bibnamefont {Knechtli}},\ }\bibfield  {title} {\enquote {\bibinfo {title}
  {{Flavor symmetry and the static potential with hypercubic blocking}},}\
  }\href {\doibase 10.1103/PhysRevD.64.034504} {\bibfield  {journal} {\bibinfo
  {journal} {Phys. Rev.}\ }\textbf {\bibinfo {volume} {D64}},\ \bibinfo {pages}
  {034504} (\bibinfo {year} {2001})},\ \Eprint
  {http://arxiv.org/abs/hep-lat/0103029} {arXiv:hep-lat/0103029 [hep-lat]}
  \BibitemShut {NoStop}%
\bibitem [{\citenamefont {DeGrand}\ \emph {et~al.}(2003)\citenamefont
  {DeGrand}, \citenamefont {Hasenfratz},\ and\ \citenamefont
  {Kovacs}}]{DeGrand:2002vu}%
  \BibitemOpen
  \bibfield  {author} {\bibinfo {author} {\bibfnamefont {Thomas~A.}\
  \bibnamefont {DeGrand}}, \bibinfo {author} {\bibfnamefont {Anna}\
  \bibnamefont {Hasenfratz}}, \ and\ \bibinfo {author} {\bibfnamefont
  {Tamas~G.}\ \bibnamefont {Kovacs}},\ }\bibfield  {title} {\enquote {\bibinfo
  {title} {{Improving the chiral properties of lattice fermions}},}\ }\href
  {\doibase 10.1103/PhysRevD.67.054501} {\bibfield  {journal} {\bibinfo
  {journal} {Phys. Rev.}\ }\textbf {\bibinfo {volume} {D67}},\ \bibinfo {pages}
  {054501} (\bibinfo {year} {2003})},\ \Eprint
  {http://arxiv.org/abs/hep-lat/0211006} {arXiv:hep-lat/0211006 [hep-lat]}
  \BibitemShut {NoStop}%
\bibitem [{\citenamefont {DeGrand}(2004)}]{DeGrand:2003in}%
  \BibitemOpen
  \bibfield  {author} {\bibinfo {author} {\bibfnamefont {Thomas~A.}\
  \bibnamefont {DeGrand}} (\bibinfo {collaboration} {MILC}),\ }\bibfield
  {title} {\enquote {\bibinfo {title} {{Kaon B parameter in quenched QCD}},}\
  }\href {\doibase 10.1103/PhysRevD.69.014504} {\bibfield  {journal} {\bibinfo
  {journal} {Phys. Rev.}\ }\textbf {\bibinfo {volume} {D69}},\ \bibinfo {pages}
  {014504} (\bibinfo {year} {2004})},\ \Eprint
  {http://arxiv.org/abs/hep-lat/0309026} {arXiv:hep-lat/0309026 [hep-lat]}
  \BibitemShut {NoStop}%
\bibitem [{\citenamefont {Durr}\ \emph {et~al.}(2004)\citenamefont {Durr},
  \citenamefont {Hoelbling},\ and\ \citenamefont {Wenger}}]{Durr:2004as}%
  \BibitemOpen
  \bibfield  {author} {\bibinfo {author} {\bibfnamefont {Stephan}\ \bibnamefont
  {Durr}}, \bibinfo {author} {\bibfnamefont {Christian}\ \bibnamefont
  {Hoelbling}}, \ and\ \bibinfo {author} {\bibfnamefont {Urs}\ \bibnamefont
  {Wenger}},\ }\bibfield  {title} {\enquote {\bibinfo {title} {{Staggered
  eigenvalue mimicry}},}\ }\href {\doibase 10.1103/PhysRevD.70.094502}
  {\bibfield  {journal} {\bibinfo  {journal} {Phys. Rev.}\ }\textbf {\bibinfo
  {volume} {D70}},\ \bibinfo {pages} {094502} (\bibinfo {year} {2004})},\
  \Eprint {http://arxiv.org/abs/hep-lat/0406027} {arXiv:hep-lat/0406027
  [hep-lat]} \BibitemShut {NoStop}%
\bibitem [{\citenamefont {Narayanan}\ and\ \citenamefont
  {Neuberger}(2006)}]{Narayanan:2006rf}%
  \BibitemOpen
  \bibfield  {author} {\bibinfo {author} {\bibfnamefont {R.}~\bibnamefont
  {Narayanan}}\ and\ \bibinfo {author} {\bibfnamefont {H.}~\bibnamefont
  {Neuberger}},\ }\bibfield  {title} {\enquote {\bibinfo {title} {{Infinite N
  phase transitions in continuum Wilson loop operators}},}\ }\href {\doibase
  10.1088/1126-6708/2006/03/064} {\bibfield  {journal} {\bibinfo  {journal}
  {JHEP}\ }\textbf {\bibinfo {volume} {0603}},\ \bibinfo {pages} {064}
  (\bibinfo {year} {2006})},\ \Eprint {http://arxiv.org/abs/hep-th/0601210}
  {arXiv:hep-th/0601210 [hep-th]} \BibitemShut {NoStop}%
\bibitem [{\citenamefont {{L\"uscher}}\ and\ \citenamefont
  {Weisz}(2011)}]{Luscher:2011bx}%
  \BibitemOpen
  \bibfield  {author} {\bibinfo {author} {\bibfnamefont {M.}~\bibnamefont
  {{L\"uscher}}}\ and\ \bibinfo {author} {\bibfnamefont {P.}~\bibnamefont
  {Weisz}},\ }\bibfield  {title} {\enquote {\bibinfo {title} {{Perturbative
  analysis of the gradient flow in non-abelian gauge theories}},}\ }\href
  {\doibase 10.1007/JHEP02(2011)051} {\bibfield  {journal} {\bibinfo  {journal}
  {JHEP}\ }\textbf {\bibinfo {volume} {1102}},\ \bibinfo {pages} {051}
  (\bibinfo {year} {2011})},\ \Eprint {http://arxiv.org/abs/1101.0963}
  {arXiv:1101.0963 [hep-th]} \BibitemShut {NoStop}%
\bibitem [{\citenamefont {{L\"uscher}}(2013)}]{Luscher:2013cpa}%
  \BibitemOpen
  \bibfield  {author} {\bibinfo {author} {\bibfnamefont {Martin}\ \bibnamefont
  {{L\"uscher}}},\ }\bibfield  {title} {\enquote {\bibinfo {title} {{Chiral
  symmetry and the Yang--Mills gradient flow}},}\ }\href {\doibase
  10.1007/JHEP04(2013)123} {\bibfield  {journal} {\bibinfo  {journal} {JHEP}\
  }\textbf {\bibinfo {volume} {1304}},\ \bibinfo {pages} {123} (\bibinfo {year}
  {2013})},\ \Eprint {http://arxiv.org/abs/1302.5246} {arXiv:1302.5246
  [hep-lat]} \BibitemShut {NoStop}%
\bibitem [{\citenamefont {Morningstar}\ and\ \citenamefont
  {Peardon}(2004)}]{Morningstar:2003gk}%
  \BibitemOpen
  \bibfield  {author} {\bibinfo {author} {\bibfnamefont {Colin}\ \bibnamefont
  {Morningstar}}\ and\ \bibinfo {author} {\bibfnamefont {Mike~J.}\ \bibnamefont
  {Peardon}},\ }\bibfield  {title} {\enquote {\bibinfo {title} {{Analytic
  smearing of SU(3) link variables in lattice QCD}},}\ }\href {\doibase
  10.1103/PhysRevD.69.054501} {\bibfield  {journal} {\bibinfo  {journal}
  {Phys.Rev.}\ }\textbf {\bibinfo {volume} {D69}},\ \bibinfo {pages} {054501}
  (\bibinfo {year} {2004})},\ \Eprint {http://arxiv.org/abs/hep-lat/0311018}
  {arXiv:hep-lat/0311018 [hep-lat]} \BibitemShut {NoStop}%
\bibitem [{\citenamefont {Del~Debbio}\ \emph {et~al.}(2013)\citenamefont
  {Del~Debbio}, \citenamefont {Patella},\ and\ \citenamefont
  {Rago}}]{DelDebbio:2013zaa}%
  \BibitemOpen
  \bibfield  {author} {\bibinfo {author} {\bibfnamefont {Luigi}\ \bibnamefont
  {Del~Debbio}}, \bibinfo {author} {\bibfnamefont {Agostino}\ \bibnamefont
  {Patella}}, \ and\ \bibinfo {author} {\bibfnamefont {Antonio}\ \bibnamefont
  {Rago}},\ }\bibfield  {title} {\enquote {\bibinfo {title} {{Space-time
  symmetries and the Yang-Mills gradient flow}},}\ }\href {\doibase
  10.1007/JHEP11(2013)212} {\bibfield  {journal} {\bibinfo  {journal} {JHEP}\
  }\textbf {\bibinfo {volume} {1311}},\ \bibinfo {pages} {212} (\bibinfo {year}
  {2013})},\ \Eprint {http://arxiv.org/abs/1306.1173} {arXiv:1306.1173
  [hep-th]} \BibitemShut {NoStop}%
\bibitem [{\citenamefont {Suzuki}(2013)}]{Suzuki:2013gza}%
  \BibitemOpen
  \bibfield  {author} {\bibinfo {author} {\bibfnamefont {Hiroshi}\ \bibnamefont
  {Suzuki}},\ }\bibfield  {title} {\enquote {\bibinfo {title}
  {{Energy--momentum tensor from the Yang-Mills gradient flow}},}\ }\href
  {\doibase 10.1093/ptep/ptt059} {\bibfield  {journal} {\bibinfo  {journal}
  {PTEP}\ }\textbf {\bibinfo {volume} {2013}},\ \bibinfo {pages} {083B03}
  (\bibinfo {year} {2013})},\ \Eprint {http://arxiv.org/abs/1304.0533}
  {arXiv:1304.0533 [hep-lat]} \BibitemShut {NoStop}%
\bibitem [{\citenamefont {Monahan}\ and\ \citenamefont
  {Orginos}(2013)}]{Monahan:2013lwa}%
  \BibitemOpen
  \bibfield  {author} {\bibinfo {author} {\bibfnamefont {Christopher}\
  \bibnamefont {Monahan}}\ and\ \bibinfo {author} {\bibfnamefont {Kostas}\
  \bibnamefont {Orginos}},\ }\bibfield  {title} {\enquote {\bibinfo {title}
  {{Finite volume renormalization scheme for fermionic operators}},}\
  }\href@noop {} {\bibfield  {journal} {\bibinfo  {journal} {PoS}\ }\textbf
  {\bibinfo {volume} {Lattice2013}},\ \bibinfo {pages} {443} (\bibinfo {year}
  {2013})},\ \Eprint {http://arxiv.org/abs/1311.2310} {arXiv:1311.2310
  [hep-lat]} \BibitemShut {NoStop}%
\bibitem [{\citenamefont {{L\"uscher}}(2014)}]{Luscher:2014kea}%
  \BibitemOpen
  \bibfield  {author} {\bibinfo {author} {\bibfnamefont {Martin}\ \bibnamefont
  {{L\"uscher}}},\ }\bibfield  {title} {\enquote {\bibinfo {title} {{Step
  scaling and the Yang-Mills gradient flow}},}\ }\href {\doibase
  10.1007/JHEP06(2014)105} {\bibfield  {journal} {\bibinfo  {journal} {JHEP}\
  }\textbf {\bibinfo {volume} {1406}},\ \bibinfo {pages} {105} (\bibinfo {year}
  {2014})},\ \Eprint {http://arxiv.org/abs/1404.5930} {arXiv:1404.5930
  [hep-lat]} \BibitemShut {NoStop}%
\bibitem [{\citenamefont {Endo}\ \emph {et~al.}(2015)\citenamefont {Endo},
  \citenamefont {Hieda}, \citenamefont {Miura},\ and\ \citenamefont
  {Suzuki}}]{Endo:2015iea}%
  \BibitemOpen
  \bibfield  {author} {\bibinfo {author} {\bibfnamefont {Tasuku}\ \bibnamefont
  {Endo}}, \bibinfo {author} {\bibfnamefont {Kenji}\ \bibnamefont {Hieda}},
  \bibinfo {author} {\bibfnamefont {Daiki}\ \bibnamefont {Miura}}, \ and\
  \bibinfo {author} {\bibfnamefont {Hiroshi}\ \bibnamefont {Suzuki}},\
  }\bibfield  {title} {\enquote {\bibinfo {title} {{Universal formula for the
  flavor non-singlet axial-vector current from the gradient flow}},}\ }\href
  {\doibase 10.1093/ptep/ptv058} {\bibfield  {journal} {\bibinfo  {journal}
  {PTEP}\ }\textbf {\bibinfo {volume} {2015}},\ \bibinfo {pages} {053B03}
  (\bibinfo {year} {2015})},\ \Eprint {http://arxiv.org/abs/1502.01809}
  {arXiv:1502.01809 [hep-lat]} \BibitemShut {NoStop}%
\bibitem [{\citenamefont {Monahan}\ and\ \citenamefont
  {Orginos}(2015)}]{Monahan:2015lha}%
  \BibitemOpen
  \bibfield  {author} {\bibinfo {author} {\bibfnamefont {Christopher}\
  \bibnamefont {Monahan}}\ and\ \bibinfo {author} {\bibfnamefont {Kostas}\
  \bibnamefont {Orginos}},\ }\bibfield  {title} {\enquote {\bibinfo {title}
  {{Locally smeared operator product expansions in scalar field theory}},}\
  }\href {\doibase 10.1103/PhysRevD.91.074513} {\bibfield  {journal} {\bibinfo
  {journal} {Phys. Rev.}\ }\textbf {\bibinfo {volume} {D91}},\ \bibinfo {pages}
  {074513} (\bibinfo {year} {2015})},\ \Eprint
  {http://arxiv.org/abs/1501.05348} {arXiv:1501.05348 [hep-lat]} \BibitemShut
  {NoStop}%
\bibitem [{\citenamefont {Monahan}\ and\ \citenamefont
  {Orginos}(2017)}]{Monahan:2016bvm}%
  \BibitemOpen
  \bibfield  {author} {\bibinfo {author} {\bibfnamefont {Chirstopher}\
  \bibnamefont {Monahan}}\ and\ \bibinfo {author} {\bibfnamefont {Kostas}\
  \bibnamefont {Orginos}},\ }\bibfield  {title} {\enquote {\bibinfo {title}
  {{Quasi parton distributions and the gradient flow}},}\ }\href {\doibase
  10.1007/JHEP03(2017)116} {\bibfield  {journal} {\bibinfo  {journal} {JHEP}\
  }\textbf {\bibinfo {volume} {03}},\ \bibinfo {pages} {116} (\bibinfo {year}
  {2017})},\ \Eprint {http://arxiv.org/abs/1612.01584} {arXiv:1612.01584
  [hep-lat]} \BibitemShut {NoStop}%
\bibitem [{\citenamefont {{L\"uscher}}(2010)}]{Luscher:2010iy}%
  \BibitemOpen
  \bibfield  {author} {\bibinfo {author} {\bibfnamefont {Martin}\ \bibnamefont
  {{L\"uscher}}},\ }\bibfield  {title} {\enquote {\bibinfo {title} {{Properties
  and uses of the Wilson flow in lattice QCD}},}\ }\href {\doibase
  10.1007/JHEP08(2010)071} {\bibfield  {journal} {\bibinfo  {journal} {JHEP}\
  }\textbf {\bibinfo {volume} {1008}},\ \bibinfo {pages} {071} (\bibinfo {year}
  {2010})},\ \Eprint {http://arxiv.org/abs/1006.4518} {arXiv:1006.4518
  [hep-lat]} \BibitemShut {NoStop}%
\bibitem [{\citenamefont {Lohmayer}\ and\ \citenamefont
  {Neuberger}(2011)}]{Lohmayer:2011si}%
  \BibitemOpen
  \bibfield  {author} {\bibinfo {author} {\bibfnamefont {Robert}\ \bibnamefont
  {Lohmayer}}\ and\ \bibinfo {author} {\bibfnamefont {Herbert}\ \bibnamefont
  {Neuberger}},\ }\bibfield  {title} {\enquote {\bibinfo {title} {{Continuous
  smearing of Wilson Loops}},}\ }\href@noop {} {\bibfield  {journal} {\bibinfo
  {journal} {PoS}\ }\textbf {\bibinfo {volume} {LATTICE2011}},\ \bibinfo
  {pages} {249} (\bibinfo {year} {2011})},\ \Eprint
  {http://arxiv.org/abs/1110.3522} {arXiv:1110.3522 [hep-lat]} \BibitemShut
  {NoStop}%
\bibitem [{\citenamefont {Syritsyn}\ and\ \citenamefont
  {Negele}(2007)}]{Syritsyn:2007mp}%
  \BibitemOpen
  \bibfield  {author} {\bibinfo {author} {\bibfnamefont {Sergey}\ \bibnamefont
  {Syritsyn}}\ and\ \bibinfo {author} {\bibfnamefont {John~W.}\ \bibnamefont
  {Negele}},\ }\bibfield  {title} {\enquote {\bibinfo {title} {{Oscillatory
  terms in the domain wall transfer matrix}},}\ }\bibfield  {booktitle} {\emph
  {\bibinfo {booktitle} {{Proceedings, 25th International Symposium on Lattice
  field theory (Lattice 2007): Regensburg, Germany, July 30-August 4, 2007}}},\
  }\href@noop {} {\bibfield  {journal} {\bibinfo  {journal} {PoS}\ }\textbf
  {\bibinfo {volume} {LAT2007}},\ \bibinfo {pages} {078} (\bibinfo {year}
  {2007})},\ \Eprint {http://arxiv.org/abs/0710.0425} {arXiv:0710.0425
  [hep-lat]} \BibitemShut {NoStop}%
\bibitem [{\citenamefont {Nicholson}\ \emph {et~al.}(2016)\citenamefont
  {Nicholson}, \citenamefont {Berkowitz}, \citenamefont {Chang}, \citenamefont
  {Clark}, \citenamefont {Joo}, \citenamefont {Kurth}, \citenamefont {Rinaldi},
  \citenamefont {Tiburzi}, \citenamefont {Vranas},\ and\ \citenamefont
  {Walker-Loud}}]{Nicholson:2016byl}%
  \BibitemOpen
  \bibfield  {author} {\bibinfo {author} {\bibfnamefont {Amy}\ \bibnamefont
  {Nicholson}}, \bibinfo {author} {\bibfnamefont {Evan}\ \bibnamefont
  {Berkowitz}}, \bibinfo {author} {\bibfnamefont {Chia~Cheng}\ \bibnamefont
  {Chang}}, \bibinfo {author} {\bibfnamefont {M.~A.}\ \bibnamefont {Clark}},
  \bibinfo {author} {\bibfnamefont {Balint}\ \bibnamefont {Joo}}, \bibinfo
  {author} {\bibfnamefont {Thorsten}\ \bibnamefont {Kurth}}, \bibinfo {author}
  {\bibfnamefont {Enrico}\ \bibnamefont {Rinaldi}}, \bibinfo {author}
  {\bibfnamefont {Brian}\ \bibnamefont {Tiburzi}}, \bibinfo {author}
  {\bibfnamefont {Pavlos}\ \bibnamefont {Vranas}}, \ and\ \bibinfo {author}
  {\bibfnamefont {Andre}\ \bibnamefont {Walker-Loud}},\ }\bibfield  {title}
  {\enquote {\bibinfo {title} {{Neutrinoless double beta decay from lattice
  QCD}},}\ }in\ \href
  {http://inspirehep.net/record/1481814/files/arXiv:1608.04793.pdf} {\emph
  {\bibinfo {booktitle} {{Proceedings, 34th International Symposium on Lattice
  Field Theory (Lattice 2016): Southampton, UK, July 24-30, 2016}}}}\ (\bibinfo
  {year} {2016})\ \Eprint {http://arxiv.org/abs/1608.04793} {arXiv:1608.04793
  [hep-lat]} \BibitemShut {NoStop}%
\bibitem [{\citenamefont {Bouchard}\ \emph {et~al.}(2017)\citenamefont
  {Bouchard}, \citenamefont {Chang}, \citenamefont {Kurth}, \citenamefont
  {Orginos},\ and\ \citenamefont {Walker-Loud}}]{Bouchard:2016heu}%
  \BibitemOpen
  \bibfield  {author} {\bibinfo {author} {\bibfnamefont {Chris}\ \bibnamefont
  {Bouchard}}, \bibinfo {author} {\bibfnamefont {Chia~Cheng}\ \bibnamefont
  {Chang}}, \bibinfo {author} {\bibfnamefont {Thorsten}\ \bibnamefont {Kurth}},
  \bibinfo {author} {\bibfnamefont {Kostas}\ \bibnamefont {Orginos}}, \ and\
  \bibinfo {author} {\bibfnamefont {Andre}\ \bibnamefont {Walker-Loud}},\
  }\bibfield  {title} {\enquote {\bibinfo {title} {{On the Feynman-Hellmann
  Theorem in Quantum Field Theory and the Calculation of Matrix Elements}},}\
  }\href {\doibase 10.1103/PhysRevD.96.014504} {\bibfield  {journal} {\bibinfo
  {journal} {Phys. Rev.}\ }\textbf {\bibinfo {volume} {D96}},\ \bibinfo {pages}
  {014504} (\bibinfo {year} {2017})},\ \Eprint
  {http://arxiv.org/abs/1612.06963} {arXiv:1612.06963 [hep-lat]} \BibitemShut
  {NoStop}%
\bibitem [{\citenamefont {Berkowitz}\ \emph {et~al.}(2017)\citenamefont
  {Berkowitz} \emph {et~al.}}]{Berkowitz:2017gql}%
  \BibitemOpen
  \bibfield  {author} {\bibinfo {author} {\bibfnamefont {Evan}\ \bibnamefont
  {Berkowitz}} \emph {et~al.},\ }\bibfield  {title} {\enquote {\bibinfo {title}
  {{An accurate calculation of the nucleon axial charge with lattice QCD}},}\
  }\href@noop {} {\  (\bibinfo {year} {2017})},\ \Eprint
  {http://arxiv.org/abs/1704.01114} {arXiv:1704.01114 [hep-lat]} \BibitemShut
  {NoStop}%
\bibitem [{\citenamefont {Basak}\ \emph
  {et~al.}(2005{\natexlab{a}})\citenamefont {Basak}, \citenamefont {Edwards},
  \citenamefont {Fleming}, \citenamefont {Heller}, \citenamefont {Morningstar},
  \citenamefont {Richards}, \citenamefont {Sato},\ and\ \citenamefont
  {Wallace}}]{Basak:2005aq}%
  \BibitemOpen
  \bibfield  {author} {\bibinfo {author} {\bibfnamefont {S.}~\bibnamefont
  {Basak}}, \bibinfo {author} {\bibfnamefont {R.~G.}\ \bibnamefont {Edwards}},
  \bibinfo {author} {\bibfnamefont {G.~T.}\ \bibnamefont {Fleming}}, \bibinfo
  {author} {\bibfnamefont {U.~M.}\ \bibnamefont {Heller}}, \bibinfo {author}
  {\bibfnamefont {C.}~\bibnamefont {Morningstar}}, \bibinfo {author}
  {\bibfnamefont {D.}~\bibnamefont {Richards}}, \bibinfo {author}
  {\bibfnamefont {I.}~\bibnamefont {Sato}}, \ and\ \bibinfo {author}
  {\bibfnamefont {S.}~\bibnamefont {Wallace}},\ }\bibfield  {title} {\enquote
  {\bibinfo {title} {{Group-theoretical construction of extended baryon
  operators in lattice QCD}},}\ }\href {\doibase 10.1103/PhysRevD.72.094506}
  {\bibfield  {journal} {\bibinfo  {journal} {Phys. Rev.}\ }\textbf {\bibinfo
  {volume} {D72}},\ \bibinfo {pages} {094506} (\bibinfo {year}
  {2005}{\natexlab{a}})},\ \Eprint {http://arxiv.org/abs/hep-lat/0506029}
  {arXiv:hep-lat/0506029 [hep-lat]} \BibitemShut {NoStop}%
\bibitem [{\citenamefont {Basak}\ \emph
  {et~al.}(2005{\natexlab{b}})\citenamefont {Basak}, \citenamefont {Edwards},
  \citenamefont {Fleming}, \citenamefont {Heller}, \citenamefont {Morningstar},
  \citenamefont {Richards}, \citenamefont {Sato},\ and\ \citenamefont
  {Wallace}}]{Basak:2005ir}%
  \BibitemOpen
  \bibfield  {author} {\bibinfo {author} {\bibfnamefont {Subhasish}\
  \bibnamefont {Basak}}, \bibinfo {author} {\bibfnamefont {Robert}\
  \bibnamefont {Edwards}}, \bibinfo {author} {\bibfnamefont {George~T.}\
  \bibnamefont {Fleming}}, \bibinfo {author} {\bibfnamefont {Urs~M.}\
  \bibnamefont {Heller}}, \bibinfo {author} {\bibfnamefont {Colin}\
  \bibnamefont {Morningstar}}, \bibinfo {author} {\bibfnamefont {David}\
  \bibnamefont {Richards}}, \bibinfo {author} {\bibfnamefont {Ikuro}\
  \bibnamefont {Sato}}, \ and\ \bibinfo {author} {\bibfnamefont {Stephen~J.}\
  \bibnamefont {Wallace}} (\bibinfo {collaboration} {Lattice Hadron Physics
  (LHPC)}),\ }\bibfield  {title} {\enquote {\bibinfo {title} {{Clebsch-Gordan
  construction of lattice interpolating fields for excited baryons}},}\ }\href
  {\doibase 10.1103/PhysRevD.72.074501} {\bibfield  {journal} {\bibinfo
  {journal} {Phys. Rev.}\ }\textbf {\bibinfo {volume} {D72}},\ \bibinfo {pages}
  {074501} (\bibinfo {year} {2005}{\natexlab{b}})},\ \Eprint
  {http://arxiv.org/abs/hep-lat/0508018} {arXiv:hep-lat/0508018 [hep-lat]}
  \BibitemShut {NoStop}%
\bibitem [{\citenamefont {Blum}\ \emph {et~al.}(2004)\citenamefont {Blum} \emph
  {et~al.}}]{Blum:2000kn}%
  \BibitemOpen
  \bibfield  {author} {\bibinfo {author} {\bibfnamefont {T.}~\bibnamefont
  {Blum}} \emph {et~al.},\ }\bibfield  {title} {\enquote {\bibinfo {title}
  {{Quenched lattice QCD with domain wall fermions and the chiral limit}},}\
  }\href {\doibase 10.1103/PhysRevD.69.074502} {\bibfield  {journal} {\bibinfo
  {journal} {Phys. Rev.}\ }\textbf {\bibinfo {volume} {D69}},\ \bibinfo {pages}
  {074502} (\bibinfo {year} {2004})},\ \Eprint
  {http://arxiv.org/abs/hep-lat/0007038} {arXiv:hep-lat/0007038 [hep-lat]}
  \BibitemShut {NoStop}%
\bibitem [{\citenamefont {Aoki}\ \emph {et~al.}(2004)\citenamefont {Aoki} \emph
  {et~al.}}]{Aoki:2002vt}%
  \BibitemOpen
  \bibfield  {author} {\bibinfo {author} {\bibfnamefont {Y.}~\bibnamefont
  {Aoki}} \emph {et~al.},\ }\bibfield  {title} {\enquote {\bibinfo {title}
  {{Domain wall fermions with improved gauge actions}},}\ }\href {\doibase
  10.1103/PhysRevD.69.074504} {\bibfield  {journal} {\bibinfo  {journal} {Phys.
  Rev.}\ }\textbf {\bibinfo {volume} {D69}},\ \bibinfo {pages} {074504}
  (\bibinfo {year} {2004})},\ \Eprint {http://arxiv.org/abs/hep-lat/0211023}
  {arXiv:hep-lat/0211023 [hep-lat]} \BibitemShut {NoStop}%
\bibitem [{\citenamefont {Lepage}(2016)}]{lsqfit}%
  \BibitemOpen
  \bibfield  {author} {\bibinfo {author} {\bibfnamefont {G.~Peter}\
  \bibnamefont {Lepage}},\ }\href {\doibase 10.5281/zenodo.60221} {\enquote
  {\bibinfo {title} {{\tt lsqfit} v8.1},}\ } (\bibinfo {year}
  {2016})\BibitemShut {NoStop}%
\bibitem [{\citenamefont {Bouchard}\ \emph {et~al.}(2014)\citenamefont
  {Bouchard}, \citenamefont {Lepage}, \citenamefont {Monahan}, \citenamefont
  {Na},\ and\ \citenamefont {Shigemitsu}}]{Bouchard:2014ypa}%
  \BibitemOpen
  \bibfield  {author} {\bibinfo {author} {\bibfnamefont {C.~M.}\ \bibnamefont
  {Bouchard}}, \bibinfo {author} {\bibfnamefont {G.~Peter}\ \bibnamefont
  {Lepage}}, \bibinfo {author} {\bibfnamefont {Christopher}\ \bibnamefont
  {Monahan}}, \bibinfo {author} {\bibfnamefont {Heechang}\ \bibnamefont {Na}},
  \ and\ \bibinfo {author} {\bibfnamefont {Junko}\ \bibnamefont {Shigemitsu}},\
  }\bibfield  {title} {\enquote {\bibinfo {title} {{$B_s \to K \ell \nu$ form
  factors from lattice QCD}},}\ }\href {\doibase 10.1103/PhysRevD.90.054506}
  {\bibfield  {journal} {\bibinfo  {journal} {Phys. Rev.}\ }\textbf {\bibinfo
  {volume} {D90}},\ \bibinfo {pages} {054506} (\bibinfo {year} {2014})},\
  \Eprint {http://arxiv.org/abs/1406.2279} {arXiv:1406.2279 [hep-lat]}
  \BibitemShut {NoStop}%
\bibitem [{\citenamefont {Borsanyi}\ \emph {et~al.}(2012)\citenamefont
  {Borsanyi} \emph {et~al.}}]{Borsanyi:2012zs}%
  \BibitemOpen
  \bibfield  {author} {\bibinfo {author} {\bibfnamefont {Szabolcs}\
  \bibnamefont {Borsanyi}} \emph {et~al.},\ }\bibfield  {title} {\enquote
  {\bibinfo {title} {{High-precision scale setting in lattice QCD}},}\ }\href
  {\doibase 10.1007/JHEP09(2012)010} {\bibfield  {journal} {\bibinfo  {journal}
  {JHEP}\ }\textbf {\bibinfo {volume} {09}},\ \bibinfo {pages} {010} (\bibinfo
  {year} {2012})},\ \Eprint {http://arxiv.org/abs/1203.4469} {arXiv:1203.4469
  [hep-lat]} \BibitemShut {NoStop}%
\bibitem [{\citenamefont {Bazavov}\ \emph
  {et~al.}(2016{\natexlab{a}})\citenamefont {Bazavov} \emph
  {et~al.}}]{Bazavov:2015yea}%
  \BibitemOpen
  \bibfield  {author} {\bibinfo {author} {\bibfnamefont {A.}~\bibnamefont
  {Bazavov}} \emph {et~al.} (\bibinfo {collaboration} {MILC}),\ }\bibfield
  {title} {\enquote {\bibinfo {title} {{Gradient flow and scale setting on MILC
  HISQ ensembles}},}\ }\href {\doibase 10.1103/PhysRevD.93.094510} {\bibfield
  {journal} {\bibinfo  {journal} {Phys. Rev.}\ }\textbf {\bibinfo {volume}
  {D93}},\ \bibinfo {pages} {094510} (\bibinfo {year} {2016}{\natexlab{a}})},\
  \Eprint {http://arxiv.org/abs/1503.02769} {arXiv:1503.02769 [hep-lat]}
  \BibitemShut {NoStop}%
\bibitem [{\citenamefont {Albanese}\ \emph {et~al.}(1987)\citenamefont
  {Albanese} \emph {et~al.}}]{Albanese:1987ds}%
  \BibitemOpen
  \bibfield  {author} {\bibinfo {author} {\bibfnamefont {M.}~\bibnamefont
  {Albanese}} \emph {et~al.} (\bibinfo {collaboration} {APE}),\ }\bibfield
  {title} {\enquote {\bibinfo {title} {{Glueball Masses and String Tension in
  Lattice QCD}},}\ }\href {\doibase 10.1016/0370-2693(87)91160-9} {\bibfield
  {journal} {\bibinfo  {journal} {Phys. Lett.}\ }\textbf {\bibinfo {volume}
  {B192}},\ \bibinfo {pages} {163--169} (\bibinfo {year} {1987})}\BibitemShut
  {NoStop}%
\bibitem [{\citenamefont {Bernard}\ \emph {et~al.}(2002)\citenamefont
  {Bernard}, \citenamefont {Datta}, \citenamefont {DeGrand}, \citenamefont
  {DeTar}, \citenamefont {Gottlieb}, \citenamefont {Heller}, \citenamefont
  {McNeile}, \citenamefont {Orginos}, \citenamefont {Sugar},\ and\
  \citenamefont {Toussaint}}]{Bernard:2002pc}%
  \BibitemOpen
  \bibfield  {author} {\bibinfo {author} {\bibfnamefont {C.}~\bibnamefont
  {Bernard}}, \bibinfo {author} {\bibfnamefont {S.}~\bibnamefont {Datta}},
  \bibinfo {author} {\bibfnamefont {Thomas~A.}\ \bibnamefont {DeGrand}},
  \bibinfo {author} {\bibfnamefont {Carleton~E.}\ \bibnamefont {DeTar}},
  \bibinfo {author} {\bibfnamefont {Steven~A.}\ \bibnamefont {Gottlieb}},
  \bibinfo {author} {\bibfnamefont {Urs~M.}\ \bibnamefont {Heller}}, \bibinfo
  {author} {\bibfnamefont {C.}~\bibnamefont {McNeile}}, \bibinfo {author}
  {\bibfnamefont {K.}~\bibnamefont {Orginos}}, \bibinfo {author} {\bibfnamefont
  {R.}~\bibnamefont {Sugar}}, \ and\ \bibinfo {author} {\bibfnamefont
  {D.}~\bibnamefont {Toussaint}} (\bibinfo {collaboration} {MILC}),\ }\bibfield
   {title} {\enquote {\bibinfo {title} {{Lattice calculation of heavy light
  decay constants with two flavors of dynamical quarks}},}\ }\href {\doibase
  10.1103/PhysRevD.66.094501} {\bibfield  {journal} {\bibinfo  {journal} {Phys.
  Rev.}\ }\textbf {\bibinfo {volume} {D66}},\ \bibinfo {pages} {094501}
  (\bibinfo {year} {2002})},\ \Eprint {http://arxiv.org/abs/hep-lat/0206016}
  {arXiv:hep-lat/0206016 [hep-lat]} \BibitemShut {NoStop}%
\bibitem [{\citenamefont {Frommer}\ \emph {et~al.}(1995)\citenamefont
  {Frommer}, \citenamefont {Nockel}, \citenamefont {Gusken}, \citenamefont
  {Lippert},\ and\ \citenamefont {Schilling}}]{Frommer:1995ik}%
  \BibitemOpen
  \bibfield  {author} {\bibinfo {author} {\bibfnamefont {Andreas}\ \bibnamefont
  {Frommer}}, \bibinfo {author} {\bibfnamefont {Bertold}\ \bibnamefont
  {Nockel}}, \bibinfo {author} {\bibfnamefont {Stephan}\ \bibnamefont
  {Gusken}}, \bibinfo {author} {\bibfnamefont {Thomas}\ \bibnamefont
  {Lippert}}, \ and\ \bibinfo {author} {\bibfnamefont {Klaus}\ \bibnamefont
  {Schilling}},\ }\bibfield  {title} {\enquote {\bibinfo {title} {{Many masses
  on one stroke: Economic computation of quark propagators}},}\ }\href
  {\doibase 10.1142/S0129183195000538} {\bibfield  {journal} {\bibinfo
  {journal} {Int. J. Mod. Phys.}\ }\textbf {\bibinfo {volume} {C6}},\ \bibinfo
  {pages} {627--638} (\bibinfo {year} {1995})},\ \Eprint
  {http://arxiv.org/abs/hep-lat/9504020} {arXiv:hep-lat/9504020 [hep-lat]}
  \BibitemShut {NoStop}%
\bibitem [{\citenamefont {Gasser}\ and\ \citenamefont
  {Leutwyler}(1985)}]{Gasser:1984gg}%
  \BibitemOpen
  \bibfield  {author} {\bibinfo {author} {\bibfnamefont {J.}~\bibnamefont
  {Gasser}}\ and\ \bibinfo {author} {\bibfnamefont {H.}~\bibnamefont
  {Leutwyler}},\ }\bibfield  {title} {\enquote {\bibinfo {title} {{Chiral
  Perturbation Theory: Expansions in the Mass of the Strange Quark}},}\ }\href
  {\doibase 10.1016/0550-3213(85)90492-4} {\bibfield  {journal} {\bibinfo
  {journal} {Nucl.Phys.}\ }\textbf {\bibinfo {volume} {B250}},\ \bibinfo
  {pages} {465} (\bibinfo {year} {1985})}\BibitemShut {NoStop}%
\bibitem [{\citenamefont {Colangelo}\ \emph {et~al.}(2005)\citenamefont
  {Colangelo}, \citenamefont {Dürr},\ and\ \citenamefont
  {Haefeli}}]{COLANGELO2005136}%
  \BibitemOpen
  \bibfield  {author} {\bibinfo {author} {\bibfnamefont {Gilberto}\
  \bibnamefont {Colangelo}}, \bibinfo {author} {\bibfnamefont {Stephan}\
  \bibnamefont {Dürr}}, \ and\ \bibinfo {author} {\bibfnamefont {Christoph}\
  \bibnamefont {Haefeli}},\ }\bibfield  {title} {\enquote {\bibinfo {title}
  {Finite volume effects for meson masses and decay constants},}\ }\href
  {\doibase http://dx.doi.org/10.1016/j.nuclphysb.2005.05.015} {\bibfield
  {journal} {\bibinfo  {journal} {Nuclear Physics B}\ }\textbf {\bibinfo
  {volume} {721}},\ \bibinfo {pages} {136 -- 174} (\bibinfo {year}
  {2005})}\BibitemShut {NoStop}%
\bibitem [{\citenamefont {Durr}\ \emph {et~al.}(2010)\citenamefont {Durr},
  \citenamefont {Fodor}, \citenamefont {Hoelbling}, \citenamefont {Katz},
  \citenamefont {Krieg}, \citenamefont {Kurth}, \citenamefont {Lellouch},
  \citenamefont {Lippert}, \citenamefont {Ramos},\ and\ \citenamefont
  {Szabo}}]{Durr:2010hr}%
  \BibitemOpen
  \bibfield  {author} {\bibinfo {author} {\bibfnamefont {S.}~\bibnamefont
  {Durr}}, \bibinfo {author} {\bibfnamefont {Z.}~\bibnamefont {Fodor}},
  \bibinfo {author} {\bibfnamefont {C.}~\bibnamefont {Hoelbling}}, \bibinfo
  {author} {\bibfnamefont {S.~D.}\ \bibnamefont {Katz}}, \bibinfo {author}
  {\bibfnamefont {S.}~\bibnamefont {Krieg}}, \bibinfo {author} {\bibfnamefont
  {T.}~\bibnamefont {Kurth}}, \bibinfo {author} {\bibfnamefont
  {L.}~\bibnamefont {Lellouch}}, \bibinfo {author} {\bibfnamefont
  {T.}~\bibnamefont {Lippert}}, \bibinfo {author} {\bibfnamefont
  {A.}~\bibnamefont {Ramos}}, \ and\ \bibinfo {author} {\bibfnamefont {K.~K.}\
  \bibnamefont {Szabo}},\ }\bibfield  {title} {\enquote {\bibinfo {title} {{The
  ratio FK/Fpi in QCD}},}\ }\href {\doibase 10.1103/PhysRevD.81.054507}
  {\bibfield  {journal} {\bibinfo  {journal} {Phys. Rev.}\ }\textbf {\bibinfo
  {volume} {D81}},\ \bibinfo {pages} {054507} (\bibinfo {year} {2010})},\
  \Eprint {http://arxiv.org/abs/1001.4692} {arXiv:1001.4692 [hep-lat]}
  \BibitemShut {NoStop}%
\bibitem [{\citenamefont {Bazavov}\ \emph
  {et~al.}(2016{\natexlab{b}})\citenamefont {Bazavov} \emph
  {et~al.}}]{Bazavov:2016nty}%
  \BibitemOpen
  \bibfield  {author} {\bibinfo {author} {\bibfnamefont {A.}~\bibnamefont
  {Bazavov}} \emph {et~al.} (\bibinfo {collaboration} {Fermilab Lattice,
  MILC}),\ }\bibfield  {title} {\enquote {\bibinfo {title} {{$B^0_{(s)}$-mixing
  matrix elements from lattice QCD for the Standard Model and beyond}},}\
  }\href {\doibase 10.1103/PhysRevD.93.113016} {\bibfield  {journal} {\bibinfo
  {journal} {Phys. Rev.}\ }\textbf {\bibinfo {volume} {D93}},\ \bibinfo {pages}
  {113016} (\bibinfo {year} {2016}{\natexlab{b}})},\ \Eprint
  {http://arxiv.org/abs/1602.03560} {arXiv:1602.03560 [hep-lat]} \BibitemShut
  {NoStop}%
\bibitem [{\citenamefont {Kim}\ and\ \citenamefont
  {Izubuchi}(2014)}]{Kim:2014mpa}%
  \BibitemOpen
  \bibfield  {author} {\bibinfo {author} {\bibfnamefont {Hyung-Jin}\
  \bibnamefont {Kim}}\ and\ \bibinfo {author} {\bibfnamefont {T.}~\bibnamefont
  {Izubuchi}},\ }\bibfield  {title} {\enquote {\bibinfo {title} {{M\"{o}bius
  domain wall fermion method on QUDA}},}\ }\bibfield  {booktitle} {\emph
  {\bibinfo {booktitle} {{Proceedings, 31st International Symposium on Lattice
  Field Theory (Lattice 2013): Mainz, Germany, July 29-August 3, 2013}}},\
  }\href@noop {} {\bibfield  {journal} {\bibinfo  {journal} {PoS}\ }\textbf
  {\bibinfo {volume} {LATTICE2013}},\ \bibinfo {pages} {033} (\bibinfo {year}
  {2014})}\BibitemShut {NoStop}%
\bibitem [{\citenamefont {Clark}\ \emph {et~al.}(2010)\citenamefont {Clark},
  \citenamefont {Babich}, \citenamefont {Barros}, \citenamefont {Brower},\ and\
  \citenamefont {Rebbi}}]{Clark:2009wm}%
  \BibitemOpen
  \bibfield  {author} {\bibinfo {author} {\bibfnamefont {M.A.}\ \bibnamefont
  {Clark}}, \bibinfo {author} {\bibfnamefont {R.}~\bibnamefont {Babich}},
  \bibinfo {author} {\bibfnamefont {K.}~\bibnamefont {Barros}}, \bibinfo
  {author} {\bibfnamefont {R.C.}\ \bibnamefont {Brower}}, \ and\ \bibinfo
  {author} {\bibfnamefont {C.}~\bibnamefont {Rebbi}},\ }\bibfield  {title}
  {\enquote {\bibinfo {title} {{Solving Lattice QCD systems of equations using
  mixed precision solvers on GPUs}},}\ }\href {\doibase
  10.1016/j.cpc.2010.05.002} {\bibfield  {journal} {\bibinfo  {journal}
  {Comput.Phys.Commun.}\ }\textbf {\bibinfo {volume} {181}},\ \bibinfo {pages}
  {1517--1528} (\bibinfo {year} {2010})},\ \Eprint
  {http://arxiv.org/abs/0911.3191} {arXiv:0911.3191 [hep-lat]} \BibitemShut
  {NoStop}%
\bibitem [{\citenamefont {Babich}\ \emph {et~al.}(2011)\citenamefont {Babich},
  \citenamefont {Clark}, \citenamefont {Joo}, \citenamefont {Shi},
  \citenamefont {Brower} \emph {et~al.}}]{Babich:2011np}%
  \BibitemOpen
  \bibfield  {author} {\bibinfo {author} {\bibfnamefont {R.}~\bibnamefont
  {Babich}}, \bibinfo {author} {\bibfnamefont {M.A.}\ \bibnamefont {Clark}},
  \bibinfo {author} {\bibfnamefont {B.}~\bibnamefont {Joo}}, \bibinfo {author}
  {\bibfnamefont {G.}~\bibnamefont {Shi}}, \bibinfo {author} {\bibfnamefont
  {R.C.}\ \bibnamefont {Brower}},  \emph {et~al.},\ }\bibfield  {title}
  {\enquote {\bibinfo {title} {{Scaling Lattice QCD beyond 100 GPUs}},}\
  }\href@noop {} {\  (\bibinfo {year} {2011})},\ \Eprint
  {http://arxiv.org/abs/1109.2935} {arXiv:1109.2935 [hep-lat]} \BibitemShut
  {NoStop}%
\bibitem [{\citenamefont {Blum}\ \emph {et~al.}()\citenamefont {Blum},
  \citenamefont {Izubuchi},\ and\ \citenamefont {Jung}}]{jung:2015latt}%
  \BibitemOpen
  \bibfield  {author} {\bibinfo {author} {\bibfnamefont {T.}~\bibnamefont
  {Blum}}, \bibinfo {author} {\bibfnamefont {T.}~\bibnamefont {Izubuchi}}, \
  and\ \bibinfo {author} {\bibfnamefont {C.}~\bibnamefont {Jung}},\ }\bibfield
  {title} {\enquote {\bibinfo {title} {{zM\"{o}bius and other recent
  developments on Domain Wall Fermions}},}\ }\bibfield  {booktitle} {\emph
  {\bibinfo {booktitle} {{private communication}}},\ }\href@noop {} {\
  }\BibitemShut {NoStop}%
\bibitem [{\citenamefont {{The HDF Group}}(1997-NNNN)}]{hdf5}%
  \BibitemOpen
  \bibfield  {author} {\bibinfo {author} {\bibnamefont {{The HDF Group}}},\
  }\href@noop {} {\enquote {\bibinfo {title} {{Hierarchical Data Format,
  version 5}},}\ } (\bibinfo {year} {1997-NNNN}),\ \bibinfo {note}
  {http://www.hdfgroup.org/HDF5/}\BibitemShut {NoStop}%
\bibitem [{\citenamefont {Kurth}\ \emph {et~al.}(2015)\citenamefont {Kurth},
  \citenamefont {Pochinsky}, \citenamefont {Sarje}, \citenamefont {Syritsyn},\
  and\ \citenamefont {Walker-Loud}}]{Kurth:2015mqa}%
  \BibitemOpen
  \bibfield  {author} {\bibinfo {author} {\bibfnamefont {Thorsten}\
  \bibnamefont {Kurth}}, \bibinfo {author} {\bibfnamefont {Andrew}\
  \bibnamefont {Pochinsky}}, \bibinfo {author} {\bibfnamefont {Abhinav}\
  \bibnamefont {Sarje}}, \bibinfo {author} {\bibfnamefont {Sergey}\
  \bibnamefont {Syritsyn}}, \ and\ \bibinfo {author} {\bibfnamefont {Andre}\
  \bibnamefont {Walker-Loud}},\ }\bibfield  {title} {\enquote {\bibinfo {title}
  {{High-Performance I/O: HDF5 for Lattice QCD}},}\ }\bibfield  {booktitle}
  {\emph {\bibinfo {booktitle} {{Proceedings, 32nd International Symposium on
  Lattice Field Theory (Lattice 2014): Brookhaven, NY, USA, June 23-28,
  2014}}},\ }\href@noop {} {\bibfield  {journal} {\bibinfo  {journal} {PoS}\
  }\textbf {\bibinfo {volume} {LATTICE2014}},\ \bibinfo {pages} {045} (\bibinfo
  {year} {2015})},\ \Eprint {http://arxiv.org/abs/1501.06992} {arXiv:1501.06992
  [hep-lat]} \BibitemShut {NoStop}%
\bibitem [{\citenamefont {Berkowitz}(2016)}]{berkowitz.metaq}%
  \BibitemOpen
  \bibfield  {author} {\bibinfo {author} {\bibfnamefont {Evan}\ \bibnamefont
  {Berkowitz}},\ }\href@noop {} {\enquote {\bibinfo {title} {\texttt{METAQ}},}\
  }\bibinfo {howpublished} {\url{https://github.com/evanberkowitz/metaq}}
  (\bibinfo {year} {2016})\BibitemShut {NoStop}%
\end{thebibliography}%


\end{document}